\documentclass[acmsmall,screen=true,anonymous=false,bookmarks=true]{acmart}

\usepackage{enumitem}
\setlist{leftmargin=5.08mm}

\usepackage{graphicx}
\usepackage{amsmath}
\usepackage{mathtools}
\usepackage{comment}
\usepackage{soul}
\usepackage[subrefformat=parens,labelformat=parens]{subfig}

\usepackage{bm,bbm}

\usepackage{amssymb}
\usepackage{multirow}
\usepackage{threeparttable,booktabs}
\usepackage{placeins}
\usepackage{flafter}
\usepackage{tabularx}
\usepackage{rotating}
\usepackage{longtable}
\usepackage{tikz}
\usepackage{pgfplots}
\usepackage{fancyvrb}
\pgfplotsset{compat=1.18}
\usepackage{forest}
\usetikzlibrary{arrows.meta,positioning,fit,calc,backgrounds,shapes.geometric}
\usepackage{xcolor}

\definecolor{sysblue}{HTML}{4F81BD}
\definecolor{sysblue2}{HTML}{DCE6F1}
\definecolor{deepblue}{HTML}{2F5597}
\definecolor{memgreen}{HTML}{70AD47}
\definecolor{memgreen2}{HTML}{E2F0D9}
\definecolor{govpurple}{HTML}{8064A2}
\definecolor{govpurple2}{HTML}{E4DFEC}
\definecolor{lightgray2}{HTML}{F3F3F3}
\definecolor{bordergray}{HTML}{C8C8C8}
\definecolor{textdark}{HTML}{333333}

\tikzset{
  titlebox/.style={font=\bfseries\large, text=textdark, align=left},
  subtitlebox/.style={font=\bfseries\normalsize, text=textdark, align=center},
  smalltitle/.style={font=\bfseries\small, text=textdark, align=center},
  bodytext/.style={font=\small, text=textdark, align=left},
  noteit/.style={font=\itshape\small, text=textdark, align=center},
  mainbox/.style={draw=bordergray, rounded corners=3mm, very thick, fill=white},
  softbox/.style={draw=bordergray, rounded corners=2.5mm, thick, fill=lightgray2},
  stage1/.style={draw=bordergray, rounded corners=3mm, thick, fill=sysblue2},
  stage2/.style={draw=bordergray, rounded corners=3mm, thick, fill=blue!10},
  stage3/.style={draw=bordergray, rounded corners=3mm, thick, fill=blue!22},
  stage4/.style={draw=bordergray, rounded corners=3mm, thick, fill=deepblue, text=white},
  layerblue/.style={draw=bordergray, rounded corners=3mm, thick, fill=sysblue2},
  layerdeep/.style={draw=bordergray, rounded corners=3mm, thick, fill=deepblue, text=white},
  layergreen/.style={draw=bordergray, rounded corners=3mm, thick, fill=memgreen2},
  layerpurple/.style={draw=bordergray, rounded corners=3mm, thick, fill=govpurple2},
  layergray/.style={draw=bordergray, rounded corners=3mm, thick, fill=lightgray2},
  sidepanel/.style={draw=bordergray, rounded corners=3mm, thick, fill=white},
  metricleft/.style={draw=bordergray, rounded corners=2.5mm, thick, fill=sysblue2},
  metricright/.style={draw=bordergray, rounded corners=2.5mm, thick, fill=deepblue, text=white},
  heatcellhi/.style={draw=white, rounded corners=1.8mm, fill=sysblue},
  heatcellmid/.style={draw=white, rounded corners=1.8mm, fill=blue!35},
  heatcelllow/.style={draw=white, rounded corners=1.8mm, fill=blue!12},
  heatcellsparse/.style={draw=white, rounded corners=1.8mm, fill=gray!15},
  flowbox/.style={draw=bordergray, rounded corners=2.8mm, thick, fill=white, minimum width=3.4cm, minimum height=1.15cm},
  pill/.style={draw=bordergray, rounded corners=5mm, thick, fill=lightgray2, inner sep=4pt, font=\small\bfseries},
  arr/.style={-{Latex[length=3mm,width=2mm]}, thick, draw=textdark},
  arr2/.style={-{Latex[length=3mm,width=2mm]}, very thick, draw=textdark},
}

\usepackage{balance}
\usepackage{courier}
\usepackage{cleveref}
\usepackage{xcolor,colortbl}
\usepackage{adjustbox}
\usepackage{makecell}
\usepackage{color}
\usepackage{diagbox}
\usepackage{pifont}
\usepackage{fontawesome}
\usepackage{array}
\usepackage{ragged2e}
\usepackage{needspace}

\newcolumntype{Y}{>{\raggedright\arraybackslash}X}
\newcolumntype{L}[1]{>{\RaggedRight\arraybackslash}p{#1}}
\newcolumntype{C}[1]{>{\centering\arraybackslash}p{#1}}

\definecolor{tableheader}{HTML}{FFFFFF}
\definecolor{tablegroup}{HTML}{FFFFFF}
\newenvironment{surveytable}{%
  \begingroup
  \scriptsize
  \setlength{\tabcolsep}{3pt}%
  \renewcommand{\arraystretch}{1.10}%
  \setlength{\emergencystretch}{1em}%
  \setlength{\LTpre}{4pt}%
  \setlength{\LTpost}{7pt}%
}{\endgroup}
\newenvironment{objectroletable}{%
  \begin{surveytable}%
  \fontsize{6.7pt}{7.25pt}\selectfont
  \setlength{\tabcolsep}{0.75pt}%
  \renewcommand{\arraystretch}{1.00}%
  \setlength{\LTpre}{2pt}%
  \setlength{\LTpost}{5pt}%
}{\end{surveytable}}

\newcommand{\tablegroup}[2]{%
  \rowcolor{tablegroup}\multicolumn{#1}{@{}l}{\strut\textbf{#2}}\\%
}

\usepackage[skip=1pt]{caption}
\newcommand{\revise}[1]{#1}
\newcommand{\minisection}[1]{\vspace{.06in}\noindent{\textbf{#1}.}}
\newcommand{\cmark}{\ding{51}}
\newcommand{\githubmark}{\textcolor{textdark}{\faGithub}}

\newcommand{\twopanelgraphic}[1]{%
  \begin{tikzpicture}
    \node[inner sep=0,outer sep=0] (panelimage)
      {#1};
    \node[anchor=north,font=\small] at
      ($(panelimage.south west)!0.25!(panelimage.south east)+(0,-1.5pt)$) {(a)};
    \node[anchor=north,font=\small] at
      ($(panelimage.south west)!0.75!(panelimage.south east)+(0,-1.5pt)$) {(b)};
  \end{tikzpicture}%
}

\theoremstyle{definition}

\definecolor{CUHKorange}{RGB}{244,106,18}
\definecolor{CUHKblue}{RGB}{0,111,190}
\definecolor{CUHKgreen}{RGB}{0,127,128}
\definecolor{CUHKred}{RGB}{228,46,36}
\definecolor{lightgray}{RGB}{230,230,230}
\definecolor{lightblue}{RGB}{219,234,254}
\definecolor{lightyellow}{RGB}{254,249,219}
\definecolor{lightgreen}{RGB}{220,252,231}

\usepackage{tcolorbox}
\tcbuselibrary{skins,breakable}
    {\endtcolorbox}

\definecolor{CUHKmiddle}{RGB}{144,44,144}   

    {\endtcolorbox}

\graphicspath{{./figs/}}
\setcopyright{acmcopyright}

\usepackage{diagbox}
\usepackage{float}
\begin{document}

\onecolumn
\setcounter{page}{1}

\title{%
    Agentic Electronic Design Automation: A Handoff Perspective%
}
\author{Jiawei Liu} \affiliation{\institution{The Chinese University of Hong Kong}\country{Hong Kong}}
\author{Peiyi Han}  \affiliation{\institution{The Chinese University of Hong Kong}\country{Hong Kong}}
\author{Yuntao Lu}  \affiliation{\institution{The Chinese University of Hong Kong}\country{Hong Kong}}
\author{Su Zheng}   \affiliation{\institution{The Chinese University of Hong Kong}\country{Hong Kong}}
\author{Fengyu Yan} \affiliation{\institution{Primarius Technologies}\country{China}}
\author{Bei Yu}     \affiliation{\institution{The Chinese University of Hong Kong}\country{Hong Kong}}
\begin{abstract}
Electronic design automation (EDA) is multi-stage and handoff-heavy, relying on transfers among humans, design artifacts, and multiple tools. LLM-based agents now participate in these transfers, yet the resulting research landscape is highly fragmented and lacks a unified perspective. This survey adopts the primary input--output pair as its organizing lens: for each agent system, we analyze its primary, supporting, and intermediate handoff objects and classify it according to the provenance of its primary input and the consumer boundary of its primary output. Intra-stage systems operate on objects within a single EDA stage, inter-stage systems transform EDA-native artifacts into forms usable by downstream stages, and extra-stage systems translate human intent into EDA artifacts. Based on this taxonomy, we survey 115 representative systems and examine them along multiple dimensions, including research trends, benchmarks, and model mechanisms. Finally, we outline an agentic EDA protocol roadmap and discuss open problems for future research.
\end{abstract}



\maketitle
\thispagestyle{plain}
\pagestyle{plain}

\section{Introduction}
\label{sec:intro}

\subsection{EDA Workflows Are Handoff-Heavy}
\label{sec:intro-handoff}

Electronic design automation (EDA) is an inherently multi-stage, handoff-heavy process. Digital designs, for example, progress through synthesis, placement, routing, and iterative timing closure~\cite{lin2019dreamplace,liu2023concurrent}, while analog designs are evaluated through schematic-level and post-layout simulation and checked by design rule checking (DRC) and layout-versus-schematic (LVS) verification. In each handoff, one or more objects are transferred from a producer to a consumer to support a subsequent engineering action. Those objects play different roles in the complete task: a principal design or verification object may be the primary input or output, constraints and collateral may be supporting inputs or outputs, and scripts, plans, reports, or checkpoints may be intermediate artifacts. Every transferred object and its metadata must remain mutually consistent across tool invocations~\cite{wu2024chateda,ghose2025orfs}. For instance, a place-and-route task may treat a netlist as its primary input, SDC, technology, and library data as supporting inputs, the implemented layout as its primary output, signoff reports as supporting outputs, and stage checkpoints as intermediate artifacts. Even when the primary design artifact is present, a stale SDC file, an incompatible library version, or an incomplete checkpoint can cause execution failures, timing violations, or non-reproducible results.

\begin{table*}[t]
\centering
\caption{Publicly announced agentic EDA offerings summarized from vendor press releases and product briefs.}
\label{tab:industry-signals}
\small
\begin{tabularx}{\textwidth}{p{0.16\textwidth} X}
\toprule
\textbf{Vendor} & \textbf{Offering and Stated Scope} \\
\midrule
Synopsys & The Synopsys agentic AI stack~\cite{synopsys_agentic_ai_2026} provides the infrastructure for agentic engineering workflows. AgentEngineer~\cite{synopsys_engineering_future_2026} supports agents that reason, plan, learn, execute, and coordinate across engineering tasks. A disclosed AgentEngineer workflow generates RTL from natural-language and formal specifications, runs lint checks, generates unit-level testbenches, and iteratively invokes EDA verification tools.\\
Cadence & ChipStack AI Super Agent~\cite{cadence_chipstack_pb_2026} and AgentStack~\cite{cadence_nvidia_agentstack_2026}: ChipStack supports agentic RTL design and verification workflows, including testbench generation, regression orchestration, debugging, and automated repair. AgentStack is a head-agent orchestration layer covering physical design, custom and analog design, migration, and system-level workflows.\\
Siemens EDA & Fuse EDA AI Agent~\cite{siemens_fuse_agent_2026} and Questa One Agentic Toolkit~\cite{siemens_questa_agentic_toolkit_2026}: Fuse provides multi-tool and multi-agent orchestration across semiconductor, 3D IC, PCB, verification, and manufacturing-signoff workflows. Questa One provides agentic workflows for RTL creation, lint, CDC analysis, verification planning, debugging, and RTL signoff, with integration into Fuse.\\
\bottomrule
\end{tabularx}
\end{table*}

\revise{\minisection{From Conventional EDA to ML-Assisted EDA and Agentic EDA}}
Conventional EDA organizes design as a sequence of deterministic algorithms, scripted tool invocations, and engineer-reviewed stage handoffs. Engineers author constraints, launch tools, inspect reports, and revise artifacts, while workflow intent and acceptance decisions remain largely encoded in scripts and expert practice. Machine learning for EDA (ML4EDA) has improved the quality and efficiency of individual tasks, including placement~\cite{lin2019dreamplace}, timing prediction and optimization~\cite{liu2023concurrent}, and HLS directive search~\cite{sun2022correlated}. These methods can be embedded directly into EDA flows, and their outputs may be consumed by downstream tools. However, they typically operate under predefined task interfaces, data schemas, technology settings, and optimization objectives. Recent large language models (LLMs) provide instruction-conditioned generation, in-context adaptation, and the ability to integrate heterogeneous information from specifications, code, logs, and tool feedback~\cite{he2025large,pan2025survey}.

Despite these capabilities, early LLM4EDA work was confined to advisory uses, such as single-turn dialogue and question answering. More recently, advances in LLM-based agent technology enable agents to plan, invoke EDA tools, and revise their actions from feedback, thereby performing specialized tasks such as repairing hardware descriptions~\cite{tsai2024rtlfixer,niu2025rechisel} and automating and optimizing multi-stage EDA flows~\cite{wu2024chateda,ghose2025orfs}. This paradigm is referred to as \textbf{agentic EDA}, and vendor materials describe related commercial products, as summarized in \Cref{tab:industry-signals}.

For agentic EDA systems, a handoff occurs when an agent delivers one or more objects to a consumer for a subsequent engineering action. The introduction of agents makes such transfers more frequent than in conventional EDA systems. Because LLM outputs may be incorrect or incomplete, the receiving tool or engineer must decide whether the transferred objects satisfy the applicable interface, semantic, configuration, and version requirements.

\begin{figure*}[t]
    \centering
    \includegraphics[width=\textwidth]{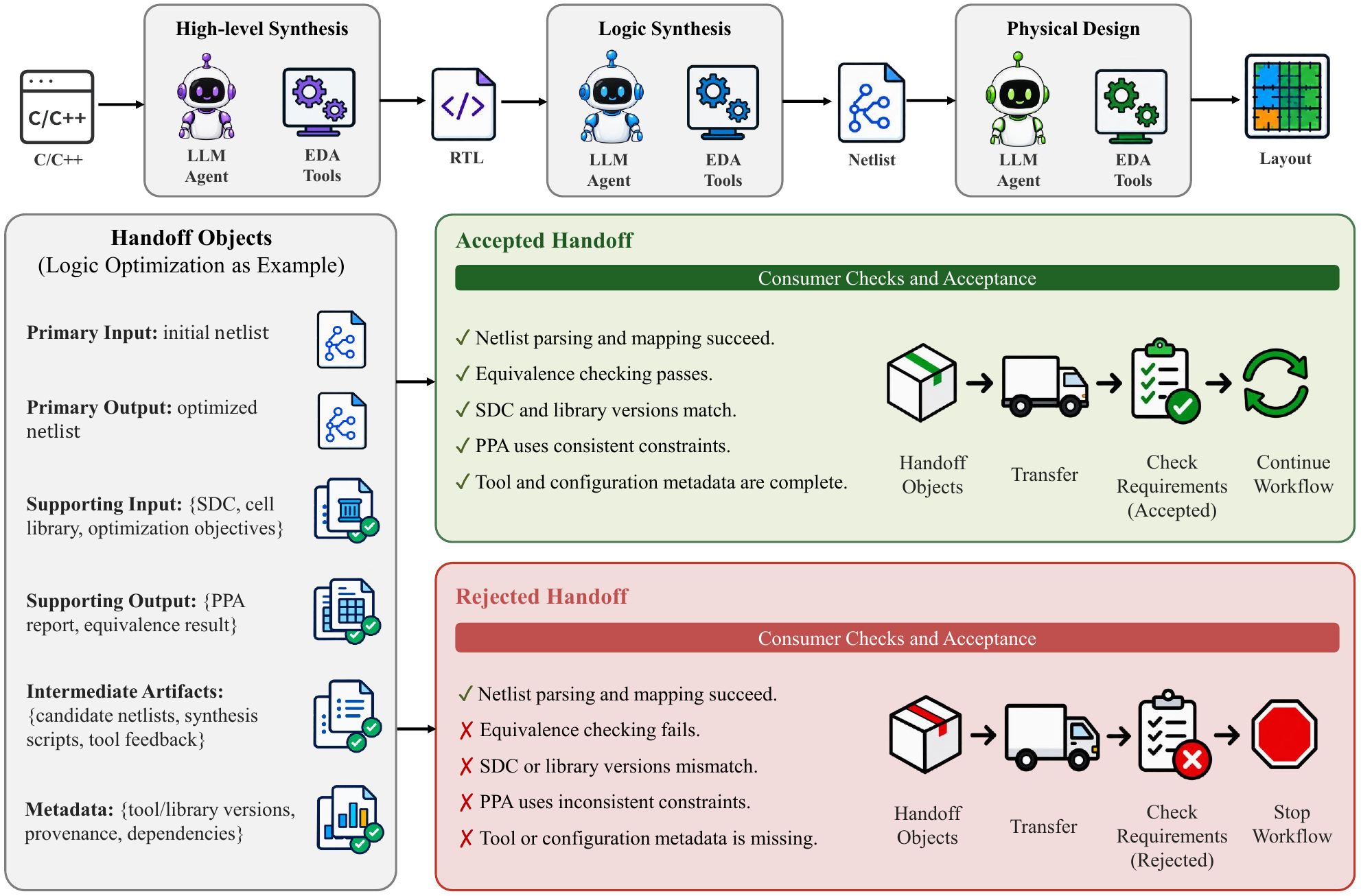}
    \caption{Consumer acceptance of a stage-local logic-optimization handoff. The upper flow situates the handoff within a C/C++$\rightarrow$RTL$\rightarrow$netlist$\rightarrow$layout workflow. The handoff objects include the initial and optimized netlists, supporting inputs and outputs, intermediate artifacts, and metadata. The two cases contrast whether the physical-design consumer accepts these mutually compatible objects under equivalence, version, constraint, and metadata checks.}
    \Description{A logic-optimization example links a C/C++-to-layout workflow to primary, supporting, intermediate, and metadata objects. Accepted and rejected branches show whether the downstream physical-design consumer can proceed with the transferred objects.}
    \label{fig:intro-concepts}
\end{figure*}

The artifacts and metadata involved in this transfer are the \textbf{handoff objects}. In the logic-optimization example in \Cref{fig:intro-concepts}, they include the primary input and output, supporting inputs and outputs, intermediate artifacts, and metadata. If equivalence checking fails, the physical-design consumer rejects the optimized netlist.

Agentic EDA systems differ in where their primary input originates and where their primary output is consumed. Three system classes follow from these two dimensions: Intra-stage, Inter-stage, and Extra-stage. Intra-stage systems consume machine-readable EDA-domain objects and produce outputs that are checked and reused within the same design stage. RTLFixer~\cite{tsai2024rtlfixer}, for example, consumes and returns RTL, and the netlist-optimization example in \Cref{fig:intro-concepts} is likewise Intra-stage because both its primary input and primary output are netlists used within logic synthesis. Inter-stage systems cross a design-stage representation boundary, so a downstream tool must accept the new representation. ChatEDA~\cite{wu2024chateda}, for example, orchestrates RTL through to GDSII. Extra-stage systems begin from human-originated text, documents, or similar materials, and their primary output is a machine-consumable EDA artifact. VerilogCoder~\cite{ho2025verilogcoder}, for example, generates RTL from a natural-language specification.

\begin{table*}[t]
\centering
\revise{\caption{Comparison of representative surveys of LLM-based EDA by scope, analytical coverage, roadmap, and organizing characteristics.}\label{tab:survey-comparison}}
\scriptsize
\setlength{\tabcolsep}{4pt}
\renewcommand{\arraystretch}{1.15}
\revise{\begin{tabularx}{\textwidth}{L{1.85cm}C{1.55cm}>{\hsize=0.82\hsize\centering\arraybackslash}X>{\hsize=0.64\hsize\centering\arraybackslash}X>{\hsize=0.90\hsize\centering\arraybackslash}X>{\hsize=0.94\hsize\centering\arraybackslash}X>{\hsize=0.72\hsize\centering\arraybackslash}X>{\hsize=1.36\hsize\centering\arraybackslash}X}
\toprule
\textbf{Survey} & \textbf{Source} & \textbf{LLM4EDA Scope} & \textbf{Agent-Centric} & \textbf{Full-Flow Scope} & \textbf{Quantitative Analysis} & \textbf{Roadmap} & \textbf{Characteristics} \\
\midrule
He et al.,~\cite{he2025large} & TODAES'25 & \ding{51} &  & \ding{51} & \ding{51} & \ding{51} & I/O modality \\
Pan et al.,~\cite{pan2025survey} & TODAES'25 & \ding{51} &  & \ding{51} &  &  & Design stages \\
Chae et al.,~\cite{chae2025generative} & ICCAD'25 & \ding{51} &  &  &  &  & Analog/RF \\
Yang et al.,~\cite{yang2025verilogcsur} & CSUR'25 & \ding{51} &  &  & \ding{51} & \ding{51} & RTL Generation \\
\midrule
Fang et al.,~\cite{fang2026circuit} & TODAES'26 & \ding{51} &  & \ding{51} & \ding{51} & \ding{51} & Circuit FMs \\
Ghose et al.,~\cite{ghose2026agenticpd} & ISPD'26 & \ding{51} & \ding{51} &  &  & \ding{51} & Physical design \\
Liu et al.,~\cite{liu2026llmverif} & ASP-DAC'26 & \ding{51} &  &  &  &  & Verification \\
\midrule
\textbf{This survey} & -- & \ding{51} & \ding{51} & \ding{51} & \ding{51} & \ding{51} & Handoff \\
\bottomrule
\end{tabularx}}
\end{table*}

\revise{\subsection{Limitations of Existing Surveys}}

\revise{Prior surveys of LLM-based EDA organize the literature by application category, input--output modality, design stage, model architecture, or design domain. None organizes agentic systems around producer-to-consumer handoffs and the objects involved in them. \Cref{tab:survey-comparison} compares representative surveys by subject scope, agent-centricity, full-flow coverage, systematic quantitative or benchmark analysis, roadmap, and defining characteristic. In the table, agent-centric denotes a survey organized around agentic multi-step workflows. Specifically, three related surveys spanning the full design flow are as follows:}

\revise{\begin{itemize}
\item He et al.~\cite{he2025large} organize existing research into code generation, verification and debugging, knowledge representation and retrieval, and optimization and modeling, and further map these applications according to \textbf{input--output modality}.
\item Pan et al.~\cite{pan2025survey} organize LLM applications by \textbf{design stage}, covering system-level design, RTL design, logic synthesis and physical design, and analog circuit design.
\item Fang et al.~\cite{fang2026circuit} survey circuit foundation models, including encoder-based models pre-trained on circuit data and LLM-based decoders, across the VLSI design flow, and benchmark RTL-generation decoders on VerilogEval~\cite{liu2023verilogeval} and RTLLM~\cite{lu2024rtllm} over time.
\end{itemize}}

\revise{These three surveys describe what LLM-based EDA systems do, where they operate in the design flow, and how they are implemented. However, none of them centers its narrative on agentic EDA. Each covers only a limited set of related work. Their organizing axes also leave implicit which consumer receives the generated artifact and which acceptance requirements apply at that boundary. In addition, domain-specific surveys refine these axes for RTL code generation~\cite{yang2025verilogcsur}, physical design~\cite{ghose2026agenticpd}, analog and RF design~\cite{chae2025generative}, and circuit verification~\cite{liu2026llmverif}. Recent focus sessions further chart agentic directions in physical design~\cite{wu2026focusllmphys}, hardware test generation~\cite{firouzi2026agentictest}, and design-rule checking~\cite{wu2026agenticdrc}. An invited DAC 2026 perspective~\cite{xu2026llmfrontend} frames the challenges and opportunities of LLMs across front-end design, and Xu et al.~\cite{xu2025revolution} benchmark the capabilities and limits of large models in hardware design.}

\revise{Unlike the above works, this survey focuses on agentic EDA and uses producer-to-consumer handoffs as its organizing principle, leading to three core questions:
\begin{itemize}
\item Q1: In agentic EDA papers, which primary, supporting, and intermediate objects are produced, consumed, and transferred, and what are the current research trends?
\item Q2: Across different primary input--output settings, how does each work perform, and which agentic techniques does it employ?
\item Q3: What limitations exist in current research, and what research questions follow?
\end{itemize}}

\revise{These questions are addressed in \Cref{sec:framework} (Q1), Sections~\ref{sec:intra-stage}--\ref{sec:extra-stage} (Q2), and Section~\ref{sec:challenges} (Q3), respectively.}

\begin{figure*}[t!]
\centering
\includegraphics[width=0.94\textwidth]{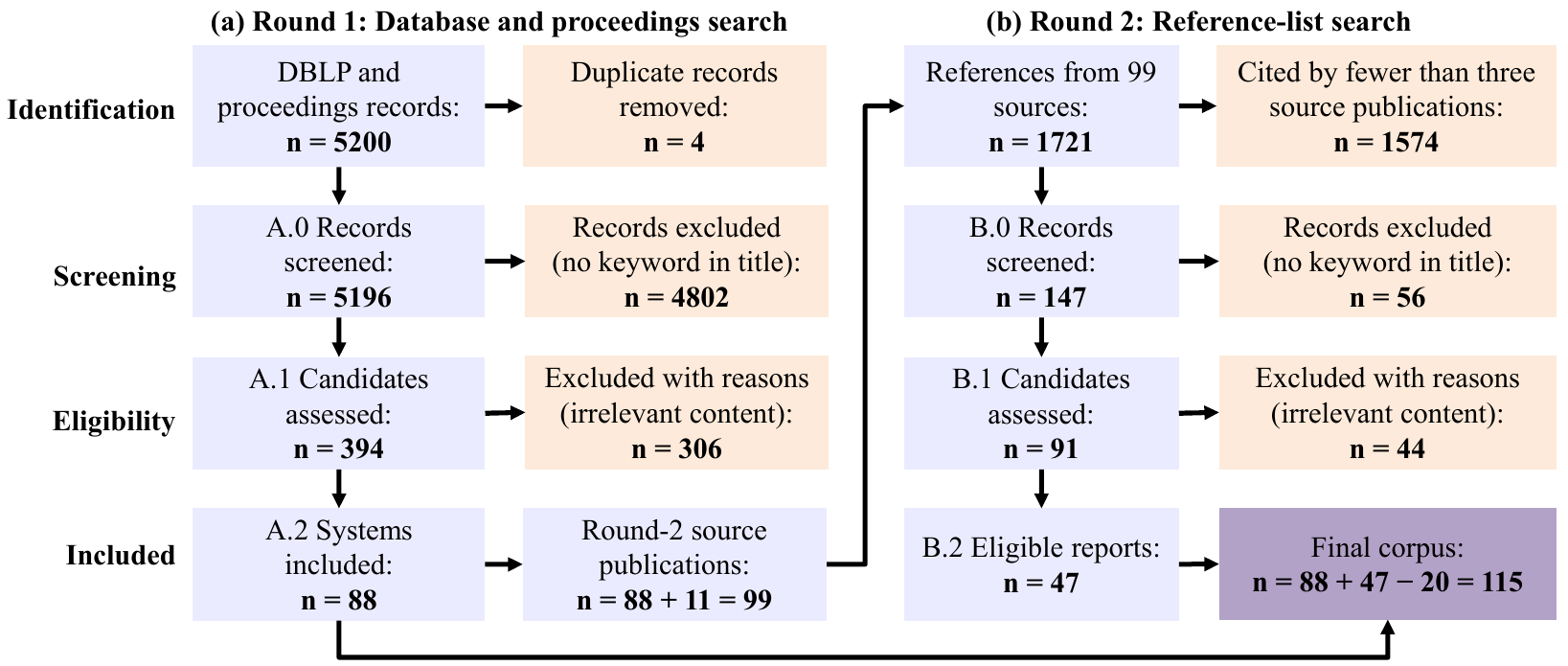}
\caption{\revise{Two-round PRISMA-style screening process. (a) Database and proceedings search. (b) Reference-list search.}}
\Description{Panel (a) reports 5,200 raw venue-search records, 5,196 after deduplication, 394 eligibility candidates, and 88 included systems. Panel (b) reports 1,721 normalized references, 147 records after citation-frequency filtering, and 91 eligibility candidates. Of these, 43 reports were excluded and one eligible companion report was consolidated into an existing system lineage, yielding 47 B.2 systems. Merging both rounds and removing 20 cross-round overlaps yields 115 distinct systems.}
\label{fig:prisma}
\end{figure*}

\revise{\subsection{Scope and Organizing Principle}}

\revise{\minisection{Corpus Construction}}
\revise{We conducted a venue-focused, two-round literature search covering work published, accepted, or publicly available by the August 31, 2026 cutoff, following the PRISMA-style procedure in \Cref{fig:prisma}. The first round searched publication venues directly. We collected DBLP records for DAC, DATE, ASP-DAC, TCAD, and TODAES over 2024--2026 and for ICCAD over 2024--2025. ICCAD 2026 proceedings were not available by the cutoff. DAC 2026 records were supplemented from its official proceedings. The resulting 5,200 raw records were reduced to A.0 = 5,196 after four true duplicates were removed. Title screening with the keywords agent, agentic, LLM, large language model, code generation, and retrieval augmented generation (RAG) excluded 4,802 records and produced A.1 = 394 eligibility candidates. We then assessed each candidate against the following inclusion and exclusion criteria.}

\revise{\minisection{Inclusion and Exclusion Criteria}}
\revise{We included a study when (1) it presented an LLM-based system for an EDA task, (2) the LLM materially affected an engineering artifact or decision, and (3) that result entered a tool-execution, multi-agent, persistent-workflow, or explicit human-review handoff. We excluded surveys, benchmarks without an implemented agentic system, conventional ML4EDA methods, advisory QA systems without a downstream engineering action, one-shot prompting demonstrations, and reports lacking sufficient technical detail. Under these criteria, the first round excluded 306 irrelevant records and retained A.2 = 88 systems.}

\revise{\minisection{Corpus Supplement}}
\revise{The 88 systems from the first round may miss widely recognized work from other venues or preprints. To supplement such work, the second round broadened coverage through citation overlap. We scanned 99 unique source publications drawn from the A.2 systems, seven survey bibliographies~\cite{he2025large,pan2025survey,chae2025generative,yang2025verilogcsur,fang2026circuit,ghose2026agenticpd,liu2026llmverif}, and five related focus-session, perspective, and panel papers~\cite{wu2026focusllmphys,firouzi2026agentictest,wu2026agenticdrc,xu2026llmfrontend,xu2025revolution}, with duplicate publications counted once. Their locally extracted reference lists contained 1,721 unique normalized references. In this round, we regard a work as representative when it is cited by at least three distinct sources. Under this criterion, the set was reduced to B.0 = 147 records. We then repeated the first round's title screening and eligibility assessment. Title screening excluded 56 records. Among the 91 reports assessed for eligibility, 43 failed the inclusion criteria. ConfiBench, an eligible journal extension of CorrectBench, was consolidated into the same system lineage rather than counted as a separate entry. These 44 exclusions or consolidations yielded B.2 = 47 systems.}

\revise{\minisection{Corpus Merging}}
\revise{Because a paper in A.2 may itself be cited multiple times and thus be hit again in B.2, we merged the system identities from both rounds and removed 20 cross-round overlaps. The final corpus therefore contains 115 distinct agentic EDA systems (88 + 47 $-$ 20). Publications describing the same system were merged unless a later version introduced a materially different agent architecture.}

\begin{figure*}[t]
  \centering
  \subfloat[\label{fig:corpus-venue}]{%
    \begin{minipage}[c]{0.59\textwidth}
    \centering
    \begin{tikzpicture}
    \begin{axis}[
      ybar,
      width=0.9\linewidth,
      height=5.1cm,
      bar width=11pt,
      ylabel={\footnotesize Number of Papers},
      symbolic x coords={DAC,ICCAD,DATE,ASP-DAC,TODAES,TCAD,Others},
      xtick=data,
      x tick label style={font=\footnotesize, rotate=25, anchor=east},
      tick label style={font=\footnotesize},
      label style={font=\footnotesize},
      ymin=0, ymax=50,
      nodes near coords,
      nodes near coords style={font=\tiny},
      enlarge x limits=0.08,
    ]
    \addplot[fill=sysblue2, draw=deepblue] coordinates {
      (DAC,38)(ICCAD,16)(DATE,12)(ASP-DAC,9)(TODAES,11)(TCAD,6)(Others,23)};
    \end{axis}
    \end{tikzpicture}
    \end{minipage}%
  }\hfill
  \subfloat[\label{fig:corpus-year}]{%
    \begin{minipage}[c]{0.35\textwidth}
    \centering
    \begin{tikzpicture}
    \fill[sysblue2, draw=white, line width=1.2pt]
      (0,0) -- (90:1.7) arc (90:86.87:1.7) -- cycle;
    \fill[sysblue!70!white, draw=white, line width=1.2pt]
      (0,0) -- (86.87:1.7) arc (86.87:8.61:1.7) -- cycle;
    \fill[deepblue!62!white, draw=white, line width=1.2pt]
      (0,0) -- (8.61:1.7) arc (8.61:-135.39:1.7) -- cycle;
    \fill[govpurple!62!white, draw=white, line width=1.2pt]
      (0,0) -- (-135.39:1.7) arc (-135.39:-270:1.7) -- cycle;
    \node[font=\footnotesize, text=textdark] at (88.44:1.15) {\revise{0.9\%}};
    \node[font=\footnotesize, text=textdark] at (47.74:1.15) {\revise{21.7\%}};
    \node[font=\footnotesize, text=textdark] at (-63.39:1.15) {\revise{40.0\%}};
    \node[font=\footnotesize, text=textdark] at (-202.70:1.15) {\revise{37.4\%}};
    \draw[fill=sysblue2, draw=bordergray, rounded corners=1pt] (-1.9,-2.35) rectangle (-1.52,-2.08);
    \node[anchor=west, font=\footnotesize, text=textdark] at (-1.47,-2.21) {2023 (\revise{1})};
    \draw[fill=sysblue!70!white,  draw=bordergray, rounded corners=1pt] (0.15,-2.35) rectangle (0.53,-2.08);
    \node[anchor=west, font=\footnotesize, text=textdark] at (0.58,-2.21) {2024 (\revise{25})};
    \draw[fill=deepblue!62!white, draw=bordergray, rounded corners=1pt] (-1.9,-2.85) rectangle (-1.52,-2.58);
    \node[anchor=west, font=\footnotesize, text=textdark] at (-1.47,-2.71) {2025 (\revise{46})};
    \draw[fill=govpurple!62!white, draw=bordergray, rounded corners=1pt] (0.15,-2.85) rectangle (0.53,-2.58);
    \node[anchor=west, font=\footnotesize, text=textdark] at (0.58,-2.71) {2026 (\revise{43})};
    \end{tikzpicture}
    \end{minipage}%
  }
  \caption{Corpus statistics across the \revise{115} surveyed systems. (a) Systems by publication venue, with venues outside the six core venues grouped as ``Others.'' (b) Year distribution of the \revise{115} surveyed systems.}
  \Description{Panel (a) is a bar chart with 38 DAC, 16 ICCAD, 12 DATE, 9 ASP-DAC, 11 TODAES, 6 TCAD, and 23 other systems. Panel (b) shows 1 system in 2023, 25 in 2024, 46 in 2025, and 43 by the August 31, 2026 cutoff.}
  \label{fig:corpus-stats}
  \end{figure*}

\revise{\minisection{Trend Observation}}
\revise{\Cref{fig:corpus-stats} documents the rapid expansion of agentic EDA research. The corpus grows from 1 system in 2023 to 25 in 2024 and 46 in 2025. The partial 2026 count reaches 43 by the August 31 cutoff. The 89 systems from 2025--2026 account for 77.4\% of the corpus. DAC contributes 38 systems, while the remaining 77 span other core and non-core venues.}

\subsection{Positioning and Contributions}

\minisection{Positioning}
This survey organizes agentic EDA systems around producer-to-consumer handoffs. The primary input--output pair determines the taxonomy class, while supporting and intermediate handoff objects describe the information used and produced along the way. This view identifies both differences among systems and infrastructure gaps shared across the corpus. These gaps point to future research directions for agentic EDA.

\minisection{Contributions}
This framing leads to four contributions. First, the survey uses producer-to-consumer handoffs to connect agent outputs with the acceptance requirements of the receiving EDA stage. Second, it introduces a three-class taxonomy determined by the primary input--output pair: Intra-stage, Inter-stage, and Extra-stage. Third, the class-specific chapters present the primary, supporting, and intermediate objects of each agentic EDA system as the main analytical dimension, complemented by model mechanisms as an auxiliary analytical dimension. Fourth, the survey outlines future research directions, poses open questions, and charts a five-layer roadmap for an agentic EDA communication protocol.

\minisection{Outline}
\Cref{sec:framework} introduces the core concepts, the three-class taxonomy, research trends, and evaluation benchmarks. \Cref{sec:intra-stage,sec:inter-stage,sec:extra-stage} present and discuss representative works by class. \Cref{sec:challenges} proposes the roadmap and future research directions. Finally, \Cref{sec:conclu} concludes the survey.

\section{Analytical Framework and Three-Class Taxonomy}
\label{sec:framework}

\revise{This section first defines handoff and related concepts used to describe each system. It then assigns systems to Intra-stage, Inter-stage, and Extra-stage classes from the primary input--output pair. Model optimization and agent harnesses provide a separate mechanism-based view, followed by the benchmark inventory. Table~\ref{tab:glossary} expands the recurring EDA and AI terms used in the definitions and throughout the survey.}

\begin{table*}[t!]
\centering
\caption{Glossary of common terms used throughout this survey, grouped into EDA and AI terms.}
\label{tab:glossary}
\fontsize{7.5pt}{8.25pt}\selectfont
\setlength{\tabcolsep}{2.5pt}
\renewcommand{\arraystretch}{1.08}
\begin{tabularx}{0.97\textwidth}{@{}C{1.05cm}>{\RaggedRight\arraybackslash\hsize=.85\hsize}X@{\hspace{8pt}}C{1.05cm}>{\RaggedRight\arraybackslash\hsize=1.15\hsize}X@{}}
\toprule
\rowcolor{tableheader}
\multicolumn{4}{@{}l}{\textbf{EDA Terms}} \\
\rowcolor{tableheader}
\textbf{Term} & \textbf{Definition} & \textbf{Term} & \textbf{Definition} \\
\midrule
ASIC & Application-Specific Integrated Circuit & DRC & Design-Rule Checking \\
DSE & Design-Space Exploration & DUT & Design Under Test \\
EDA & Electronic Design Automation & ERC & Electrical-Rule Checking \\
FPGA & Field-Programmable Gate Array & GDSII & Graphic Data System II Stream Format \\
HDL & Hardware Description Language & HLS & High-Level Synthesis \\
HPWL & Half-Perimeter Wirelength & IP & Intellectual Property \\
LVS & Layout Versus Schematic & NL & Natural Language \\
PCB & Printed Circuit Board & PDK & Process Design Kit \\
PPA & Power, Performance, And Area & QoR & Quality Of Results \\
RF & Radio Frequency & RTL & Register-Transfer Level \\
SDC & Synopsys Design Constraints & SPICE & Simulation Program With Integrated Circuit Emphasis \\
SVA & SystemVerilog Assertions & TCAD & Technology Computer-Aided Design \\
UVM & Universal Verification Methodology & VHDL & VHSIC Hardware Description Language \\
\midrule
\rowcolor{tableheader}
\multicolumn{4}{@{}l}{\textbf{AI Terms}} \\
\rowcolor{tableheader}
\textbf{Term} & \textbf{Definition} & \textbf{Term} & \textbf{Definition} \\
\midrule
AI & Artificial Intelligence & VLM & Vision-Language Model \\
LLM & Large Language Model & ML & Machine Learning \\
ML4EDA & Machine Learning For EDA & RAG & Retrieval-Augmented Generation \\
RL & Reinforcement Learning & SFT & Supervised Fine-Tuning \\
\bottomrule
\end{tabularx}
\end{table*}

\subsection{Handoff and Related Concepts}
\label{sec:framework-handoff}

\revise{For corpus coding and comparison, we distinguish the handoff event, the objects involved in it, and the primary input--output pair used for classification.}

\minisection{Handoff}
\revise{A handoff is an event in which a producer makes one or more objects available to a consumer for a subsequent engineering action. Here, the producer is typically an agent, and the consumer is an EDA tool. The consumer accepts the handoff when the transferred objects satisfy its interface, semantic, configuration, and version requirements. Tool checks or human review provide acceptance evidence, while provenance and version metadata determine whether that evidence applies to the delivered objects and execution environment. Producer-side success alone does not establish consumer acceptance.}

\minisection{Handoff Objects}
\revise{A handoff may involve multiple objects that serve different roles. We represent the complete set of handoff objects as $\mathcal{R}=(I_p,O_p,I_s,O_s,A_m;M)$, where:}

\begin{figure*}[t!]
\centering
\twopanelgraphic{\includegraphics[width=\textwidth]{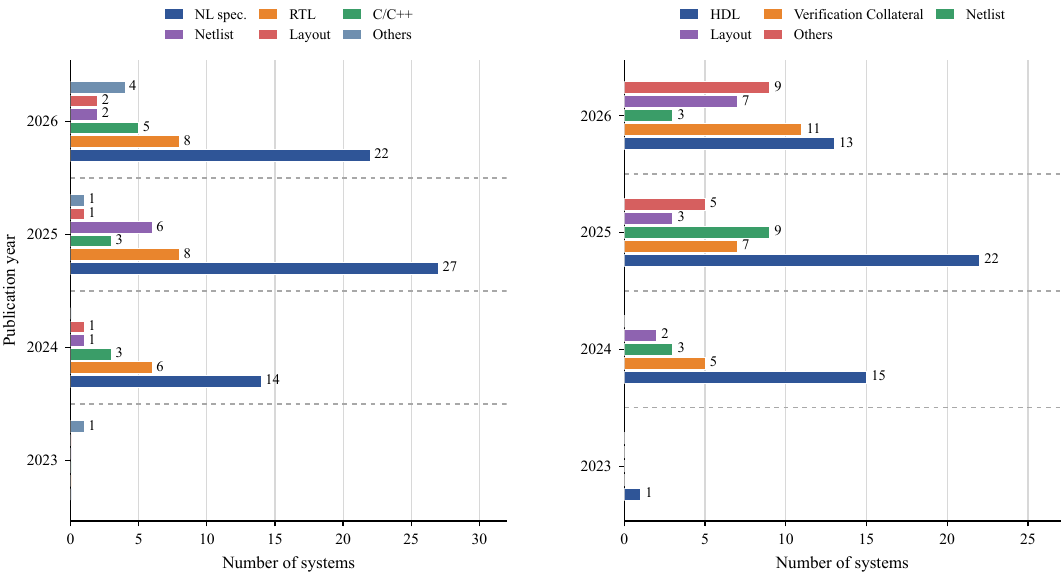}}
\caption{Temporal evolution of primary inputs and outputs in the 115 surveyed systems. Both panels are generated directly from the $I_p\rightarrow O_p$ entries in the handoff-object tables in Sections~\ref{sec:intra-stage}--\ref{sec:extra-stage}. (a) Primary-input types, aggregating types that appear at most twice as Others. (b) Primary outputs aggregated into HDL, Verification Collateral, Netlist, Layout, and Others; the last category aggregates the remaining output types.}
\Description{Two side-by-side grouped horizontal bar charts show annual counts from 2023 to 2026. The left groups primary inputs by representation type. The right groups primary outputs into HDL, verification collateral, netlist, layout, and others.}
\label{fig:primary-io-trends}
\end{figure*}

\begin{itemize}[leftmargin=1.5em,topsep=2pt,itemsep=1pt]
    \item \revise{The \emph{primary input} $I_p$ defines the task instance and the accepted \emph{primary output} $O_p$ closes the task. The pair $(I_p,O_p)$ is the \emph{primary input--output pair}. It characterizes the core task of an agentic EDA system and determines its taxonomy class. The types of $I_p$ and $O_p$ are diverse, as shown in \Cref{fig:primary-io-trends}. Other internal transfers do not change the class label unless they implement a separate task objective.}
    \item \revise{\emph{Supporting inputs} $I_s$ provide additional information needed to perform or assess the task. Depending on their use, they are categorized as constraints (Constr., e.g., timing targets or design rules), references (Ref., e.g., testbenches or golden models), or knowledge sources (Knowl., e.g., tool manuals or prior design records).}
    \item \revise{\emph{Supporting outputs} $O_s$ provide evidence or records produced alongside the primary output. Depending on their use, they are categorized as checking results (Check, e.g., compilation, simulation, formal-verification, or DRC results), quality-of-result measurements (QoR, e.g., power, performance, area, or resource use), or records (Record, e.g., execution logs, provenance records, or human-review records).}
    \item \revise{\emph{Intermediate artifacts} $A_m$ are produced and consumed within the workflow. Depending on their use, they are categorized as retrieved context (Retr., e.g., retrieved documentation or prior examples), candidates (Cand., e.g., candidate RTL, netlists, layouts, or configurations), or coordination state (Coord., e.g., plans, role messages, or checkpoints).}
    \item \revise{$M$ records version, provenance, producer, consumer, and dependency metadata.}
\end{itemize}

\revise{In practice, the pair $(I_p,O_p)$ is identified first, after which the remaining objects are assigned by function. The same format may therefore occupy different roles. Candidate RTL belongs to $A_m$ during revision and becomes $O_p$ only when delivered as the accepted result. ChatEDA~\cite{wu2024chateda} illustrates the assignment. Its primary pair is RTL$\rightarrow$implemented layout/GDSII. The natural-language objective and platform settings belong to $I_s$, PPA and validation reports to $O_s$, and plans, scripts, intermediate netlists, and checkpoints to $A_m$.}

\noindent\revise{\textbf{Trend Observation.} \Cref{fig:primary-io-trends} shows that natural-language specifications remain the largest primary-input category, although by 2026 they are nearly matched by all other input types combined (22 versus 21 systems). Output diversification is more pronounced: HDL declines from 60\% of the 2024 systems to 30\% in 2026, while Verification Collateral increases from 5 to 11 systems and Layout from 2 to 7.}

\subsection{The Three-Class Taxonomy}
\label{sec:framework-boundaries}

\begin{figure*}[t!]
\centering
\includegraphics[width=\textwidth]{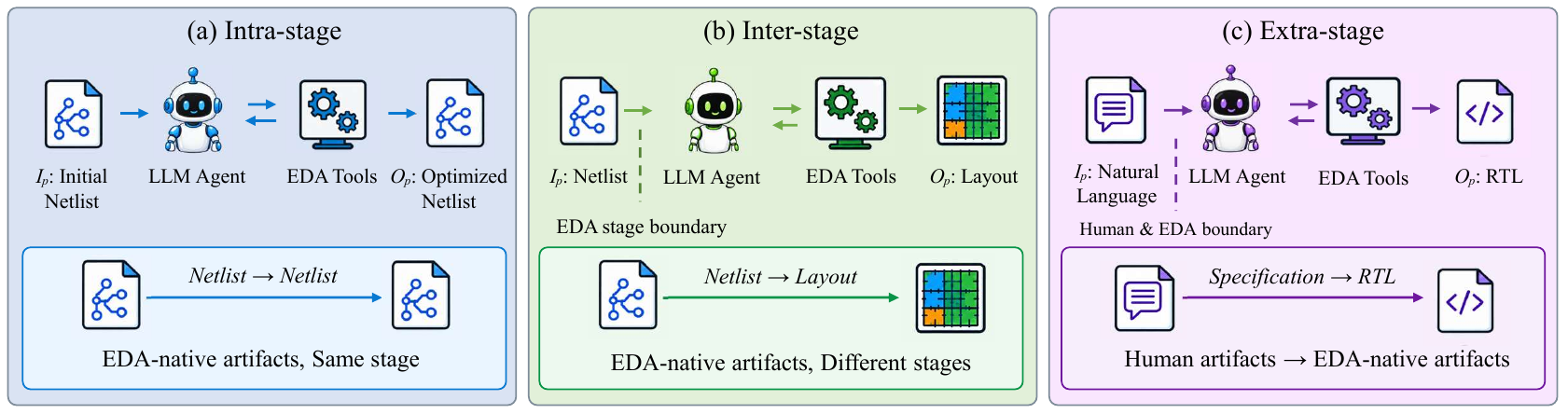}
\caption{Three-class taxonomy of agentic EDA systems: (a) Intra-stage, (b) Inter-stage, and (c) Extra-stage systems, distinguished by the primary input--output pair.}
\Description{Three panels show representative primary input--output pairs and consumer-side acceptance evidence for same-stage EDA-native refinement, cross-stage EDA-native transformation, and human-originated specification to an EDA-native artifact.}
\label{fig:framework-taxonomy}
\end{figure*}

\revise{Building on the primary input--output pair defined in \Cref{sec:framework-handoff}, the taxonomy uses two properties: the provenance of $I_p$ and the consumer boundary of $O_p$. The first distinguishes machine-readable EDA-domain objects from human-originated intent or knowledge. The second distinguishes same-stage reuse from a representation change accepted by a downstream stage. Supporting inputs and outputs, intermediate artifacts, metadata, LLM architecture, mechanisms, benchmarks, and evidence strength remain descriptive dimensions and do not determine the primary class.}

\begin{samepage}
\minisection{Assignment Rules}
\revise{We apply the following decision procedure.}
\begin{itemize}[leftmargin=1.5em,topsep=2pt,itemsep=1pt]
    \item \revise{\textbf{Rule 1 (Input provenance).} Determine whether the primary input is a machine-readable EDA-domain object (a design artifact, EDA-tool code, or device-characterization measurements) or human-originated intent or knowledge (a natural-language specification, query, or document).}
    \item \revise{\textbf{Rule 2 (Consumer boundary).} If the primary input is human-originated, assign \emph{Extra-stage}. If it is an EDA-domain object, assign \emph{Inter-stage} only when the output changes a design representation and is accepted by a downstream design stage. Otherwise, assign \emph{Intra-stage} when the output is checked and reused within the same stage.}
    \item \revise{\textbf{Rule 3 (Secondary labels).} Systems that additionally instantiate a second class receive a secondary label, allowing one system to represent multiple classes.}
\end{itemize}
\end{samepage}

\revise{\Cref{fig:framework-taxonomy} illustrates the three classes with representative primary input--output pairs: same-stage refinement, cross-stage representation change, and the translation of human-originated intent into an EDA artifact.}

\minisection{Intra-stage Systems}
These systems take a machine-readable EDA-domain object as their primary input. Such objects include a design artifact, EDA-tool code, or device-characterization measurements. The output is checked and reused within the same stage. It commonly preserves or locally refines the input representation, but may instead be verification collateral or an analysis product consumed by that stage. Consumer acceptance relies on a stage-local checker or optimizer: compilation, simulation, fitting error, equivalence, PPA evaluation, lint/DRC, coverage, or an equivalent local acceptance condition. Representative systems include RTL repair and debug loops (RTLFixer~\cite{tsai2024rtlfixer}), analog sizing (ADO-LLM~\cite{yin2024ado}), EDA-tool code evolution (MappingEvolve~\cite{fu2026mappingevolve}), and device-model parameter extraction (ModelGen~\cite{wei2025modelgen}).

\minisection{Inter-stage Systems}
These systems consume an EDA-native representation and produce a different design-stage representation that a downstream tool must accept and consume. Acceptance relies on cross-tool execution, downstream validation, QoR, or checkpoint coherence across a representation boundary. Representative systems include high-level synthesis (HLSPilot~\cite{chenwei2024hlspilotllmbasedhigh}), ASIC layout (HyperPlace~\cite{magi2025hyperplaceharnessing}), PCB layout (PCBAgent~\cite{chen2025pcbagent}), and RTL-to-GDSII orchestration (ChatEDA~\cite{wu2024chateda}).

\minisection{Extra-stage Systems}
These systems take human intent or knowledge as their primary input, including natural-language specifications, queries, or documents, and translate it into a machine-consumable EDA artifact. Acceptance depends on specification completeness, provenance, and attribution to inspectable sources with an appropriate scope. Representative systems are VerilogCoder~\cite{ho2025verilogcoder} for RTL generation, AssertLLM~\cite{yan2025assertllm} for verification collateral, and ChatLS~\cite{zheng2025chatls} for retrieval-based script generation.

\begin{figure*}[t!]
\centering
\twopanelgraphic{\includegraphics[width=\textwidth]{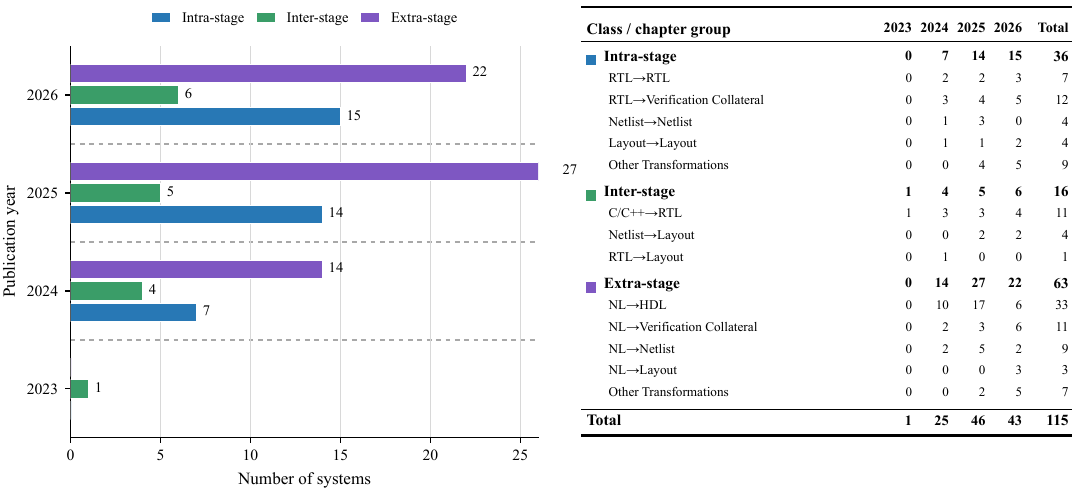}}
\caption{(a) Temporal evolution and (b) distribution by class and chapter group of the 115 surveyed systems.}
\Description{A grouped horizontal bar chart and a six-column table summarize 115 systems. The chart compares annual counts for the Intra-stage, Inter-stage, and Extra-stage classes from 2023 to 2026. The table reports 2023, 2024, 2025, 2026, and total counts for 11 task families and two heterogeneous residual groups.}
\label{fig:task-family-overview}
\end{figure*}

\FloatBarrier

\noindent\revise{\textbf{Trend Observation.} \Cref{fig:task-family-overview} shows expansion in all three classes within a strongly uneven corpus. Intra-stage systems increase from none in 2023 to 7 in 2024, 14 in 2025, and 15 in 2026. Inter-stage systems grow more gradually from 1 to 4, 5, and 6. Extra-stage systems account for 63 of the 115 systems and dominate from 2024 onward, rising from 14 in 2024 to 27 in 2025 and 22 by the August 31, 2026 cutoff.}

\subsection{LLM-Based Agents and Related Concepts}
\label{sec:framework-mechanisms}

\begin{figure*}[t!]
\centering
\twopanelgraphic{\includegraphics[width=\textwidth]{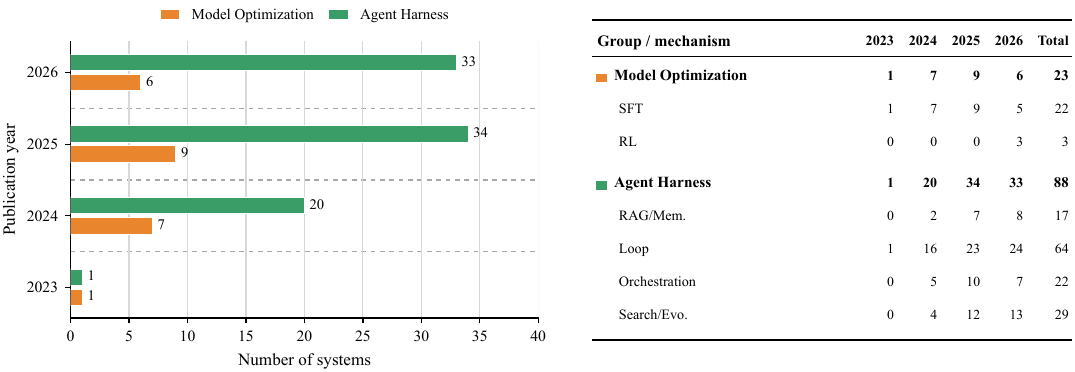}}
\caption{(a) Temporal evolution and (b) distribution of model-optimization and agent-harness mechanisms in the 115 surveyed systems.}
\Description{A grouped horizontal bar chart and a mechanism table summarize the surveyed systems from 2023 to 2026. Model optimization appears in 1, 7, 9, and 6 systems by year, while harnesses grow from 1 system in 2023 to 33 in 2026. The table gives annual counts and totals for two model-optimization and four harness mechanisms.}
\label{fig:mechanism-trends}
\end{figure*}

\revise{Having classified systems by their primary input--output pairs, we next characterize how LLM-based agents produce, evaluate, and coordinate the associated objects. In this survey, an LLM-based agent embeds an LLM in an EDA workflow that uses model outputs to act on artifacts or invoke tools. The workflow may also retrieve sources, maintain state, and revise outputs from feedback. We describe a system as VLM-based when visual inputs and multimodal reasoning are integral to these actions. \Cref{fig:mechanism-trends} organizes six mechanisms used by both types into model optimization and agent harnesses.}

\minisection{Model Optimization}
\revise{Model optimization changes the parameters of the LLM or VLM used for reasoning or generation. \emph{Supervised fine-tuning (SFT)} updates those parameters from labeled examples or validated agent trajectories. \emph{Reinforcement learning (RL)} instead updates the LLM/VLM policy from task rewards, including rewards derived from EDA-tool outcomes. A non-LLM controller or optimizer trained separately and used alongside the language model is a specialized backend, because its training does not modify the LLM/VLM.}

\minisection{Agent Harness}
\revise{Following the agent-harness formulation in current agent-system practice, we use \emph{agent harness} for the runtime layer around a model that supplies context and memory, control flow, tool access, orchestration, and candidate selection~\cite{trivedy2026anatomy}. The four harness mechanisms in \Cref{fig:mechanism-trends} modify runtime behavior without changing model parameters. \emph{Retrieval/memory} supplies external knowledge, prior examples, persistent episodic memory, or workflow state to a later decision. A \emph{loop} implements a proposal--action--check--feedback--revision cycle within a task. \emph{Orchestration} uses a controller, scheduler, or coordination protocol to allocate work and transfer state among agents, tools, or workflow stages. \emph{Search/evolution} generates, evaluates, and selects among alternative artifacts, configurations, or workflows through methods such as Monte Carlo tree search (MCTS), Bayesian optimization, concolic exploration, or evolutionary operators. It is coded as a harness mechanism only when the agent-controlled workflow governs this process. Optimization performed entirely inside an invoked EDA tool is not counted.}

\revise{These flags may co-occur because they describe different operations. A compile--repair cycle uses a loop. Ranking several repairs additionally uses search/evolution. Routing tasks and intermediate state adds orchestration. Updating the LLM/VLM from tool-derived rewards adds RL. In \Cref{fig:mechanism-trends}, the harness columns proceed from information state (retrieval/memory), through temporal control (loop) and organizational control (orchestration), to candidate-space control (search/evolution).}

\noindent\revise{\textbf{Trend Observation.} \Cref{fig:mechanism-trends} shows that mechanism growth is concentrated in agent harnesses, which appear in 88 of the 115 systems, compared with 23 systems using model optimization. Loops are the most common mechanism, increasing from 1 system in 2023 to 16, 23, and 24 in 2024--2026. Search/evolution grows from none in 2023 to 4, 12, and 13 systems. Retrieval/memory grows from none to 2, 7, and 8, while orchestration grows from none to 5, 10, and 7. Model optimization has annual coverage of 1, 7, 9, and 6 systems: SFT accounts for 22 systems overall, whereas LLM/VLM policy optimization through RL appears only in three 2026 systems. The contrast shows that recent systems more often add external runtime structure than update language-model parameters.}

\subsection{Evaluation Resources}
\label{sec:framework-resources}

\minisection{Reusable Benchmarks and Evaluation Scope}
\revise{The benchmark inventory includes open resources introduced or cited by a corpus paper. \emph{Task-direct benchmarks} are organized by their primary task family. \emph{Multi-task benchmarks} span several tasks, while \emph{reused workloads} provide circuits, kernels, or programs evaluated under study-specific protocols. \emph{Training-oriented datasets} support model fitting or prediction. A held-out task and an independently specified acceptance criterion are required for benchmark use.}

\revise{\Cref{tab:benchmark-landscape} organizes task-direct resources by their dominant evaluated use, first by the three handoff classes and then by task family. Multi-task benchmarks, reused workloads, and training-oriented datasets form cross-cutting groups. Public-artifact status is verified from a repository, dataset card, or original release page. Resources without a confirmed public artifact are discussed in the task-local text and excluded from the table. General-purpose circuit/design suites such as ISCAS~\cite{brglez1985iscas,brglez1989iscas}, EPFL~\cite{amaru2015epfl}, OpenCores~\cite{opencores}, and Chipyard~\cite{amid2020chipyard} also remain in the task-local discussions.}

\begin{table*}[t!]
\centering
\begin{surveytable}
\fontsize{6.65pt}{7.25pt}\selectfont
\setlength{\tabcolsep}{1.8pt}
\caption{Open benchmark and evaluation resources associated with the 115-system corpus, organized by dominant evaluated use. Counts follow the cited paper or linked release and retain the source's native unit. A dash denotes that the source does not define a fixed instance count or does not state one unambiguously. Each GitHub icon links to a verified public artifact.}\label{tab:benchmark-landscape}
\begin{tabular}{@{}L{3.10cm}C{0.52cm}C{1.15cm}L{2.45cm}L{5.60cm}@{}}
\toprule
\rowcolor{tableheader}
\textbf{Benchmark} & \textbf{Link} & \textbf{Count} & \textbf{Unit} & \textbf{Primary evaluation target} \\
\midrule
\tablegroup{5}{Intra-stage: RTL$\rightarrow$RTL}
RTLFixer repair set~\cite{tsai2024rtlfixer} & \href{https://github.com/NVlabs/RTLFixer}{\githubmark} & 212 & programs & Syntax and functional repair of corrupted RTL \\
MEIC debug set~\cite{xu2024meic} & \href{https://anonymous.4open.science/r/Verilog-Auto-Debug-6E7F/}{\githubmark} & 178 & variants & Functional debugging on RTLLM-derived faults \\
\tablegroup{5}{Intra-stage: RTL$\rightarrow$Verification Collateral}
AutoBench~\cite{autobench2024} & \href{https://github.com/AutoBench/AutoBench}{\githubmark} & 156 & tasks & Generated-testbench correctness at Eval0--Eval2 \\
VerifLLMBench~\cite{murthy2025verifllmbench} & \href{https://github.com/uma-ece-sartorilabs/Verif-LLM}{\githubmark} & 30 & DUTs & Functional correctness of RTL-conditioned testbenches \\
\tablegroup{5}{Intra-stage: Other Transformations}
RTL-BenchMT~\cite{wang2026rtlbenchmt} & \href{https://github.com/hkust-zhiyao/RTL-BenchMT}{\githubmark} & 47 & revised cases & Executable RTL benchmark maintenance and provenance \\
\tablegroup{5}{Inter-stage: C/C++$\rightarrow$RTL}
MachSuite~\cite{reagen2014machsuite} & \href{https://github.com/breagen/MachSuite}{\githubmark} & 19 & benchmarks & HLS generation, optimization, and DSE \\
PolyBench/C~\cite{pouchet2016polybench} & \href{https://web.cse.ohio-state.edu/~pouchet.2/software/polybench/}{\githubmark} & 30 & kernels & Numerical-kernel optimization and DSE \\
S2CBench v2.2~\cite{schafer2014s2cbench} & \href{https://sourceforge.net/projects/s2cbench/files/}{\githubmark} & 20 & programs & Synthesis of SystemC designs \\
\tablegroup{5}{Inter-stage: RTL$\rightarrow$Layout / Full-flow}
ChatEDA-Bench~\cite{wu2024chateda} & \href{https://github.com/wuhy68/ChatEDA}{\githubmark} & 50 & tasks & Full-flow execution and parameter tuning \\
\tablegroup{5}{Extra-stage: NL$\rightarrow$HDL}
VerilogEval v1/v2~\cite{liu2023verilogeval} & \href{https://github.com/NVlabs/verilog-eval}{\githubmark} & 156/156 & problems & Syntax and functional correctness of generated RTL \\
RTLLM v1.1/v2~\cite{lu2024rtllm} & \href{https://github.com/hkust-zhiyao/RTLLM}{\githubmark} & 29/50 & designs & RTL correctness and post-synthesis PPA \\
ArchXBench~\cite{purini2025archxbench} & \href{https://github.com/sureshpurini/ArchXBench}{\githubmark} & 51 & designs & RTL generation from architectural specifications \\
GenBen~\cite{wan2025genben} & \href{https://github.com/ChatDesignVerification/GenBen}{\githubmark} & 324 & tests & RTL generation from executable NL specifications \\
RealBench~\cite{jin2025realbench} & \href{https://github.com/IPRC-DIP/RealBench}{\githubmark} & 60+4 & tasks & RTL generation for production-style IP \\
ResBench~\cite{guo2025resbench} & \href{https://github.com/jultrishyyy/ResBench}{\githubmark} & 56 & problems & RTL correctness and FPGA resource use \\
RTL-Repo~\cite{allam2024rtlrepo} & \href{https://github.com/AUCOHL/RTL-Repo}{\githubmark} & $>$4,000 & samples & Repository-context RTL generation and completion \\
PyHDL-Eval~\cite{batten2024pyhdleval} & \href{https://github.com/cornell-brg/pyhdl-eval}{\githubmark} & 168 & problems & Executable generation through Python HDL DSLs \\
CreativEval~\cite{delorenzo2024creativeval} & \href{https://github.com/matthewdelorenzo/CreativEval}{\githubmark} & 120 & prompts & RTL correctness under underspecified prompts \\
ChipGPT-V~\cite{chang2024naturallanguage} & \href{https://github.com/aichipdesign/chipgptv}{\githubmark} & 33 & designs & Multimodal Verilog generation and completion \\
\tablegroup{5}{Extra-stage: NL$\rightarrow$Verification Collateral}
CorrectBench~\cite{qiu2025correctbench} & \href{https://github.com/AutoBench/CorrectBench}{\githubmark} & 156 & tasks & Functional correctness of generated testbenches \\
ConfiBench~\cite{ruidi2026confibenchautomatict} & \href{https://github.com/AutoBench/ConfiBench}{\githubmark} & 156 & tasks & Confidence-guided testbench generation \\
AssertEval/AssertLLM~\cite{liu2025openllmrtl} & \href{https://github.com/hkust-zhiyao/AssertLLM}{\githubmark} & 18 & designs & Assertion generation from NL specifications \\
SDCBench/LLM4SDC~\cite{han2026llm4sdc} & \href{https://github.com/PiyiHan/LLM4DC}{\githubmark} & 22 & designs & SDC generation from specification, RTL, and testbench \\
ReflectBench / ReflectVDB~\cite{liu2026reflectbench} & \href{https://gitee.com/sdu-aes-lab/reflectbench}{\githubmark} & 53 & designs & System-level testbench generation and retrieval \\
\tablegroup{5}{Extra-stage: NL$\rightarrow$Netlist}
PICBench~\cite{wu2025picbench} & \href{https://github.com/PICDA/PICBench}{\githubmark} & 24 & problems & Photonic netlist construction and simulation \\
\tablegroup{5}{Cross-cutting Resources: Multi-task Benchmarks}
CVDP~\cite{pinckney2025cvdp} & \href{https://github.com/NVlabs/cvdp_benchmark}{\githubmark} & 783 & problems & RTL generation, verification, debugging, and QA \\
MMCircuitEval~\cite{zhao2025mmcircuiteval} & \href{https://github.com/cure-lab/MMCircuitEval}{\githubmark} & 3,614 & QA pairs & Multimodal circuit understanding and generation \\
\tablegroup{5}{Cross-cutting Resources: Reused Workloads}
CoreMark~\cite{galon2009coremark} & \href{https://github.com/eembc/coremark}{\githubmark} & 1 & program & Processor verification with a standard workload \\
MiBench~\cite{guthaus2001mibench} & \href{https://vhosts.eecs.umich.edu/mibench/}{\githubmark} & 35 & programs & Processor verification with embedded workloads \\
Embench IoT 2.0~\cite{patterson2025embench} & \href{https://github.com/embench/embench-iot}{\githubmark} & 19 & benchmarks & Processor verification with embedded workloads \\
\tablegroup{5}{Cross-cutting Resources: Training-oriented Datasets}
VeriThoughtsBenchmark~\cite{yubeaton2025verithoughts} & \href{https://huggingface.co/datasets/wilyub/VeriThoughtsBenchmark}{\githubmark} & 291 & tasks & Held-out evaluation of Verilog reasoning \\
MG-Verilog~\cite{zhang2024mgverilog} & \href{https://github.com/GATECH-EIC/mg-verilog}{\githubmark} & $>$11,000 & samples & Module- and repository-level Verilog modeling \\
VGen / VeriGen data~\cite{thakur2023vgen,thakur2023verigen} & \href{https://github.com/shailja-thakur/VGen}{\githubmark} & $\sim$50K + 70 & Verilog files / textbooks & Training and evaluation for Verilog generation \\
\bottomrule
\end{tabular}
\end{surveytable}
\end{table*}

\FloatBarrier

\section{Intra-stage Systems}
\label{sec:intra-stage}

Intra-stage systems consume a machine-readable EDA-domain primary input and produce a primary output that is checked and reused within the same design stage. Most preserve or locally refine a design representation. Others produce stage-local verification collateral, revise an EDA tool, or calibrate a device model from measurements. Consumer acceptance relies on a local checker or optimizer: compilation, simulation, fitting error, equivalence, PPA evaluation, lint/DRC, coverage, or an equivalent acceptance condition. \Crefrange{tab:intra-objects-rtl-repair}{tab:intra-objects-auxiliary} report the primary input--output pair and the supporting and intermediate handoff-object roles by chapter group, and \Cref{tab:intra-mechanisms} reports public code together with model-optimization and agent-harness mechanisms.

\begin{table*}[t!]
\begin{objectroletable}
\centering
\caption{Roles of handoff objects for RTL$\rightarrow$RTL in Intra-stage systems. The Primary I/O column reports $I_p\rightarrow O_p$. Grouped columns use the role-first categories defined in \Cref{sec:framework-handoff}. A checkmark denotes at least one explicitly reported object in that category. A blank denotes none.}\label{tab:intra-objects-rtl-repair}
\begin{tabular}{@{}L{2.15cm}L{3.45cm}*{9}{C{0.695cm}}@{}}
\toprule
\rowcolor{tableheader}
\textbf{System} & \textbf{Primary I/O pair ($I_p \rightarrow O_p$)} & \multicolumn{3}{c}{$\boldsymbol{I_s}$: \textbf{Supporting Inputs}} & \multicolumn{3}{c}{$\boldsymbol{O_s}$: \textbf{Supporting Outputs}} & \multicolumn{3}{c}{$\boldsymbol{A_m}$: \textbf{Intermediate Artifacts}} \\
\cmidrule(lr){3-5}\cmidrule(lr){6-8}\cmidrule(lr){9-11}
\rowcolor{tableheader}
& & {\tiny\textbf{Constr.}} & \textbf{Ref.} & {\tiny\textbf{Knowl.}} & \textbf{Check} & \textbf{QoR} & {\tiny\textbf{Record}} & \textbf{Retr.} & \textbf{Cand.} & \textbf{Coord.} \\
\midrule
RTLFixer~\cite{tsai2024rtlfixer} & RTL$\rightarrow$RTL &  &  & \cmark & \cmark &  &  & \cmark & \cmark &  \\
MEIC~\cite{xu2024meic} & RTL$\rightarrow$RTL &  &  &  & \cmark &  &  &  &  & \cmark \\
\revise{HDLdebugger}~\cite{hdldebugger2024} & RTL$\rightarrow$RTL &  &  & \cmark & \cmark &  &  & \cmark &  &  \\
\revise{AutoVeriFix}~\cite{yan2026autoverifixautomatic} & RTL$\rightarrow$RTL &  &  &  & \cmark &  &  &  & \cmark &  \\
\revise{BluesFL}~\cite{liu2026debuglikehuman} & RTL$\rightarrow$RTL &  &  &  & \cmark &  &  &  &  &  \\
\revise{Stitch}~\cite{zonta2026stitch} & RTL$\rightarrow$RTL &  & \cmark &  & \cmark &  &  &  & \cmark &  \\
AssertSolver~\cite{jie2025insightsfromrightsan} & RTL$\rightarrow$RTL & \cmark &  &  & \cmark &  &  &  & \cmark &  \\
\bottomrule
\end{tabular}
\end{objectroletable}
\end{table*}


\subsection{RTL-to-RTL}
These systems consume buggy or low-quality RTL and return the same representation after a stage-local repair loop. The repaired RTL is accepted through compilation, simulation, or lint checks. No downstream representation boundary is crossed. \Cref{tab:intra-objects-rtl-repair} summarizes these systems and their object roles.

\minisection{RTL Syntax and Functional Repair}
RTL syntax and functional repair locates and corrects malformed or behaviorally incorrect RTL while preserving the intended module interface and specification.
\textit{Typical systems.} Syntax-gate systems concentrate on compiler diagnostics. RTLFixer~\cite{tsai2024rtlfixer} combines compiler feedback with ReAct-style decomposition~\cite{yao2023react} and retrieved human guidance, while HDLdebugger~\cite{hdldebugger2024} couples supervised fine-tuning with retrieval. Functional-gate systems target behavioral correctness. AssertSolver~\cite{jie2025insightsfromrightsan} adapts the model using successful repairs, with the failing assertion and testbench defining the correction target at inference time. MEIC~\cite{xu2024meic} and AutoVeriFix~\cite{yan2026autoverifixautomatic} span both gates: MEIC separates log analysis, localization, candidate repair, and simulation-based scoring, whereas AutoVeriFix regenerates after either compilation or simulation failure. Across these systems, diagnostics are supporting outputs, retrieved guidance is an intermediate context artifact, and proposed patches remain candidates until the applicable gate accepts the repaired RTL. Runtime feedback loops revise candidates using external diagnostics and retained state, while supervised adaptation internalizes patterns from earlier repair attempts. Template- and synthesis-based repair in CirFix and RTL-Repair~\cite{ahmad2022cirfix,laeufer2024rtlrepair} provide non-agentic comparisons based on symbolically constrained patches.
\textit{Validation and metrics.} The repaired RTL must pass two ordered gates. A compiler or linter first establishes executability, measured by syntax pass@1. Simulation and regression tests then establish behavioral preservation, measured by functional fix rate or pass@$k$. Repair iterations and debugging time report the cost of reaching the accepted output.
\textit{Reported results.} The reported rates should be read at the corresponding gate. Against the one-shot LLM setting used by VerilogEval~\cite{liu2023verilogeval}, RTLFixer raises syntax pass@1 from 73\% to 93\%. MEIC reports a 93\% syntax pass rate, a 78\% functional fix rate, and up to 48$\times$ lower debugging time than experienced engineers.

\begin{table*}[t!]
\begin{objectroletable}
\centering
\caption{Roles of handoff objects for RTL$\rightarrow$Verification Collateral in Intra-stage systems. The Primary I/O column reports $I_p\rightarrow O_p$. Grouped columns use the role-first categories defined in \Cref{sec:framework-handoff}. A checkmark denotes at least one explicitly reported object in that category. A blank denotes none.}\label{tab:intra-objects-verification-collateral}
\begin{tabular}{@{}L{2.15cm}L{3.45cm}*{9}{C{0.695cm}}@{}}
\toprule
\rowcolor{tableheader}
\textbf{System} & \textbf{Primary I/O pair ($I_p \rightarrow O_p$)} & \multicolumn{3}{c}{$\boldsymbol{I_s}$: \textbf{Supporting Inputs}} & \multicolumn{3}{c}{$\boldsymbol{O_s}$: \textbf{Supporting Outputs}} & \multicolumn{3}{c}{$\boldsymbol{A_m}$: \textbf{Intermediate Artifacts}} \\
\cmidrule(lr){3-5}\cmidrule(lr){6-8}\cmidrule(lr){9-11}
\rowcolor{tableheader}
& & {\tiny\textbf{Constr.}} & \textbf{Ref.} & {\tiny\textbf{Knowl.}} & \textbf{Check} & \textbf{QoR} & {\tiny\textbf{Record}} & \textbf{Retr.} & \textbf{Cand.} & \textbf{Coord.} \\
\midrule
\revise{VerilogReader}~\cite{ruiyang2024verilogreaderllmaide} & RTL$\rightarrow$Test vectors &  &  &  & \cmark &  &  &  & \cmark &  \\
\revise{COTIA}~\cite{tan2025cotia} & RTL$\rightarrow$Test vectors &  &  &  & \cmark &  &  &  & \cmark &  \\
LLM4DV~\cite{zhang2025llm4dv} & RTL$\rightarrow$Test stimuli & \cmark &  &  & \cmark &  & \cmark &  & \cmark &  \\
\revise{AutoBench}~\cite{autobench2024} & RTL$\rightarrow$Testbench &  &  &  & \cmark &  &  &  & \cmark &  \\
UVLLM~\cite{hu2025uvllm} & RTL$\rightarrow$UVM testbench &  &  &  & \cmark &  &  &  &  & \cmark \\
\revise{Ye et al.}~\cite{junhao2025fromconcepttopractic} & RTL$\rightarrow$UVM testbench &  &  &  & \cmark &  &  &  & \cmark &  \\
\revise{PRO-V-R1}~\cite{zhao2026provr1} & RTL$\rightarrow$Testbench &  &  &  & \cmark &  &  &  & \cmark &  \\
\revise{UVMarvel}~\cite{ye2026uvmarvel} & RTL$\rightarrow$UVM testbench &  &  &  & \cmark &  &  &  & \cmark & \cmark \\
ChatTest~\cite{wan2026chattest} & RTL$\rightarrow$Testbench &  &  & \cmark & \cmark &  &  & \cmark & \cmark & \cmark \\
\revise{ChipModeler}~\cite{ye2026chipmodeler} & RTL$\rightarrow$Reference model &  &  &  & \cmark &  &  &  &  &  \\
\revise{Qayyum et al.}~\cite{qayyum2024incrementalproof} & RTL$\rightarrow$Proof &  &  &  & \cmark &  &  &  & \cmark &  \\
\revise{STELLAR}~\cite{rajabi2026stellar} & RTL$\rightarrow$SVA &  &  & \cmark & \cmark &  &  & \cmark &  &  \\
\bottomrule
\end{tabular}
\end{objectroletable}
\end{table*}

\minisection{RTL Fault Localization}
RTL fault localization identifies the statement, block, or execution slice responsible for an observed design failure before a patch is generated.
\textit{Typical systems.} BluesFL~\cite{liu2026debuglikehuman} constructs instruction-execution paths and narrows faults to block-level slices before repair. This preserves the causal path from a failing architectural behavior to the implicated RTL blocks and presents the LLM with semantically meaningful execution units. Whole-module prompting lacks this localization, while Strider~\cite{yang2024strider} uses a fixed signal-transition repair template. Separate localization accuracy and final repair success reveal whether the gain comes from representation and search-space reduction or from patch generation.
\textit{Validation and metrics.} The downstream repair step can use a reported fault location only when it identifies the responsible block or statement with a sufficiently small search space. Ground-truth localization measures this handoff directly, while final repair success and runtime measure the combined localization--patching workflow.
\textit{Reported results.} On a 19K-line RISC-V core, BluesFL localizes 24 bugs at Top-1, compared with seven for the prior baseline, and reports an average cost of \$0.257 per bug~\cite{liu2026debuglikehuman}.

\minisection{Protocol and Interface Repair}
Protocol and interface repair restores temporal and transactional conformance between communicating RTL blocks under explicit protocol properties.
\textit{Typical systems.} Stitch~\cite{zonta2026stitch} uses a failing assertion and its trace to guide patching of on-chip protocol implementations. The relevant baselines are assertion-guided symbolic repair such as RTL-Repair~\cite{laeufer2024rtlrepair} and transition-guided repair such as Strider~\cite{yang2024strider}. Unconstrained code generation does not use equivalent diagnostic information. The independent protocol property and regression suite support acceptance because compilation or a single passing stimulus cannot capture a temporal acceptance condition.
\textit{Validation and metrics.} The repaired interface is accepted by the protocol checker only after the governing properties pass and prior regression behavior is preserved. Compilation is only an executability gate. Repaired-interface correctness, iterations, and checker calls describe closure quality and cost.
\textit{Reported results.} On 100 violations across 11 implementations of five protocols, Stitch reaches 61\% patch success after five iterations, up from 19\% in its initial four-model setting~\cite{zonta2026stitch}.
\subsection{RTL-to-Verification Collateral}
These systems consume RTL and produce collateral used to check it. Their primary outputs fall into four groups: test vectors and stimuli, testbenches and UVM environments, behavioral reference models, and assertions or proofs. Specifications, tool configurations, and independently prepared reference artifacts remain supporting inputs. Because generated collateral may inherit the generator's interpretation of the task, its validation depends on both the relevant checker and the behavioral or property scope exercised. \Cref{tab:intra-objects-verification-collateral} summarizes these systems and their object roles.

\minisection{Test Vectors and Stimuli}
Test-vector and stimulus generation produces input sequences that exercise selected RTL behaviors. Stimuli may additionally specify timing, driving behavior, and a coverage plan.
\textit{Typical systems.} VerilogReader~\cite{ruiyang2024verilogreaderllmaide} uses Verilator coverage to revise generated test vectors, while COTIA~\cite{tan2025cotia} uses concolic path exploration to select the next vector from branches that remain uncovered. LLM4DV~\cite{zhang2025llm4dv} generates broader test stimuli from an explicit coverage plan and retains the run history so that simulation and coverage feedback determine the next inputs. In object-role terms, proposed vectors are candidate artifacts, simulator and coverage reports are supporting outputs, and LLM4DV's accumulated history is a workflow record.
\textit{Validation and metrics.} The simulator must execute the generated inputs against the DUT. Coverage indicates which behaviors were exercised, while agreement with independently specified outcomes and injected-fault detection determine whether those inputs expose relevant behavior. Vector count, simulation count, and correction rounds measure generation cost.
\textit{Reported results.} VerilogReader reaches nearly 100\% line coverage on a 16-state FSM after 20 generated inputs~\cite{ruiyang2024verilogreaderllmaide}. COTIA reaches full coverage on several smaller benchmarks and reports nearly 95\% lower execution time on b11 and 71\% lower memory use on b12 than conventional concolic testing~\cite{tan2025cotia}.

\minisection{Testbench and UVM Environments}
Testbench and UVM-environment generation produces executable verification code that drives the DUT, observes its responses, and checks expected behavior.
\textit{Typical systems.} The primary output separates three standard verification artifacts. (a) \textit{Testbenches.} AutoBench~\cite{autobench2024} uses Icarus Verilog feedback for DUT-facing testbench code and reference behavior, while ChatTest~\cite{wan2026chattest} uses divide-and-conquer agents to revise testbenches toward coverage closure. (b) \textit{UVM environments.} UVLLM~\cite{hu2025uvllm} connects specialized generation and debugging roles to simulators, synthesizers, and linters, Ye et al.~\cite{junhao2025fromconcepttopractic} close a three-role construction loop with VCS coverage, and UVMarvel~\cite{ye2026uvmarvel} separates planning, generation, execution, and repair. (c) \textit{Testbenches with reference outputs.} PRO-V-R1~\cite{zhao2026provr1} pairs generated input vectors with cycle-aligned expected outputs from sampled Python functional reference models in a final \texttt{testbench.json}. Its sampled models and intermediate testbenches remain candidates until the simulation driver accepts the paired artifact. Constrained-random stimuli, manually authored environments, and same-backbone direct prompting under matched tool access provide the corresponding comparisons.
\textit{Validation and metrics.} The simulator must compile and execute the generated environment before it can assess the DUT. Agreement with independently specified behavior and detection of injected faults measure functional checking, while coverage, simulation count, correction rounds, and tool cost characterize verification scope and production effort.
\textit{Reported results.} On 20 complex RTL designs, ChatTest reports 1.46$\times$ toggle coverage and 2.28$\times$ line coverage relative to its evaluated baseline, together with a 24.23\% functional-coverage gain~\cite{wan2026chattest}. UVMarvel reports 95.65\% average code coverage and reduces the reported verification time from days to 4.5 hours~\cite{ye2026uvmarvel}. PRO-V-R1 reaches 57.7\% functional correctness and 34.0\% robust fault detection, compared with 25.7\% and 21.8\% for its evaluated baseline~\cite{zhao2026provr1}.

\minisection{Behavioral Reference Models}
Behavioral-reference-model generation produces an executable, higher-level account of expected behavior for comparison with the RTL implementation.
\textit{Typical systems.} ChipModeler~\cite{ye2026chipmodeler} generates an executable reference model and compares it with RTL through co-simulation. The reference model is the primary output, while co-simulation results are supporting outputs passed to the verification flow. Shared omissions remain possible if the same prompt and model generate both the RTL and its reference.
\textit{Validation and metrics.} The verification flow accepts a behavioral reference only if it executes and supports co-simulation against the RTL. Agreement, exercised behaviors, and divergence detection measure its checking value. These results are strongest when the reference is independently derived, and construction effort records the cost of producing it.
\textit{Reported results.} Across 300 designs, ChipModeler reports a peak performance gain of 58.99\% over the evaluated alternatives, 9.18$\times$ greater reference-model generation capacity, and 7.11$\times$ faster design-and-validation cycles than manual development~\cite{ye2026chipmodeler}.

\begin{table*}[t!]
\begin{objectroletable}
\centering
\caption{Roles of handoff objects for Netlist$\rightarrow$Netlist in Intra-stage systems. The Primary I/O column reports $I_p\rightarrow O_p$. Grouped columns use the role-first categories defined in \Cref{sec:framework-handoff}. A checkmark denotes at least one explicitly reported object in that category. A blank denotes none.}\label{tab:intra-objects-netlist-analog}
\begin{tabular}{@{}L{2.15cm}L{3.45cm}*{9}{C{0.695cm}}@{}}
\toprule
\rowcolor{tableheader}
\textbf{System} & \textbf{Primary I/O pair ($I_p \rightarrow O_p$)} & \multicolumn{3}{c}{$\boldsymbol{I_s}$: \textbf{Supporting Inputs}} & \multicolumn{3}{c}{$\boldsymbol{O_s}$: \textbf{Supporting Outputs}} & \multicolumn{3}{c}{$\boldsymbol{A_m}$: \textbf{Intermediate Artifacts}} \\
\cmidrule(lr){3-5}\cmidrule(lr){6-8}\cmidrule(lr){9-11}
\rowcolor{tableheader}
& & {\tiny\textbf{Constr.}} & \textbf{Ref.} & {\tiny\textbf{Knowl.}} & \textbf{Check} & \textbf{QoR} & {\tiny\textbf{Record}} & \textbf{Retr.} & \textbf{Cand.} & \textbf{Coord.} \\
\midrule
ADO-LLM~\cite{yin2024ado} & Netlist$\rightarrow$Netlist & \cmark &  &  & \cmark & \cmark &  &  & \cmark &  \\
AnaFlow~\cite{ahmadzadeh2025anaflow} & Netlist$\rightarrow$Netlist & \cmark &  &  & \cmark & \cmark &  &  & \cmark &  \\
\revise{ASTRA}~\cite{xing2025astra} & Netlist$\rightarrow$Netlist &  &  &  & \cmark & \cmark &  &  & \cmark &  \\
LLM-USO~\cite{somayaji2025llm} & Netlist$\rightarrow$Netlist &  &  &  & \cmark & \cmark &  &  & \cmark &  \\
\bottomrule
\end{tabular}
\end{objectroletable}
\end{table*}

\minisection{Assertions and Proofs}
Assertion and proof generation translates design intent or observed behavior into properties and invariants that a formal engine can check.
\textit{Typical systems.} Qayyum et al.~\cite{qayyum2024incrementalproof} propose invariants incrementally and admit each candidate only after Z3 proof checking, so a mechanically independent solver gates progress. STELLAR~\cite{rajabi2026stellar} retrieves structurally relevant assertions before generation, using structure to reduce the property search space, but likewise requires coverage evidence to show that the resulting property set is not merely provable. The relevant non-agentic baseline is static-analysis and data-mining assertion generation such as GoldMine~\cite{vasudevan2010goldmine}, alongside human-authored properties and the same conventional formal engine. An agent is useful when a counterexample changes the next candidate. Proof success establishes only the stated property, not completeness of the property set.
\textit{Validation and metrics.} A formal engine can consume only syntactically valid properties, but proof success alone does not show that they express the intended behavior. Semantic agreement and cone/property coverage address relevance and scope, while counterexample-guided iterations and solver calls measure the effort required to obtain an acceptable property set.
\textit{Reported results.} Qayyum et al. demonstrate incremental proof construction on ripple-carry-adder and Dadda-tree-multiplier case studies~\cite{qayyum2024incrementalproof}. For GPT-4o mini, STELLAR raises SVA syntax correctness from 84.9\% to 99.8\%~\cite{rajabi2026stellar}.

\begin{table*}[t!]
\begin{objectroletable}
\centering
\caption{Roles of handoff objects for Layout$\rightarrow$Layout in Intra-stage systems. The Primary I/O column reports $I_p\rightarrow O_p$. Grouped columns use the role-first categories defined in \Cref{sec:framework-handoff}. A checkmark denotes at least one explicitly reported object in that category. A blank denotes none.}\label{tab:intra-objects-layout-drc}
\begin{tabular}{@{}L{2.15cm}L{3.45cm}*{9}{C{0.695cm}}@{}}
\toprule
\rowcolor{tableheader}
\textbf{System} & \textbf{Primary I/O pair ($I_p \rightarrow O_p$)} & \multicolumn{3}{c}{$\boldsymbol{I_s}$: \textbf{Supporting Inputs}} & \multicolumn{3}{c}{$\boldsymbol{O_s}$: \textbf{Supporting Outputs}} & \multicolumn{3}{c}{$\boldsymbol{A_m}$: \textbf{Intermediate Artifacts}} \\
\cmidrule(lr){3-5}\cmidrule(lr){6-8}\cmidrule(lr){9-11}
\rowcolor{tableheader}
& & {\tiny\textbf{Constr.}} & \textbf{Ref.} & {\tiny\textbf{Knowl.}} & \textbf{Check} & \textbf{QoR} & {\tiny\textbf{Record}} & \textbf{Retr.} & \textbf{Cand.} & \textbf{Coord.} \\
\midrule
LayoutCopilot~\cite{liu2025layoutcopilot} & Layout$\rightarrow$Layout & \cmark &  &  & \cmark &  & \cmark &  & \cmark &  \\
\revise{drcAgent}~\cite{ma2026drcagent} & Layout$\rightarrow$Layout &  &  &  & \cmark &  &  &  & \cmark &  \\
\revise{Wu et al.}~\cite{wu2026agenticdrc} & Layout$\rightarrow$Layout & \cmark &  &  & \cmark &  &  &  & \cmark & \cmark \\
Ho and Ren~\cite{ho2024large} & Layout$\rightarrow$Layout & \cmark &  &  & \cmark & \cmark &  &  & \cmark &  \\
\bottomrule
\end{tabular}
\end{objectroletable}
\end{table*}

\subsection{Netlist-to-Netlist}
These four systems optimize analog/AMS circuit parameters without crossing a design-stage boundary. SPICE-class simulation establishes feasibility and QoR. \Cref{tab:intra-objects-netlist-analog} summarizes these systems and their object roles.

\minisection{Sizing and Multi-Objective Optimization}
Circuit sizing and multi-objective optimization select device or netlist parameters that satisfy constraints while improving competing performance objectives.
\textit{Typical systems.} ADO-LLM~\cite{yin2024ado} couples one GPT-3.5-Turbo proposer with Gaussian-process Bayesian optimization and evaluates candidates in HSPICE using a commercial 90 nm CMOS PDK. Its zero-shot initialization and in-context proposals target faster discovery of feasible regions than GP-BO alone. AnaFlow~\cite{ahmadzadeh2025anaflow} uses several reasoning stages within one reported LLM-driven workflow. It makes operating-region reasoning explicit and calls NGSPICE on the open 45 nm PTM, enabling a reproducible comparison with numerical sizing under the same simulation budget. ASTRA~\cite{xing2025astra} uses GPT-4o and combines MCP-guided design initialization with Bayesian optimization focused on critical design variables. Its circuit simulator and process technology are unspecified. LLM-USO~\cite{somayaji2025llm} separates summarization, critique, and optimization into three LLM roles and closes its uncertainty-guided exploration loop with an unspecified SPICE simulator and PDK. In these systems, the LLM supplies topology-aware hypotheses, while batch BO~\cite{lyu2018batchbo}, transferable GCN-RL~\cite{wang2020gcnrl}, AutoCkt~\cite{settaluri2020autockt}, SMAC~\cite{hutter2011smac}, or evolutionary search performs quantitative selection. Their central claim is sample efficiency, so simulator calls and feasibility rate are as important as the final figure of merit.
\textit{Validation and metrics.} A circuit simulator accepts a candidate only when it is executable and satisfies the required specifications. PPA, figure of merit, and yield rank the feasible outputs, while simulator calls, convergence, and wall-clock time measure the cost of selecting the returned netlist.
\textit{Reported results.} Across its analog-sizing experiments, ASTRA reports 1.40--4.35$\times$ faster convergence and 1.29--2.36$\times$ performance improvement over the evaluated optimization methods~\cite{xing2025astra}.
\subsection{Layout-to-Layout}
These systems consume a layout and return a repaired or refined layout. Acceptance depends on DRC, geometric legality, and, when performance is affected, extraction and post-layout simulation. \Cref{tab:intra-objects-layout-drc} summarizes these systems and their object roles.

\minisection{DRC Repair}
DRC repair identifies geometric rule violations and applies layout edits until the physical-design checker accepts the result.
\textit{Typical systems.} Wu et al.~\cite{wu2026agenticdrc} evaluate tool-using LLM agents on distinct DRV-detection and repair tasks, with KLayout~\cite{klayout} and OpenROAD~\cite{ajayi2019openroad} providing layout manipulation and physical-design feedback. drcAgent~\cite{ma2026drcagent} uses two roles in an adversarial relationship: a violation generator creates hard cases while a Qwen2.5-VL-based fixer learns from Calibre DRC feedback. The first method emphasizes benchmarked diagnosis and repair in a digital implementation context. The second emphasizes robustness to generated violation distributions and an industrial signoff checker. Compared with fixed rule-based repair, both can choose context-dependent edits. Compared with unconstrained visual reasoning, tool-mediated geometry keeps candidates executable. A clean DRC report establishes legality. Connectivity, timing, and parasitic checks cover electrical effects of the edit.
\textit{Validation and metrics.} A layout checker first rejects candidates with remaining DRC violations. Connectivity and downstream PPA checks then determine whether a legal edit preserves the design, while repaired-case rate, edit--check iterations, and signoff runtime measure closure coverage and cost.
\textit{Reported results.} drcAgent reports average repair rates of 76\%, 60\%, and 25\% for Tier-1, Tier-2, and Tier-3 rules, respectively~\cite{ma2026drcagent}. The focus-session study likewise finds strong performance on small structured cases and a marked decline as design scale and rule interaction increase~\cite{wu2026agenticdrc}.

\minisection{Analog-Layout Refinement}
Analog-layout refinement converts designer intent into placement and routing edits while retaining human control over electrically sensitive trade-offs.
\textit{Typical systems.} LayoutCopilot~\cite{liu2025layoutcopilot} decomposes a designer request into specialized placement and routing commands and keeps the human in the acceptance loop. Its comparison between a monolithic agent and specialized agents attributes improved sanity-check and command success to role decomposition. Non-agentic layout synthesis such as MAGICAL~\cite{chen2021magical} and interactive placement/routing tools provide the relevant automation reference, while human preference captures intent that DRC cannot express. The trade-off is reduced autonomy: the handoff preserves the request, command history, and rejected alternatives for designer review.
\textit{Validation and metrics.} The layout environment must execute the generated commands and pass sanity checks before the designer can review the result. DRC and electrical preservation support technical acceptance. Task completion, interaction count, and human time show how much control remains with the designer.
\textit{Reported results.} LayoutCopilot demonstrates both direct command execution and multi-agent refinement of abstract requests in real analog-layout case studies~\cite{liu2025layoutcopilot}.

\begin{table*}[t!]
\begin{objectroletable}
\centering
\caption{Roles of handoff objects for Other Transformations in Intra-stage systems. The Primary I/O column reports $I_p\rightarrow O_p$. Grouped columns use the role-first categories defined in \Cref{sec:framework-handoff}. A checkmark denotes at least one explicitly reported object in that category. A blank denotes none.}\label{tab:intra-objects-auxiliary}
\begin{tabular}{@{}L{2.15cm}L{3.45cm}*{9}{C{0.695cm}}@{}}
\toprule
\rowcolor{tableheader}
\textbf{System} & \textbf{Primary I/O pair ($I_p \rightarrow O_p$)} & \multicolumn{3}{c}{$\boldsymbol{I_s}$: \textbf{Supporting Inputs}} & \multicolumn{3}{c}{$\boldsymbol{O_s}$: \textbf{Supporting Outputs}} & \multicolumn{3}{c}{$\boldsymbol{A_m}$: \textbf{Intermediate Artifacts}} \\
\cmidrule(lr){3-5}\cmidrule(lr){6-8}\cmidrule(lr){9-11}
\rowcolor{tableheader}
& & {\tiny\textbf{Constr.}} & \textbf{Ref.} & {\tiny\textbf{Knowl.}} & \textbf{Check} & \textbf{QoR} & {\tiny\textbf{Record}} & \textbf{Retr.} & \textbf{Cand.} & \textbf{Coord.} \\
\midrule
\revise{CROP}~\cite{jingyu2025cropcircuitretrieval} & RTL$\rightarrow$Flow parameters & \cmark &  & \cmark & \cmark & \cmark &  & \cmark &  &  \\
ORFS-Agent~\cite{ghose2025orfs} & RTL$\rightarrow$Flow parameters & \cmark &  &  & \cmark & \cmark &  &  & \cmark & \cmark \\
\revise{RTL-BenchMT}~\cite{wang2026rtlbenchmt} & RTL suite$\rightarrow$RTL suite & \cmark &  & \cmark & \cmark &  & \cmark &  & \cmark & \cmark \\
\revise{HLSRewriter}~\cite{xu2026hlsrewriter} & C/\allowbreak C++$\rightarrow$C/\allowbreak C++ & \cmark &  &  &  & \cmark &  &  &  &  \\
\revise{YieldAgent}~\cite{qin2025multi} & Netlist$\rightarrow$Yield report & \cmark &  &  & \cmark &  &  &  & \cmark & \cmark \\
\revise{ModelGen}~\cite{wei2025modelgen} & Measured characteristics$\rightarrow$Model parameters & \cmark &  & \cmark &  & \cmark &  &  &  &  \\
\revise{LLMVA}~\cite{su2026llmva} & Verilog-A$\rightarrow$Verilog-A & \cmark &  &  & \cmark &  &  &  & \cmark &  \\
\revise{MappingEvolve}~\cite{fu2026mappingevolve} & Tool code$\rightarrow$Tool code & \cmark & \cmark &  & \cmark & \cmark &  &  & \cmark &  \\
\revise{Yu and Ren}~\cite{yu2026autonomous} & Tool code$\rightarrow$Tool code & \cmark & \cmark &  & \cmark & \cmark &  &  & \cmark &  \\
\bottomrule
\end{tabular}
\end{objectroletable}
\end{table*}

\minisection{Standard-Cell Layout Optimization}
Standard-cell layout optimization revises a cell layout under design rules and electrical objectives while preserving its logic function.
\textit{Typical systems.} Ho and Ren~\cite{ho2024large} use a ReAct-style agent to inspect tool results, revise geometric choices, and explore alternative layouts. A design-rule checker filters invalid candidates, while characterization results rank valid alternatives. Both are required because DRC cleanliness alone does not establish layout quality.
\textit{Validation and metrics.} DRC is the minimum gate for returning a manufacturable cell layout. Characterized area, delay, and power distinguish legal candidates, and iteration cost records the effort required to select the final geometry.
\textit{Reported results.} Across 17 complex sequential cells in a 2~nm technology, Ho and Ren report up to 19.4\% smaller cell area and 100\% LVS/DRC-clean layouts~\cite{ho2024large}.

\begin{figure*}[t!]
\centering
\includegraphics[width=\textwidth]{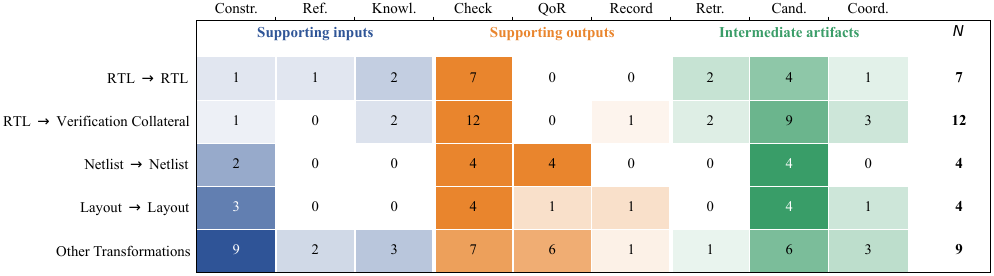}
\caption{Handoff-object role incidence across Intra-stage chapter groups. Each cell gives the number of systems reporting that role, and darker cells indicate a larger share of the systems in the corresponding row.}
\Description{An annotated heatmap compares nine supporting-input, supporting-output, and intermediate-artifact categories across five Intra-stage chapter groups containing 36 systems.}
\label{fig:intra-object-role-trends}
\end{figure*}


\begin{table*}[t!]
\begin{surveytable}
\centering
\caption{Model-optimization and agent-harness mechanisms in Intra-stage systems. A checkmark denotes explicit implementation evidence under the definitions in \Cref{sec:framework-mechanisms}. A linked GitHub icon denotes public code.}\label{tab:intra-mechanisms}
\begin{tabular}{@{}L{2.50cm}C{0.65cm}C{0.76cm}C{0.76cm}C{1.20cm}C{0.82cm}C{1.15cm}C{1.25cm}L{1.55cm}@{}}
\toprule
\rowcolor{tableheader}
\textbf{System} & \textbf{Code} & \multicolumn{2}{c}{\textbf{Model Optimization}} & \multicolumn{4}{c}{\textbf{Agent Harness}} & \textbf{Source} \\
\cmidrule(lr){3-4}\cmidrule(lr){5-8}
\rowcolor{tableheader}
& & \textbf{SFT} & \textbf{RL} & \textbf{RAG/Mem.} & \textbf{Loop} & \textbf{Orchestr.} & \textbf{Search/Evo.} & \\
\midrule
\tablegroup{9}{RTL$\rightarrow$RTL}
RTLFixer~\cite{tsai2024rtlfixer} & \href{https://github.com/NVlabs/RTLFixer}{\githubmark} &  &  & \cmark & \cmark &  &  & DAC'24 \\
MEIC~\cite{xu2024meic} & -- &  &  &  &  & \cmark &  & ICCAD'24 \\
\revise{HDLdebugger}~\cite{hdldebugger2024} & -- & \cmark &  & \cmark &  &  &  & \revise{TODAES'25} \\
\revise{AutoVeriFix}~\cite{yan2026autoverifixautomatic} & \href{https://github.com/HKUSTGZ-MICS-LYU/AutoVeriFix}{\githubmark} &  &  &  & \cmark &  &  & \revise{ASP-DAC'26} \\
\revise{BluesFL}~\cite{liu2026debuglikehuman} & \href{https://github.com/pointerliu/bluesfl}{\githubmark} &  &  &  &  &  &  & \revise{DAC'26} \\
\revise{Stitch}~\cite{zonta2026stitch} & -- &  &  &  & \cmark &  &  & \revise{DAC'26} \\
AssertSolver~\cite{jie2025insightsfromrightsan} & \href{https://github.com/SEU-ACAL/reproduce-AssertSolver-DAC-25}{\githubmark} & \cmark &  &  &  &  &  & DAC'25 \\
\tablegroup{9}{RTL$\rightarrow$Verification Collateral}
\revise{Qayyum et al.}~\cite{qayyum2024incrementalproof} & -- &  &  &  & \cmark &  &  & \revise{DAC'24} \\
\revise{VerilogReader}~\cite{ruiyang2024verilogreaderllmaide} & -- &  &  &  & \cmark &  &  & \revise{LAD'24} \\
\revise{AutoBench}~\cite{autobench2024} & \href{https://github.com/AutoBench/AutoBench}{\githubmark} &  &  &  & \cmark &  &  & \revise{MLCAD'24} \\
UVLLM~\cite{hu2025uvllm} & -- &  &  &  &  & \cmark &  & DAC'25 \\
\revise{COTIA}~\cite{tan2025cotia} & -- &  &  &  & \cmark &  & \cmark & \revise{ICCAD'25} \\
\revise{Ye et al.}~\cite{junhao2025fromconcepttopractic} & -- &  &  &  & \cmark &  &  & \revise{ICCAD'25} \\
\revise{ChipModeler}~\cite{ye2026chipmodeler} & -- &  &  &  &  &  &  & \revise{DAC'26} \\
\revise{PRO-V-R1}~\cite{zhao2026provr1} & \href{https://github.com/stable-lab/Pro-V}{\githubmark} & \cmark & \cmark &  & \cmark &  &  & \revise{DAC'26} \\
\revise{STELLAR}~\cite{rajabi2026stellar} & -- &  &  & \cmark &  &  &  & \revise{DAC'26} \\
\revise{UVMarvel}~\cite{ye2026uvmarvel} & \href{https://github.com/SEU-ACAL/reproduce-UVMarvel-DAC-26}{\githubmark} &  &  &  & \cmark & \cmark &  & \revise{DAC'26} \\
ChatTest~\cite{wan2026chattest} & -- &  &  & \cmark & \cmark & \cmark &  & DATE'26 \\
LLM4DV~\cite{zhang2025llm4dv} & \href{https://github.com/ZixiBenZhang/ml4dv}{\githubmark} &  &  &  & \cmark &  &  & FCCM'25 \\
\tablegroup{9}{Netlist$\rightarrow$Netlist}
ADO-LLM~\cite{yin2024ado} & -- &  &  &  & \cmark &  & \cmark & ICCAD'24 \\
AnaFlow~\cite{ahmadzadeh2025anaflow} & -- &  &  &  & \cmark &  &  & ICCAD'25 \\
\revise{ASTRA}~\cite{xing2025astra} & \href{https://github.com/IceLab-X/ASTRA}{\githubmark} &  &  &  & \cmark &  & \cmark & \revise{ICCAD'25} \\
LLM-USO~\cite{somayaji2025llm} & -- &  &  &  & \cmark &  & \cmark & TCAD'25 \\
\tablegroup{9}{Layout$\rightarrow$Layout}
LayoutCopilot~\cite{liu2025layoutcopilot} & -- &  &  &  & \cmark &  &  & TCAD'25 \\
\revise{drcAgent}~\cite{ma2026drcagent} & -- &  & \cmark &  & \cmark &  &  & \revise{DAC'26} \\
\revise{Wu et al.}~\cite{wu2026agenticdrc} & -- &  &  &  & \cmark & \cmark &  & \revise{DAC'26} \\
Ho and Ren~\cite{ho2024large} & -- &  &  &  & \cmark &  & \cmark & LAD'24 \\
\tablegroup{9}{Other Transformations}
\revise{CROP}~\cite{jingyu2025cropcircuitretrieval} & -- &  &  & \cmark &  &  &  & \revise{ICCAD'25} \\
ORFS-Agent~\cite{ghose2025orfs} & \href{https://github.com/ABKGroup/ORFS-Agent}{\githubmark} &  &  &  & \cmark & \cmark & \cmark & MLCAD'25 \\
\revise{RTL-BenchMT}~\cite{wang2026rtlbenchmt} & -- &  &  & \cmark & \cmark &  &  & \revise{DAC'26} \\
\revise{HLSRewriter}~\cite{xu2026hlsrewriter} & \href{https://github.com/code-source1/catapult}{\githubmark} &  &  &  &  &  &  & \revise{TODAES'26} \\
\revise{YieldAgent}~\cite{qin2025multi} & -- &  &  &  & \cmark & \cmark & \cmark & DAC'25 \\
\revise{ModelGen}~\cite{wei2025modelgen} & -- &  &  &  &  &  &  & \revise{TODAES'25} \\
\revise{LLMVA}~\cite{su2026llmva} & -- & \cmark &  &  & \cmark &  &  & \revise{DAC'26} \\
\revise{MappingEvolve}~\cite{fu2026mappingevolve} & \href{https://github.com/algorithmicsuperintelligence/openevolve}{\githubmark} &  &  &  & \cmark &  & \cmark & \revise{DAC'26} \\
\revise{Yu and Ren}~\cite{yu2026autonomous} & -- &  &  &  & \cmark &  & \cmark & DAC'26 \\
\bottomrule
\end{tabular}
\end{surveytable}
\end{table*}

\subsection{Other Transformations}
This residual group contains nine systems across six low-frequency Intra-stage tasks: flow configuration, RTL benchmark maintenance, HLS source refactoring, yield estimation, device-model construction and refinement, and EDA-tool self-evolution. Their distinct input--output representations are analyzed separately within the shared boundary. \Cref{tab:intra-objects-auxiliary} summarizes these systems and their object roles.

\minisection{Flow Configuration}
Flow configuration maps an RTL design and prior implementation experience to tool parameters that improve downstream QoR without changing the design representation.
\textit{Typical systems.} CROP~\cite{jingyu2025cropcircuitretrieval} retrieves semantically similar synthesis runs and proposes parameters whose QoR is measured by Design Compiler. ORFS-Agent~\cite{ghose2025orfs} schedules parallel OpenROAD-flow-scripts runs, reads intermediate timing, power, and wirelength metrics, and iteratively modifies \texttt{config.mk} and SDC parameters. Its LLM remains frozen while the harness explores candidates directly or invokes Bayesian-optimization tools. The relevant baseline for both systems is parameter search under the same tool, design, technology, parameter domain, and execution budget.
\textit{Validation and metrics.} The configured flow must execute successfully before its downstream QoR can be compared with the starting or searched configurations. Valid-configuration rate, evaluated configurations, and runtime measure closure efficiency. For CROP, retrieval relevance also shows whether prior flow records supplied useful guidance.
\textit{Reported results.} Against OR-AutoTuner under the reported budgets, ORFS-Agent's 12-parameter tool-using configuration uses 600 rather than 1,000 iterations and reports average improvements above 13\% for the optimized routed-wirelength or effective-clock-period objective.

\minisection{RTL Benchmark Maintenance}
RTL benchmark maintenance revises executable benchmark cases while preserving behavior, provenance, and simulator compatibility.
\textit{Typical systems.} RTL-BenchMT~\cite{wang2026rtlbenchmt} uses three roles and both VCS and Icarus Verilog to preserve revision provenance while cross-checking executable cases. Static human curation under the same cases and simulator matrix is the relevant baseline.
\textit{Validation and metrics.} A revised benchmark case is reusable only if the supported simulators can compile and execute it with consistent expected behavior. Cross-simulator agreement, provenance completeness, accepted revisions, and maintenance effort therefore determine whether the revised suite can be handed to later evaluations.
\textit{Reported results.} RTL-BenchMT identifies and revises 47 ambiguous cases drawn from six upstream benchmark suites, producing simulator-checked replacements for subsequent model evaluation~\cite{wang2026rtlbenchmt}.

\minisection{HLS Source Refactoring}
HLS source refactoring rewrites C/C++ into a form accepted by an HLS compiler before the representation-changing synthesis step.
\textit{Typical systems.} HLSRewriter~\cite{xu2026hlsrewriter} uses one LLM role to repair or restructure unsupported source forms. Compiler transformation and direct diagnostic-guided repair under the same HLS tool and source set provide the relevant baselines.
\textit{Validation and metrics.} The HLS tool can consume a refactored program only after it compiles and preserves the source behavior. Synthesis success and resulting QoR determine whether the rewrite serves its downstream purpose. Repair iterations and runtime report the cost of producing it.
\textit{Reported results.} HLSRewriter raises the refactoring pass rate by 30.84\% over direct LLM rewriting and reports average reductions of 36.58\% in latency, 17.78\% in area, and 7.52\% in power~\cite{xu2026hlsrewriter}.

\minisection{Yield Estimation}
Yield estimation characterizes circuit robustness under process and operating variation and returns a statistical report rather than a revised netlist.
\textit{Typical systems.} YieldAgent~\cite{qin2025multi} coordinates Monte Carlo exploration toward rare failure regions. The netlist and yield constraints define the primary analysis input, sampled designs and scores remain candidate artifacts, and simulation results support the returned yield report. Its loop, orchestration, and search mechanisms select informative experiments. Conventional Monte Carlo sampling under the same simulator, variation model, and sample budget provides the direct baseline.
\textit{Validation and metrics.} Estimation error, calibration, and detected rare failures determine whether the report characterizes the stated variation space. Simulator calls and runtime record the evidence budget used to produce it.
\textit{Reported results.} Across 12~nm and 40~nm process settings, YieldAgent reports up to 2.9$\times$ lower computational overhead while maintaining or improving the evaluated estimation accuracy~\cite{qin2025multi}.

\minisection{Device-Model Construction and Refinement}
Device-model construction and refinement either fits model parameters to measured characteristics or revises an executable behavioral model against simulator-observed behavior.
\textit{Typical systems.} ModelGen~\cite{wei2025modelgen} converts measured device-characterization curves into calibrated BSIM or ASM-HEMT parameters by minimizing the discrepancy between simulated and measured characteristics. LLMVA~\cite{su2026llmva} uses a supervised model and an analog-simulation loop to revise Verilog-A directly. For ModelGen, measurements and model definitions constrain the fitted parameters. For LLMVA, the proposed Verilog-A revisions remain candidate artifacts until Cadence Virtuoso simulation accepts them. Conventional parameter fitting or manual model editing under the same data, model form, and simulator provides the corresponding baseline.
\textit{Validation and metrics.} The returned model must compile or simulate and reproduce the target behavior with low error while remaining physically consistent across the stated operating regions. Measurement provenance, model version, environment versions, simulation calls, and runtime qualify that result.
\textit{Reported results.} ModelGen improves pass@1, pass@3, and pass@5 by 26.8--33.1\% over its evaluated baseline~\cite{wei2025modelgen}. LLMVA reports an approximately 20\% accuracy gain and a 12$\times$ reduction in model-correction time~\cite{su2026llmva}.

\minisection{EDA Tool Self-Evolution}
EDA tool self-evolution modifies the implementation of an EDA optimizer and accepts changes only after regression, equivalence, and QoR checks.
\textit{Typical systems.} MappingEvolve~\cite{fu2026mappingevolve} evolves technology-mapping code through a population-based LLM search under compilation, logical-equivalence, and QoR feedback. Yu and Ren~\cite{yu2026autonomous} likewise edit ABC's own source code under compilation, combinational-equivalence, and QoR feedback. Their task-specific baseline is the unmodified ABC optimization flow~\cite{brayton2010abc} under the same benchmarks and command budget. Because their primary input--output pairs connect versions of a machine-readable EDA tool within the same synthesis stage, these systems are Intra-stage. Source-code provenance and full regression evidence remain necessary because the modified producer affects later transformations.
\textit{Validation and metrics.} The modified tool must build and pass its regression and equivalence checks before any QoR gain is admissible. QoR change then measures improvement, while failed edits, commands or evaluations, and wall-clock time report the cost of delivering the revised implementation.
\textit{Reported results.} MappingEvolve reports 10.04\% area reduction relative to ABC and 7.93\% relative to mockturtle~\cite{fu2026mappingevolve}. Yu and Ren demonstrate correctness-preserving evolution on the full million-line ABC codebase across multiple benchmark suites~\cite{yu2026autonomous}.
\FloatBarrier

\subsection{Discussion}
\label{sec:intra-discussion}

\noindent\revise{\textbf{Handoff Object Observation.} \Cref{fig:intra-object-role-trends} shows checker outputs in 34 of 36 systems and candidate artifacts in 27, but explicit reference inputs in only three. All 12 RTL-to-Verification Collateral systems report checker outputs, and all four Netlist-to-Netlist sizing systems combine checking and candidate revision with QoR evaluation. Intra-stage workflows therefore most often expose the results and candidates produced while an existing machine-readable artifact is checked and revised.}

\begin{table*}[t!]
\begin{objectroletable}
\centering
\caption{Roles of handoff objects for C/C++$\rightarrow$RTL in Inter-stage systems. The Primary I/O column reports $I_p\rightarrow O_p$. Grouped columns use the role-first categories defined in \Cref{sec:framework-handoff}. A checkmark denotes at least one explicitly reported object in that category. A blank denotes none.}\label{tab:inter-objects-hls}
\begin{tabular}{@{}L{2.15cm}L{3.45cm}*{9}{C{0.695cm}}@{}}
\toprule
\rowcolor{tableheader}
\textbf{System} & \textbf{Primary I/O pair ($I_p \rightarrow O_p$)} & \multicolumn{3}{c}{$\boldsymbol{I_s}$: \textbf{Supporting Inputs}} & \multicolumn{3}{c}{$\boldsymbol{O_s}$: \textbf{Supporting Outputs}} & \multicolumn{3}{c}{$\boldsymbol{A_m}$: \textbf{Intermediate Artifacts}} \\
\cmidrule(lr){3-5}\cmidrule(lr){6-8}\cmidrule(lr){9-11}
\rowcolor{tableheader}
& & {\tiny\textbf{Constr.}} & \textbf{Ref.} & {\tiny\textbf{Knowl.}} & \textbf{Check} & \textbf{QoR} & {\tiny\textbf{Record}} & \textbf{Retr.} & \textbf{Cand.} & \textbf{Coord.} \\
\midrule
\revise{GPT4AIGChip}~\cite{fu2023gpt4aigchip} & HLS template$\rightarrow$RTL & \cmark &  &  &  & \cmark &  &  & \cmark &  \\
\revise{SynthAI}~\cite{synthai2024} & C/\allowbreak C++$\rightarrow$RTL & \cmark &  &  & \cmark & \cmark &  &  & \cmark & \cmark \\
\revise{HLSPilot}~\cite{chenwei2024hlspilotllmbasedhigh} & C/\allowbreak C++$\rightarrow$RTL & \cmark &  &  &  & \cmark &  &  &  &  \\
\revise{Xu et al.}~\cite{kangwei2024automatedccprogramre} & C/\allowbreak C++$\rightarrow$RTL & \cmark &  &  &  & \cmark &  &  & \cmark &  \\
\revise{C2HLSC}~\cite{c2hlsc2024} & C/\allowbreak C++$\rightarrow$RTL & \cmark &  &  &  & \cmark &  &  & \cmark &  \\
LHS~\cite{reddy2025lhs} & C/\allowbreak C++$\rightarrow$RTL &  &  &  &  & \cmark &  &  &  &  \\
\revise{Yao et al.}~\cite{yao2025high} & C/\allowbreak C++$\rightarrow$RTL & \cmark &  &  &  & \cmark &  &  &  &  \\
\revise{CHiRM-DSE}~\cite{shi2026chirmdse} & C/\allowbreak C++$\rightarrow$RTL & \cmark &  &  &  & \cmark &  &  & \cmark &  \\
\revise{ParetoPilot}~\cite{xiong2026paretopilot} & C/\allowbreak C++$\rightarrow$RTL &  &  &  &  & \cmark &  &  & \cmark &  \\
\revise{Xiao et al.}~\cite{xiao2026llmhlsarch} & C/\allowbreak C++$\rightarrow$RTL & \cmark &  &  & \cmark & \cmark &  &  & \cmark &  \\
\revise{MPM-LLM4DSE}~\cite{lei2026mpmllm4dsereachingth} & C/\allowbreak C++$\rightarrow$RTL &  &  &  &  & \cmark &  &  & \cmark &  \\
\bottomrule
\end{tabular}
\end{objectroletable}
\end{table*}

\noindent\revise{\textbf{Mechanism Observation.} Table~\ref{tab:intra-mechanisms} shows that 31 of 36 Intra-stage systems (86.1\%) use at least one agent harness mechanism and 26 (72.2\%) implement an explicit loop, whereas five (13.9\%) optimize model parameters. Harness use is concentrated in tasks with repeatable executable feedback from compilation, simulation, formal checking, DRC, fitting, or circuit analysis. Retrieval or memory appears where previous flow and benchmark records can be reused. Twenty-three systems use one LLM-driven role and 13 use multiple roles, generally around a specialized EDA backend.}

\noindent\revise{\textbf{Validation Observation.} A successful check establishes only the property that the checker observes. RTLFixer uses compilation to establish well-formedness, MEIC adds simulation over exercised traces, Qayyum et al. admit invariants through formal proof, and drcAgent uses DRC to establish geometric legality~\cite{tsai2024rtlfixer,xu2024meic,qayyum2024incrementalproof,ma2026drcagent}. These checks do not establish properties outside their respective scope. Independence also depends on collateral provenance: ChipModeler, for example, generates a reference model for RTL co-simulation, making independently derived reference behavior important to the strength of the comparison~\cite{ye2026chipmodeler}.}

\section{Inter-stage Systems}
\label{sec:inter-stage}

Inter-stage systems consume an EDA-native primary input and produce a different design-stage primary output that a downstream tool must accept and consume. Consumer acceptance relies on cross-tool execution, semantic preservation, downstream QoR, or checkpoint coherence across a representation boundary. \Crefrange{tab:inter-objects-hls}{tab:inter-objects-rtl-layout} report the primary input--output pair and the supporting and intermediate handoff-object roles by chapter group, and \Cref{tab:inter-mechanisms} reports public code together with model-optimization and agent-harness mechanisms.


\subsection{C/C++-to-RTL}
These systems consume high-level C/C++ and produce synthesizable RTL through high-level synthesis. The downstream implementation flow accepts the result after successful HLS synthesis, with area, latency, throughput, or resource usage providing QoR evidence. The handoff is therefore a representation boundary from software to hardware. \Cref{tab:inter-objects-hls} summarizes these systems and their object roles.

\minisection{HLS Generation}
HLS generation translates a behavioral C/C++ description into synthesizable RTL through an HLS compiler and its scheduling, allocation, and binding decisions.
\textit{Typical systems.} SynthAI~\cite{synthai2024} coordinates software analysis, hardware generation, and checking through a five-tool pipeline, with plans and generated modules retained before the final RTL handoff. HLSPilot~\cite{chenwei2024hlspilotllmbasedhigh} uses one planner to identify module boundaries and transform existing code for Vitis HLS on an Alveo U280, whereas LHS~\cite{reddy2025lhs} feeds CNN-derived C code into a GPT-4o and Vivado-HLS accelerator flow. C2HLSC~\cite{c2hlsc2024} concentrates on repairing unsupported software constructs in response to compiler feedback. Established HLS compilation under the same test harness and synthesis budget provides the common non-agentic reference.
\textit{Validation and metrics.} The downstream RTL flow accepts an HLS result only after C/RTL co-simulation or equivalence establishes semantic preservation and synthesis establishes executability. Latency, throughput, area or resource use, and power rank accepted implementations. Tool calls and runtime measure handoff cost. \revise{Transformation can change interface signatures, bit widths, numerical behavior, memory-access patterns, and concurrency. C-level simulation checks software behavior, while HLS synthesis checks realizability. Neither alone establishes cycle-accurate or bit-level faithfulness, so the handoff requires a stated software-to-hardware correspondence and RTL co-simulation against an independent software reference or equivalence checking where available.}
\textit{Reported results.} LHS reports a 507$\times$ average latency improvement and a maximum of 2,690$\times$ across its evaluated CNN kernels, outperforming hls4ml on four of six cases~\cite{reddy2025lhs}. HLSPilot produces implementations comparable with the reported human-optimized designs on its evaluated workloads~\cite{chenwei2024hlspilotllmbasedhigh}.

\begin{table*}[t!]
\begin{objectroletable}
\centering
\caption{Roles of handoff objects in the Netlist$\rightarrow$Layout group of Inter-stage systems, covering ASIC and PCB targets. The Primary I/O column gives each system's exact $I_p\rightarrow O_p$ pair. Grouped columns use the role-first categories defined in \Cref{sec:framework-handoff}. A checkmark denotes at least one explicitly reported object in that category. A blank denotes none.}\label{tab:inter-objects-layout}
\begin{tabular}{@{}L{2.15cm}L{3.45cm}*{9}{C{0.695cm}}@{}}
\toprule
\rowcolor{tableheader}
\textbf{System} & \textbf{Primary I/O pair ($I_p \rightarrow O_p$)} & \multicolumn{3}{c}{\fontsize{6pt}{6pt}\selectfont $\boldsymbol{I_s}$: \textbf{Supporting Inputs}} & \multicolumn{3}{c}{\fontsize{6pt}{6pt}\selectfont $\boldsymbol{O_s}$: \textbf{Supporting Outputs}} & \multicolumn{3}{c}{\fontsize{6pt}{6pt}\selectfont $\boldsymbol{A_m}$: \textbf{Intermediate Artifacts}} \\
\cmidrule(lr){3-5}\cmidrule(lr){6-8}\cmidrule(lr){9-11}
\rowcolor{tableheader}
& & {\tiny\textbf{Constr.}} & {\tiny\textbf{Ref.}} & {\tiny\textbf{Knowl.}} & {\tiny\textbf{Check}} & {\tiny\textbf{QoR}} & {\tiny\textbf{Record}} & {\tiny\textbf{Retr.}} & {\tiny\textbf{Cand.}} & {\tiny\textbf{Coord.}} \\
\midrule
\revise{HyperPlace}~\cite{magi2025hyperplaceharnessing} & Netlist$\rightarrow$ASIC layout &  &  &  &  & \cmark &  &  & \cmark &  \\
\revise{CHIP-MAP}~\cite{yiming2026chipmapacollaborativ} & Netlist$\rightarrow$ASIC layout & \cmark &  &  &  & \cmark &  &  & \cmark & \cmark \\
PCBAgent~\cite{chen2025pcbagent} & Netlist$\rightarrow$PCB layout &  &  &  & \cmark & \cmark &  &  & \cmark &  \\
\revise{Li et al.}~\cite{li2026coarsepcb} & Netlist$\rightarrow$PCB layout &  &  & \cmark & \cmark & \cmark &  & \cmark &  &  \\
\bottomrule
\end{tabular}
\end{objectroletable}
\end{table*}

\minisection{HLS Repair and Optimization}
HLS repair and optimization rewrites source code or directives so that the program becomes synthesizable and achieves improved latency, throughput, area, or power.
\textit{Typical systems.} Xu et al.~\cite{kangwei2024automatedccprogramre} use one GPT-4-Turbo role to interpret compiler failures and make local source repairs before Design Compiler evaluation. Yao et al.~\cite{yao2025high} address pragma configurations with a fine-tuned model and Vivado HLS on a Xilinx Zynq-7000 device. Static compiler rewrites and direct prompting provide repair baselines for the former. Enumeration, Bayesian optimization, and evolutionary search under the same directive space and synthesis budget provide optimization baselines for the latter. In both cases, the agent's plausible advantage is fewer, semantically related evaluations, but only matched tool budgets establish that advantage.
\textit{Validation and metrics.} The HLS compiler first determines whether a repaired program or directive set synthesizes successfully. Behavioral agreement and resulting latency, throughput, area, and power determine whether the generated RTL is acceptable, while evaluations and time to feasibility measure the repair or optimization cost.
\textit{Reported results.} Xu et al. report average reductions of 36.57\% in area, 33.03\% in power, and 29.08\% in minimum clock period after LLM-assisted rewriting and bit-width optimization~\cite{kangwei2024automatedccprogramre}. Yao et al. improve normalized ADRS by 15\%~\cite{yao2025high}.

\begin{table*}[t!]
\begin{objectroletable}
\centering
\caption{Roles of handoff objects for RTL$\rightarrow$Layout in Inter-stage systems. The Primary I/O column reports $I_p\rightarrow O_p$. Grouped columns use the role-first categories defined in \Cref{sec:framework-handoff}. A checkmark denotes at least one explicitly reported object in that category. A blank denotes none.}\label{tab:inter-objects-rtl-layout}
\begin{tabular}{@{}L{2.15cm}L{3.45cm}*{9}{C{0.695cm}}@{}}
\toprule
\rowcolor{tableheader}
\textbf{System} & \textbf{Primary I/O pair ($I_p \rightarrow O_p$)} & \multicolumn{3}{c}{$\boldsymbol{I_s}$: \textbf{Supporting Inputs}} & \multicolumn{3}{c}{$\boldsymbol{O_s}$: \textbf{Supporting Outputs}} & \multicolumn{3}{c}{$\boldsymbol{A_m}$: \textbf{Intermediate Artifacts}} \\
\cmidrule(lr){3-5}\cmidrule(lr){6-8}\cmidrule(lr){9-11}
\rowcolor{tableheader}
& & {\tiny\textbf{Constr.}} & \textbf{Ref.} & {\tiny\textbf{Knowl.}} & \textbf{Check} & \textbf{QoR} & {\tiny\textbf{Record}} & \textbf{Retr.} & \textbf{Cand.} & \textbf{Coord.} \\
\midrule
ChatEDA~\cite{wu2024chateda} & RTL$\rightarrow$GDSII & \cmark &  &  & \cmark & \cmark &  &  & \cmark & \cmark \\
\bottomrule
\end{tabular}
\end{objectroletable}
\end{table*}

\minisection{HLS Architecture Exploration and DSE}
HLS architecture exploration and design-space exploration select microarchitectural structures and pragma configurations under a fixed synthesis and evaluation budget.
\textit{Typical systems.} The optimized design object separates three groups. (a) \textit{Architecture structure.} Xiao et al.~\cite{xiao2026llmhlsarch} modify both the architecture graph and its directives, comparing with a handcrafted implementation, HLSPilot, and the hierarchical HIDA compiler~\cite{ye2024hida}. (b) \textit{Directives and pragmas.} MPM-LLM4DSE~\cite{lei2026mpmllm4dsereachingth} combines program representations with QoR predictors and compares against GNN-DSE~\cite{sohrabizadeh2022gnndse} and ProgSG~\cite{qin2024progsg}. CHiRM-DSE~\cite{shi2026chirmdse} couples mined optimization rules to a CodeT5+ predictor~\cite{wang2023codet5plus} and compares with random and HARP-based exploration~\cite{sohrabizadeh2023harp}, while ParetoPilot~\cite{xiong2026paretopilot} allocates a fixed Vitis-HLS evaluation budget against NSGA-II~\cite{deb2002nsga2}, Lattice traversal~\cite{ferretti2018lattice}, and HGBO-DSE~\cite{kuang2023hgbodse}. (c) \textit{Accelerator templates.} GPT4AIGChip~\cite{fu2023gpt4aigchip} iteratively evaluates template configurations with Vivado HLS on a ZCU104. Synthesis under a fixed evaluation budget supplies the common ranking evidence across these design objects.
\textit{Validation and metrics.} Synthesizable RTL candidates form the admissible output set for the downstream implementation flow. Under a fixed evaluation budget, Pareto distance, hypervolume, and diversity characterize the selected design set. Predictor error, final PPA, synthesis count, and search time qualify the evidence used to choose it.
\textit{Reported results.} ParetoPilot reports a 49.5\% improvement over HGBO-DSE and reaches the reference ADRS with fewer syntheses~\cite{xiong2026paretopilot}.
\subsection{Netlist-to-Layout}
These systems transform a gate-, macro-, or component-level netlist into an ASIC or PCB layout. Their primary input--output pairs map connectivity to physical geometry, while the validation criteria remain target-specific. ASIC layout covers floorplanning, placement, routing, and final GDSII generation. PCB placement additionally handles irregular and rotatable components, board outlines, keep-outs, heterogeneous spacing rules, connector locations, surface-layer wiring, and designer preferences. Routed wirelength and via count complement board-specific legality. \Cref{tab:inter-objects-layout} summarizes these systems and their object roles.

\minisection{Standard-Cell Placer Tuning}
Standard-cell placer tuning chooses placement-engine parameters that improve wirelength, density, congestion, timing, or routability for a given netlist.
\textit{Typical systems.} HyperPlace~\cite{magi2025hyperplaceharnessing} does not replace a numerical placer. Its LLM agent proposes and explains hyperparameter configurations for DREAMPlace and Xplace. Its baselines are the default non-agentic placers~\cite{lin2019dreamplace,liu2022xplace} and a tree-structured Parzen estimator (TPE) Bayesian-optimization method~\cite{bergstra2011tpe}. The agent contributes configuration search, while legality and coordinate optimization remain the placer's responsibility.
\textit{Validation and metrics.} The placement flow rejects illegal candidates before comparing their quality. HPWL, congestion, timing, routability, and routed PPA rank legal outputs, while evaluated configurations and wall-clock time measure the cost of delivering the selected placement.
\textit{Reported results.} HyperPlace reports up to 1.66\% HPWL reduction on ISPD 2005 and, under ISPD 2015 fence constraints, up to 22.24\% over DREAMPlace 3.0 and 19.28\% over Xplace 2.0~\cite{magi2025hyperplaceharnessing}.

\begin{figure*}[t!]
\centering
\includegraphics[width=\textwidth]{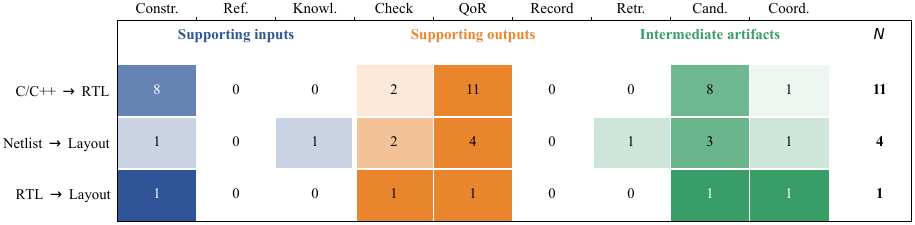}
\caption{Handoff-object role incidence across Inter-stage chapter groups. Each cell gives the number of systems reporting that role, and darker cells indicate a larger share of the systems in the corresponding row.}
\Description{An annotated heatmap compares nine supporting-input, supporting-output, and intermediate-artifact categories across three Inter-stage chapter groups containing 16 systems.}
\label{fig:inter-object-role-trends}
\end{figure*}

\minisection{Macro Placement and Floorplanning}
Macro placement and floorplanning assign legal block locations and orientations while balancing connectivity, congestion, timing, and boundary constraints.
\textit{Typical systems.} CHIP-MAP~\cite{yiming2026chipmapacollaborativ} addresses macro placement. Standard-cell placement remains in the numerical flow. Analysis, grouping, floorplan, and verification agents exchange structured JSON objects and refine a macro floorplan using timing, wirelength, congestion, and DRC feedback. Its baselines include simulated annealing, DREAMPlace~\cite{lin2019dreamplace}, ChiPFormer~\cite{lai2023chipformer}, and OpenROAD's Hier-RTLMP~\cite{kahng2024hierrtlmp}. The multi-agent design adds function-aware grouping and an auditable correction loop. The numerical flow still performs standard-cell placement, clock-tree synthesis, and routing. End-to-end PPA therefore provides stronger evidence than macro-coordinate quality and requires identical downstream OpenROAD settings for a fair comparison.
\textit{Validation and metrics.} The downstream OpenROAD flow accepts a macro placement only when it is legal and routable under the same technology and constraint settings. Wirelength, congestion, timing, DRC, and routed PPA compare accepted layouts. Correction rounds and tool runtime measure closure effort.
\textit{Reported results.} CHIP-MAP reports up to 1.5\% area reduction and repairs 61.6\% of total negative slack on average across the evaluated processor designs~\cite{yiming2026chipmapacollaborativ}.

\minisection{PCB Placement and Legalization}
PCB placement and legalization convert component connectivity, board constraints, and design intent into a legal placement accepted by the target routing and implementation flow.
\textit{Typical systems.} PCBAgent~\cite{chen2025pcbagent} combines two separately trained components: a decision-transformer-style RL placement policy, trained and refined from state--reward--action trajectories, and an LLM interaction agent that translates user requests into optimization commands. The RL updates the placement policy, not the LLM. It is a specialized backend rather than LLM optimization. Its baselines include MaskPlace~\cite{lai2022maskplace}, ChiPFormer~\cite{lai2023chipformer}, and a MILP post-optimizer. Li et al.~\cite{li2026coarsepcb} instead use multimodal serialization, topology-enriched retrieval, two grouping agents, group-aware global placement, and constraint-graph/MILP legalization. The PCB layout output is accompanied by grouping, orientation, and spacing constraints as supporting inputs and legalization evidence as a supporting output. An unlegalized coordinate set is not acceptable even when its estimated HPWL is low.
\textit{Validation and metrics.} The PCB placement or routing backend first requires full legality. Normalized HPWL, surface-layer wiring, routed wirelength, via count, and requested-category improvement rank legal outputs, while orientation or spacing violations, legalization cost, and runtime expose unresolved constraints and handoff effort.
\textit{Reported results.} PCBAgent is the only evaluated method to produce fully legal results on all 17 high-density tasks. Its normalized HPWL and surface-layer-wire count reach 1.04 and 0.94 relative to manual layouts, and interactive refinement improves the requested wire categories by 39.6\% on average~\cite{chen2025pcbagent}. Some baselines attain shorter wirelength by violating spacing or boundary constraints, so the result primarily reflects constraint completion and intent translation. On six industrial boards routed by the same commercial router, Li et al. report average routed-wirelength/via improvements of 55\%/73\% over the commercial placer, 34\%/26\% over NS-Place~\cite{cheng2022nsplace}, and 15\%/13\% over Quilter~\cite{quilter2026,li2026coarsepcb}.

\subsection{RTL-to-Layout}
ChatEDA~\cite{wu2024chateda} is the sole system in the corpus whose primary input--output pair spans RTL to a final ASIC layout. It traverses synthesis, floorplanning, placement, and routing before producing GDSII, so its input path and validation path differ from the netlist-originated placement systems in Section~4.2. \Cref{tab:inter-objects-rtl-layout} summarizes this system and its object roles.

\minisection{Full-Flow Planning and Execution}
Full-flow planning and execution decomposes an RTL-to-layout objective into ordered tool actions whose outputs become the inputs of later implementation stages.
\textit{Typical systems.} ChatEDA maps a natural-language objective onto synthesis, floorplanning, placement, and related OpenROAD operations, generates scripts, executes them, and uses reports to revise the next action. Its reported final artifact is a concrete GDSII layout. Unlike placement agents, ChatEDA preserves several representations and tool states simultaneously. Its non-agentic baseline is a fixed OpenROAD flow~\cite{ajayi2019openroad} using the same design, platform, and optimization settings. Script executability is necessary, but an end-to-end handoff also requires stage completion, constraint continuity, artifact identity, and comparable PPA.
ChatEDA also makes plan recovery a concrete full-flow concern: its planner produces executable task steps, its executor returns tool reports, and those results condition later OpenROAD actions~\cite{wu2024chateda}. Once transferred between components, a plan or checkpoint is an intermediate artifact $A_m$ whose metadata identifies its version, dependencies, referenced artifacts, tool configuration, authorization, producer, and consumer. A tool result that invalidates an upstream assumption requires a new plan version and selective revalidation of dependent steps. Failure-injection experiments expose recovery success, stale-state detection, recomputed versus reused steps, extra tool calls, and recovery time.
\textit{Validation and metrics.} Each implementation stage must execute the generated script and deliver an artifact consumable by the next stage. Task and stage completion establish workflow continuity, final PPA and signoff reports support the layout handoff, and tool calls, recovery iterations, and wall-clock time measure orchestration cost.
\textit{Reported results.} ChatEDA reports successful completion of representative scripting, error-correction, and clock-period-minimization tasks by AutoMage2, with generated actions executed through its EDA interface~\cite{wu2024chateda}.
\FloatBarrier

\subsection{Discussion}
\label{sec:inter-discussion}

\noindent\revise{\textbf{Handoff Object Observation.} All 16 systems in \Cref{fig:inter-object-role-trends} report QoR outputs, while 12 expose candidate artifacts and five report checker outputs. Explicit reference inputs do not occur in this class. The emphasis on QoR appears in both C/C++-to-RTL systems evaluated through downstream synthesis and layout systems evaluated through legality, timing, area, or related implementation measures.}


\begin{table*}[t!]
\begin{surveytable}
\centering
\caption{Model-optimization and agent-harness mechanisms in Inter-stage systems. A checkmark denotes explicit implementation evidence under the definitions in \Cref{sec:framework-mechanisms}. A linked GitHub icon denotes public code.}\label{tab:inter-mechanisms}
\begin{tabular}{@{}L{2.50cm}C{0.65cm}C{0.76cm}C{0.76cm}C{1.20cm}C{0.82cm}C{1.15cm}C{1.25cm}L{1.55cm}@{}}
\toprule
\rowcolor{tableheader}
\textbf{System} & \textbf{Code} & \multicolumn{2}{c}{\textbf{Model Optimization}} & \multicolumn{4}{c}{\textbf{Agent Harness}} & \textbf{Source} \\
\cmidrule(lr){3-4}\cmidrule(lr){5-8}
\rowcolor{tableheader}
& & \textbf{SFT} & \textbf{RL} & \textbf{RAG/Mem.} & \textbf{Loop} & \textbf{Orchestr.} & \textbf{Search/Evo.} & \\
\midrule
\tablegroup{9}{C/C++$\rightarrow$RTL}
\revise{GPT4AIGChip}~\cite{fu2023gpt4aigchip} & -- & \cmark &  &  & \cmark &  &  & \revise{ICCAD'23} \\
\revise{SynthAI}~\cite{synthai2024} & -- &  &  &  &  & \cmark &  & \revise{arXiv'24} \\
\revise{HLSPilot}~\cite{chenwei2024hlspilotllmbasedhigh} & -- &  &  &  &  &  &  & \revise{ICCAD'24} \\
\revise{Xu et al.}~\cite{kangwei2024automatedccprogramre} & -- &  &  &  & \cmark &  &  & \revise{MLCAD'24} \\
\revise{C2HLSC}~\cite{c2hlsc2024} & -- &  &  &  & \cmark &  &  & \revise{TODAES'25} \\
LHS~\cite{reddy2025lhs} & -- &  &  &  &  &  &  & TODAES'25 \\
\revise{Yao et al.}~\cite{yao2025high} & -- & \cmark &  &  &  &  &  & TODAES'25 \\
\revise{CHiRM-DSE}~\cite{shi2026chirmdse} & \href{https://github.com/ScopeHLS/CHiRM-DSE}{\githubmark} & \cmark &  &  & \cmark &  & \cmark & \revise{DAC'26} \\
\revise{ParetoPilot}~\cite{xiong2026paretopilot} & \href{https://github.com/GEAR-SEU/ParetoPilot-DAC-26}{\githubmark} &  &  &  & \cmark &  & \cmark & \revise{DAC'26} \\
\revise{Xiao et al.}~\cite{xiao2026llmhlsarch} & \href{https://github.com/Szzer1/HLSArchOpt}{\githubmark} &  &  &  & \cmark &  & \cmark & \revise{DAC'26} \\
\revise{MPM-LLM4DSE}~\cite{lei2026mpmllm4dsereachingth} & \href{https://github.com/wslcccc/MPM-LLM4DSE}{\githubmark} &  &  &  &  &  & \cmark & \revise{DATE'26} \\
\tablegroup{9}{Netlist$\rightarrow$Layout}
\revise{HyperPlace}~\cite{magi2025hyperplaceharnessing} & -- &  &  &  & \cmark &  & \cmark & \revise{TODAES'25} \\
\revise{CHIP-MAP}~\cite{yiming2026chipmapacollaborativ} & -- &  &  &  & \cmark & \cmark &  & \revise{DATE'26} \\
PCBAgent~\cite{chen2025pcbagent} & -- &  &  &  & \cmark &  &  & ASP-DAC'25 \\
\revise{Li et al.}~\cite{li2026coarsepcb} & -- &  &  & \cmark &  &  &  & \revise{DAC'26} \\
\tablegroup{9}{RTL$\rightarrow$Layout}
ChatEDA~\cite{wu2024chateda} & \href{https://github.com/wuhy68/ChatEDA}{\githubmark} & \cmark &  &  & \cmark & \cmark &  & TCAD'24 \\
\bottomrule
\end{tabular}
\end{surveytable}
\end{table*}

\noindent\revise{\textbf{Mechanism Observation.} Table~\ref{tab:inter-mechanisms} reports model optimization in four of 16 systems, search/\allowbreak evolution in five, at least one agent harness mechanism in 13, and an explicit loop in ten. Thirteen systems use a single LLM-driven role, so the prevalent structure couples one model to an HLS, placement, or implementation backend. Multi-role systems instead divide planning, grouping, or implementation across agents. PCBAgent is distinct: its RL component trains a placement policy, not the LLM interaction agent.}

\noindent\revise{\textbf{Validation Observation.} Acceptance by the next tool does not establish that the originating intent survived a representation change. HLS systems therefore pair successful synthesis with software/RTL co-simulation or equivalence evidence~\cite{chenwei2024hlspilotllmbasedhigh,c2hlsc2024}. CHIP-MAP similarly evaluates macro placement through legality and downstream implementation QoR~\cite{yiming2026chipmapacollaborativ}. In a multi-stage flow, ChatEDA shows that plans, scripts, reports, and checkpoints pass between components and condition later actions~\cite{wu2024chateda}. Their versions and dependencies determine which downstream results remain applicable after a change.}

\section{Extra-stage Systems}
\label{sec:extra-stage}

Extra-stage systems take a natural-language specification, query, or document as their primary input and translate this human-originated intent or knowledge into a machine-consumable EDA artifact. Consumer acceptance depends on specification completeness, source applicability, and checks performed in the target EDA environment. \Crefrange{tab:extra-objects-spec-rtl}{tab:extra-objects-other} report the primary input--output pair and the supporting and intermediate handoff-object roles by chapter group, and \Cref{tab:extra-mechanisms} reports public code together with model-optimization and agent-harness mechanisms.


\subsection{NL-to-HDL}
These systems take a natural-language specification as their primary input and generate a hardware description, including Verilog/SystemVerilog RTL and Chisel that elaborates to RTL. The generated design is accepted through compilation, simulation, synthesis, or PPA checks. The 33 systems are organized by the task and delivered artifact: general-purpose RTL generation, architecture- and application-specific RTL/IP generation, and Chisel-based hardware generation. \Cref{tab:extra-objects-spec-rtl} summarizes these systems and their object roles.

\begin{table*}[t!]
\begin{objectroletable}
\centering
\caption{Roles of handoff objects for NL$\rightarrow$HDL in Extra-stage systems. The Primary I/O column reports $I_p\rightarrow O_p$. Grouped columns use the role-first categories defined in \Cref{sec:framework-handoff}. A checkmark denotes at least one explicitly reported object in that category. A blank denotes none. $^{\dagger}$ denotes a system also discussed under another primary class.}\label{tab:extra-objects-spec-rtl}
\begin{tabular}{@{}L{2.15cm}L{3.45cm}*{9}{C{0.695cm}}@{}}
\toprule
\rowcolor{tableheader}
\textbf{System} & \textbf{Primary I/O pair ($I_p \rightarrow O_p$)} & \multicolumn{3}{c}{$\boldsymbol{I_s}$: \textbf{Supporting Inputs}} & \multicolumn{3}{c}{$\boldsymbol{O_s}$: \textbf{Supporting Outputs}} & \multicolumn{3}{c}{$\boldsymbol{A_m}$: \textbf{Intermediate Artifacts}} \\
\cmidrule(lr){3-5}\cmidrule(lr){6-8}\cmidrule(lr){9-11}
\rowcolor{tableheader}
& & {\tiny\textbf{Constr.}} & \textbf{Ref.} & {\tiny\textbf{Knowl.}} & \textbf{Check} & \textbf{QoR} & {\tiny\textbf{Record}} & \textbf{Retr.} & \textbf{Cand.} & \textbf{Coord.} \\
\midrule
\revise{DeLorenzo et al.}~\cite{delorenzo2024makeeverymove} & NL spec.$\rightarrow$RTL &  &  &  & \cmark & \cmark &  &  & \cmark &  \\
\revise{VeriAssist}~\cite{huang2024veriassist} & NL spec.$\rightarrow$RTL &  &  &  & \cmark &  &  &  & \cmark &  \\
\revise{Chang et al.}~\cite{chang2024data} & NL spec.$\rightarrow$RTL &  &  &  & \cmark & \cmark &  &  &  &  \\
ChatCPU~\cite{wang2024chatcpu} & NL spec.$\rightarrow$CPU RTL & \cmark &  &  & \cmark &  &  &  &  &  \\
LLM-AID$^{\dagger}$~\cite{firouzi2024llm} & NL spec.$\rightarrow$HDL &  &  &  & \cmark & \cmark &  &  & \cmark & \cmark \\
\revise{OriGen}~\cite{cui2024origen} & NL spec.$\rightarrow$RTL &  &  &  & \cmark & \cmark &  &  & \cmark &  \\
\revise{AutoVCoder}~\cite{mingzhe2024autovcoderasystemati} & NL spec.$\rightarrow$RTL &  &  &  & \cmark & \cmark &  &  &  &  \\
\revise{ChatChisel}~\cite{liu2024chatchisel} & NL spec.$\rightarrow$Chisel & \cmark &  &  & \cmark &  &  &  & \cmark &  \\
\revise{Rome}~\cite{rome2024} & NL spec.$\rightarrow$RTL &  &  &  & \cmark &  &  &  &  &  \\
Blocklove et al.~\cite{blocklove2024evaluating} & NL spec.$\rightarrow$RTL &  &  &  & \cmark &  &  &  & \cmark &  \\
VerilogCoder~\cite{ho2025verilogcoder} & NL spec.$\rightarrow$RTL &  &  &  & \cmark &  &  &  & \cmark & \cmark \\
RTLSquad~\cite{wang2025rtlsquad} & NL spec.$\rightarrow$RTL &  & \cmark &  & \cmark & \cmark & \cmark & \cmark & \cmark & \cmark \\
MAGE~\cite{zhao2025mage} & NL spec.$\rightarrow$RTL &  &  &  & \cmark &  &  &  &  & \cmark \\
ReChisel~\cite{niu2025rechisel} & NL spec.$\rightarrow$Chisel &  &  &  & \cmark &  &  &  & \cmark &  \\
\revise{AIvril2}~\cite{ul2025eda} & NL spec.$\rightarrow$RTL &  &  &  & \cmark & \cmark &  &  & \cmark &  \\
\revise{HaVen}~\cite{yiyao2025havenhallucinationmi} & NL spec.$\rightarrow$RTL &  &  &  & \cmark &  &  &  &  &  \\
\revise{LLM4GV}~\cite{dingyang2025llm4gvanllmbasedflex} & NL spec.$\rightarrow$GEMM RTL &  &  &  & \cmark & \cmark &  &  &  &  \\
\revise{Firouzi et al.}~\cite{firouzi2025promptfabflex} & NL spec.$\rightarrow$Flexible design &  &  &  &  & \cmark &  &  & \cmark &  \\
MAHL$^{\dagger}$~\cite{tang2025mahl} & NL spec.$\rightarrow$Chiplet IP & \cmark &  &  & \cmark &  &  &  &  & \cmark \\
VeriOpt~\cite{tasnia2025veriopt} & NL spec.$\rightarrow$RTL &  &  &  &  & \cmark &  &  & \cmark &  \\
ITERTL~\cite{wu2025itertl} & NL spec.$\rightarrow$RTL &  &  &  & \cmark &  &  &  & \cmark &  \\
RTLCoder~\cite{liu2024rtlcoder} & NL spec.$\rightarrow$RTL &  &  &  & \cmark &  &  &  & \cmark &  \\
VRank~\cite{zhao2025vrank} & NL spec.$\rightarrow$RTL &  &  &  & \cmark &  & \cmark &  & \cmark &  \\
\revise{HiVeGen}~\cite{hivegen2024} & NL spec.$\rightarrow$RTL & \cmark &  &  &  & \cmark &  &  &  &  \\
\revise{AutoChip}~\cite{blocklove2025autochip} & NL spec.$\rightarrow$RTL &  &  &  & \cmark &  &  &  & \cmark &  \\
AutoSilicon~\cite{li2025autosilicon} & NL spec.$\rightarrow$RTL & \cmark &  &  & \cmark &  &  &  &  &  \\
\revise{Chang et al.}~\cite{chang2025data} & NL spec.$\rightarrow$RTL &  &  &  &  & \cmark &  &  & \cmark &  \\
REvolution~\cite{min2026revolution} & NL spec.$\rightarrow$RTL & \cmark &  &  &  & \cmark &  &  & \cmark &  \\
VFlow~\cite{wei2026vflow} & NL spec.$\rightarrow$RTL &  &  &  &  &  &  &  & \cmark &  \\
\revise{Lu et al.}~\cite{lu2026twotierverilog} & NL spec.$\rightarrow$RTL & \cmark &  &  & \cmark & \cmark &  &  &  &  \\
\revise{LLM4PQC}~\cite{perera2026llm4pqc} & NL spec.$\rightarrow$PQC RTL & \cmark &  &  & \cmark & \cmark &  &  &  &  \\
\revise{CodeV}~\cite{yang2026codevempoweringllmsw} & NL spec.$\rightarrow$RTL &  &  &  & \cmark & \cmark &  &  &  &  \\
VeriRAG~\cite{thangellamudi2026verirag} & NL spec.$\rightarrow$Verilog &  & \cmark & \cmark & \cmark &  & \cmark & \cmark & \cmark &  \\
\bottomrule
\end{tabular}
\end{objectroletable}
\end{table*}

\minisection{General-Purpose RTL Generation}
General-purpose RTL generation translates module-level functional specifications into Verilog or SystemVerilog without restricting the task to one processor architecture or application domain. The systems differ mainly in how they obtain and validate candidate RTL.
\textit{Typical systems.} All systems in this group deliver RTL. The comparison follows the strongest check reported for that RTL. (a) \textit{Compilation and simulation.} VeriAssist~\cite{huang2024veriassist}, OriGen~\cite{cui2024origen}, AutoChip~\cite{blocklove2025autochip}, VerilogCoder~\cite{ho2025verilogcoder}, MAGE~\cite{zhao2025mage}, AutoVCoder~\cite{mingzhe2024autovcoderasystemati}, ITERTL, and RTLCoder~\cite{wu2025itertl,liu2024rtlcoder} use compiler or simulator outcomes to filter, revise, or accept generated RTL. VeriRAG~\cite{thangellamudi2026verirag}, VRank~\cite{zhao2025vrank}, and VFlow~\cite{wei2026vflow} add retrieved context, candidate ranking, or workflow selection before the same functional checks, while HaVen, Rome, Lu et al., and CodeV~\cite{yiyao2025havenhallucinationmi,rome2024,lu2026twotierverilog,yang2026codevempoweringllmsw} report functional evaluation of specialized generators. (b) \textit{Synthesis and PPA.} DeLorenzo et al.~\cite{delorenzo2024makeeverymove} use MCTS~\cite{browne2012mcts} with Yosys, AIvril2~\cite{ul2025eda} uses Vivado feedback, and VeriOpt, HiVeGen, AutoSilicon, RTLSquad, REvolution, and Chang et al.~\cite{tasnia2025veriopt,hivegen2024,li2025autosilicon,wang2025rtlsquad,min2026revolution,chang2025data} report synthesis or PPA evidence in addition to functional acceptance. (c) \textit{Physical implementation.} Blocklove et al.~\cite{blocklove2024evaluating} carry accepted RTL through OpenROAD in SkyWater 130~nm, while Chang et al.~\cite{chang2024data} use OpenLane and SiliconCompiler together with earlier compiler and simulator filters.
\textit{Validation and metrics.} A compiler first determines whether generated HDL is executable, and simulation against the task tests whether it implements the specification. Synthesis then supplies area, power, and timing evidence for downstream use. Samples, iterations, tokens, tool calls, and wall-clock time define the cost of reaching the accepted design.
\textit{Reported results.} On VerilogEval~\cite{liu2023verilogeval} Human-v2, MAGE and VerilogCoder report 95.7\% and 94.2\% functional success, respectively~\cite{zhao2025mage,ho2025verilogcoder}. VeriAssist reports 75.9\% functional pass@5 on RTLLM~\cite{lu2024rtllm}, while HaVen, OriGen, CodeV, and AutoVCoder report 66.0\%, 65.5\%, 55.2\%, and 51.7\% functional success on their stated v1.1 settings~\cite{huang2024veriassist,yiyao2025havenhallucinationmi,cui2024origen,yang2026codevempoweringllmsw,mingzhe2024autovcoderasystemati}. AutoChip reports 34.2\% lower generation cost and 5.8\% more solved designs than its direct-generation setting~\cite{blocklove2025autochip}. VFlow reports a 20--30\% pass@1 gain at 10.9$\times$ return on inference cost~\cite{wei2026vflow}. REvolution reports a 24.0-point gain to 95.5\%~\cite{min2026revolution}.

\minisection{Architecture- and Application-Specific RTL/IP Generation}
This task variant generates RTL or reusable IP for a named processor, accelerator, security primitive, integration architecture, or fabrication context.
\textit{Typical systems.} The target IP domain determines the supporting context and downstream checks. (a) Processor generation is represented by ChatCPU~\cite{wang2024chatcpu}, which assigns CPU construction to specialized LLMs and validates the RTL through Verilator and OpenLane with SkyWater 130~nm. (b) Accelerator generators deliver application-specific datapaths: LLM-AID~\cite{firouzi2024llm} coordinates controller, code-generation, and synthesis-optimization functions through Vitis HLS and Design Compiler, while LLM4GV~\cite{dingyang2025llm4gvanllmbasedflex} generates GEMM RTL evaluated by Design Compiler and Vivado. (c) MAHL~\cite{tang2025mahl} targets chiplet IP through six roles and validates it with Icarus Verilog, Design Compiler, and OpenROAD~\cite{ajayi2019openroad} under separate TSMC 28~nm and SkyWater 130~nm contexts. (d) Specialized technology or security targets require domain-specific libraries and tests: LLM4PQC~\cite{perera2026llm4pqc} generates post-quantum-cryptography RTL checked by ModelSim and Vivado, while Firouzi et al.~\cite{firouzi2025promptfabflex} generate flexible-electronics Verilog validated by task-aligned testbenches and the PragmatIC FlexICs Helvellyn 2.1.0 library.
\textit{Validation and metrics.} The target compiler and simulator first determine whether the generated processor, accelerator, chiplet component, or IP block implements its specification. Synthesis or FPGA implementation supplies area, timing, power, resource, and domain-specific evidence, while tool calls, correction rounds, and wall-clock time record closure effort.
\textit{Reported results.} ChatCPU reports a 3.81$\times$ average design-iteration speedup and tapes out its generated CPU in SkyWater 130~nm~\cite{wang2024chatcpu}. For chiplet-IP generation, MAHL raises average pass@1 by 44.67\% and increases pass@5 from zero to 0.72 on its evaluated tasks~\cite{tang2025mahl}.

\minisection{Chisel Generation}
Chisel-based generation produces generator source that must elaborate through the Chisel/FIRRTL toolchain before downstream RTL checks apply.
\textit{Typical systems.} ChatChisel~\cite{liu2024chatchisel} generates Chisel with a single GPT-3.5-Turbo role and uses Verilator feedback after elaboration. ReChisel~\cite{niu2025rechisel} iteratively repairs generated Chisel using FIRRTL-compiler and simulation feedback. Their evaluation therefore reports both source-level success and the correctness of the elaborated hardware, which distinguishes this task from direct Verilog/SystemVerilog generation.
\textit{Validation and metrics.} The Chisel toolchain accepts a source artifact only if it compiles and elaborates through FIRRTL. Simulation of the elaborated RTL establishes functional behavior, downstream synthesis measures implementation quality, and repair iterations and tool calls report the cost of obtaining the accepted module.
\textit{Reported results.} ChatChisel reports an average 31.86\% improvement over its Verilog-based LLM workflow~\cite{liu2024chatchisel}. ReChisel raises Chisel-generation success to a level comparable with the evaluated state-of-the-art Verilog agents~\cite{niu2025rechisel}.

\begin{table*}[t!]
\begin{objectroletable}
\centering
\caption{Roles of handoff objects for NL$\rightarrow$Verification Collateral in Extra-stage systems. The Primary I/O column reports $I_p\rightarrow O_p$. Grouped columns use the role-first categories defined in \Cref{sec:framework-handoff}. A checkmark denotes at least one explicitly reported object in that category. A blank denotes none.}\label{tab:extra-objects-verification-collateral}
\begin{tabular}{@{}L{2.15cm}L{3.45cm}*{9}{C{0.695cm}}@{}}
\toprule
\rowcolor{tableheader}
\textbf{System} & \textbf{Primary I/O pair ($I_p \rightarrow O_p$)} & \multicolumn{3}{c}{$\boldsymbol{I_s}$: \textbf{Supporting Inputs}} & \multicolumn{3}{c}{$\boldsymbol{O_s}$: \textbf{Supporting Outputs}} & \multicolumn{3}{c}{$\boldsymbol{A_m}$: \textbf{Intermediate Artifacts}} \\
\cmidrule(lr){3-5}\cmidrule(lr){6-8}\cmidrule(lr){9-11}
\rowcolor{tableheader}
& & {\tiny\textbf{Constr.}} & \textbf{Ref.} & {\tiny\textbf{Knowl.}} & \textbf{Check} & \textbf{QoR} & {\tiny\textbf{Record}} & \textbf{Retr.} & \textbf{Cand.} & \textbf{Coord.} \\
\midrule
\revise{AssertLLM}~\cite{yan2025assertllm} & NL spec.$\rightarrow$SVA &  &  &  & \cmark &  &  &  & \cmark &  \\
CorrectBench/\allowbreak ConfiBench~\cite{qiu2025correctbench,ruidi2026confibenchautomatict} & NL spec.$\rightarrow$Testbench &  &  &  & \cmark &  &  &  & \cmark &  \\
LAAG-RV~\cite{laagrv2024} & NL spec.$\rightarrow$SVA & \cmark &  &  & \cmark &  &  &  & \cmark &  \\
Hassan et al.~\cite{muhammad2024llmguidedformalverif} & NL spec.$\rightarrow$Invariants & \cmark &  &  & \cmark &  &  &  & \cmark &  \\
\revise{SuperSAGA}~\cite{paul2026supersaga} & NL spec.$\rightarrow$SVA &  &  &  & \cmark &  &  &  & \cmark &  \\
ChatSVA~\cite{fu2026chatsva} & NL spec.$\rightarrow$SVA &  &  &  & \cmark &  &  &  &  &  \\
\revise{LLM4SDC}~\cite{han2026llm4sdc} & NL spec.$\rightarrow$SDC &  &  &  & \cmark & \cmark &  &  &  &  \\
\revise{Qing et al.}~\cite{qing2026llmstl} & NL spec.$\rightarrow$STL &  &  &  & \cmark &  &  &  & \cmark &  \\
\revise{ReflectBench}~\cite{liu2026reflectbench} & NL spec.$\rightarrow$Testbench &  &  &  & \cmark &  &  &  & \cmark &  \\
\revise{CoverAssert}~\cite{wang2026coverassert} & NL spec.$\rightarrow$SVA &  &  &  & \cmark &  &  &  & \cmark &  \\
DRC-Coder~\cite{chang2025drc} & NL spec.$\rightarrow$DRC code & \cmark &  &  & \cmark &  &  &  &  &  \\
\bottomrule
\end{tabular}
\end{objectroletable}
\end{table*}

\subsection{NL-to-Verification Collateral}
These systems generate assertions, testbenches, timing constraints, temporal properties, or rule-checking code from a natural-language specification, often with RTL or design rules as supporting context. Because the generated artifact is itself used to judge a design, its evidential strength depends on whether the specification, reference behavior, or property set is independent of the generator. \Cref{tab:extra-objects-verification-collateral} summarizes these systems and their object roles.

\minisection{Assertions and Temporal Properties}
Assertion and temporal-property generation translates natural-language requirements into executable SVA or STL properties for formal or simulation-based checking.
\textit{Typical systems.} The primary property artifact separates three groups. (a) \textit{SVA.} AssertLLM~\cite{yan2025assertllm} separates specification analysis, assertion generation, and JasperGold checking, while ChatSVA~\cite{fu2026chatsva} carries a specification through plan and checkpoint artifacts before producing SVA. SuperSAGA~\cite{paul2026supersaga} and CoverAssert~\cite{wang2026coverassert} use coverage feedback to extend the property set, LAAG-RV~\cite{laagrv2024} revises SVA derived from requirements and RTL summaries using VCS feedback, and VeriRAG~\cite{thangellamudi2026verirag} evaluates SVA as a secondary output after retrieving hardware knowledge. (b) \textit{STL.} Qing et al.~\cite{qing2026llmstl} separate temporal-property drafting from Z3-based refinement. (c) \textit{Invariants.} Hassan et al.~\cite{muhammad2024llmguidedformalverif} refine candidate invariants with Python/Z3 diagnostics and test them on Yosys-generated mutants. GoldMine's static-analysis/data-mining pipeline~\cite{vasudevan2010goldmine}, human-authored properties, and the same formal engines provide non-agentic references across the three artifact types.
\textit{Validation and metrics.} A formal engine first parses and proves the generated assertion, but the receiving verification flow also needs evidence that the property expresses the intended requirement. Semantic correctness and cone/property coverage measure relevance and scope. Counterexample-guided iterations and formal-tool calls measure generation cost.
\textit{Reported results.} AssertLLM reports 88\% correctness with 97\% cone-of-influence coverage~\cite{yan2025assertllm}. ChatSVA reports 98.66\% syntax, 96.12\% functional correctness, and 82.50\% functional coverage, improving correctness by 33.3 points over its prior baseline~\cite{fu2026chatsva}.

\minisection{Testbench Generation}
Testbench generation turns behavioral intent into executable stimuli, drivers, and expected-result logic.
\textit{Typical systems.} Beyond CorrectBench's~\cite{qiu2025correctbench} self-validation loop, ReflectBench~\cite{liu2026reflectbench} adds multi-agent proposals, consensus, and reflection. These systems are evaluated against direct prompting, human-written collateral, and established simulators. Syntactic success and coverage can improve through iteration, but semantic correctness still depends on an independently specified expected result.
ConfiBench~\cite{ruidi2026confibenchautomatict} extends the CorrectBench system lineage with confidence-based scenario masking and testbench ensembles.
\textit{Validation and metrics.} The simulator must compile and execute the generated testbench before it can assess the design. Agreement with reference behavior, injected-error detection, and coverage determine whether the testbench checks the intended behavior. Correction rounds and simulator cost measure handoff effort.
\textit{Reported results.} Under the GPT-4o setting reported by CorrectBench, self-validation and correction raise the final testbench pass ratio from 33.33\% for direct generation and 52.18\% for AutoBench to 70.13\%. Its 88.85\% self-validation success describes an intermediate step rather than the final accepted output~\cite{qiu2025correctbench}.

\minisection{Timing-Constraint Generation}
Timing-constraint generation translates clock, interface, and timing intent into executable SDC constraints for static timing analysis and implementation tools.
\textit{Typical systems.} LLM4SDC~\cite{han2026llm4sdc} applies generation and checking to timing constraints rather than simulation stimuli. Direct prompting, human-written SDC, and the same timing backend provide the relevant comparisons. A syntactically valid constraint set remains incomplete if it omits intended clocks, exceptions, or interface relationships.
\textit{Validation and metrics.} Static timing and implementation tools must parse the generated SDC before using it. Agreement with reference constraints, coverage of intended clocks and interfaces, and timing-analysis outcomes determine whether it constrains the design correctly. Correction rounds and tool calls measure production cost.
\textit{Reported results.} LLM4SDC reports approximately 1.00 syntactic correctness, 0.93 semantic fidelity, and 0.67 false-path accuracy. The resulting constraints improve slack by 29.4\% with a 1.15\% power reduction and a 0.27\% area increase~\cite{han2026llm4sdc}.

\minisection{Design-Rule Checker Generation}
Design-rule checker generation translates textual geometric constraints into executable code that detects layout violations.
\textit{Typical systems.} DRC-Coder~\cite{chang2025drc} produces verification-side code rather than physical geometry. Reference violations test whether the generated program detects the intended geometry. Program execution alone does not establish rule completeness.
\textit{Validation and metrics.} The checker backend must execute the generated code and detect reference violations that instantiate the textual rule. Rule-variant coverage measures the scope of the delivered checker, while repair iterations and runtime measure the cost of obtaining it.
\textit{Reported results.} DRC-Coder reports perfect precision, recall, and F1 on its standard-cell-layout rules and a 37\% higher F1 score than standard prompting over the full evaluation~\cite{chang2025drc}.

\begin{table*}[t!]
\begin{objectroletable}
\centering
\caption{Roles of handoff objects for NL$\rightarrow$Netlist in Extra-stage systems. The Primary I/O column reports $I_p\rightarrow O_p$. Grouped columns use the role-first categories defined in \Cref{sec:framework-handoff}. A checkmark denotes at least one explicitly reported object in that category. A blank denotes none.}\label{tab:extra-objects-netlists}
\begin{tabular}{@{}L{2.15cm}L{3.45cm}*{9}{C{0.695cm}}@{}}
\toprule
\rowcolor{tableheader}
\textbf{System} & \textbf{Primary I/O pair ($I_p \rightarrow O_p$)} & \multicolumn{3}{c}{$\boldsymbol{I_s}$: \textbf{Supporting Inputs}} & \multicolumn{3}{c}{$\boldsymbol{O_s}$: \textbf{Supporting Outputs}} & \multicolumn{3}{c}{$\boldsymbol{A_m}$: \textbf{Intermediate Artifacts}} \\
\cmidrule(lr){3-5}\cmidrule(lr){6-8}\cmidrule(lr){9-11}
\rowcolor{tableheader}
& & {\tiny\textbf{Constr.}} & \textbf{Ref.} & {\tiny\textbf{Knowl.}} & \textbf{Check} & \textbf{QoR} & {\tiny\textbf{Record}} & \textbf{Retr.} & \textbf{Cand.} & \textbf{Coord.} \\
\midrule
Artisan~\cite{chen2024artisan} & NL spec.$\rightarrow$Op-amp netlist & \cmark &  &  & \cmark & \cmark &  &  &  & \cmark \\
\revise{LADAC}~\cite{ladac2024} & NL spec.$\rightarrow$Analog circuit & \cmark &  & \cmark & \cmark & \cmark &  & \cmark & \cmark &  \\
AnalogCoder~\cite{lai2025analogcoder} & NL spec.$\rightarrow$Analog netlist & \cmark &  &  & \cmark & \cmark &  &  &  &  \\
\revise{Somayaji and Li}~\cite{ns2025ai} & NL spec.$\rightarrow$Analog sizing & \cmark &  & \cmark & \cmark & \cmark & \cmark & \cmark &  &  \\
\revise{Wang et al.}~\cite{wang2025multiagentgenerative} & NL spec.$\rightarrow$Analog/\allowbreak RF netlist & \cmark &  &  & \cmark & \cmark &  &  &  & \cmark \\
Atelier~\cite{shen2025atelier} & NL spec.$\rightarrow$Analog netlist & \cmark &  &  & \cmark & \cmark &  &  & \cmark & \cmark \\
MenTeR~\cite{chen2025menter} & NL spec.$\rightarrow$Analog/\allowbreak RF netlist & \cmark &  & \cmark & \cmark & \cmark & \cmark & \cmark & \cmark & \cmark \\
HRL-Topo~\cite{vijayaraghavan2026hrltopo} & NL spec.$\rightarrow$Analog netlist & \cmark &  &  & \cmark & \cmark &  &  &  & \cmark \\
\revise{PCBgen}~\cite{zhang2026pcbgen} & NL spec.$\rightarrow$PCB schematic & \cmark &  &  & \cmark &  &  &  &  & \cmark \\
\bottomrule
\end{tabular}
\end{objectroletable}
\end{table*}

\subsection{NL-to-Netlist}
This family groups natural-language-to-connectivity artifacts across circuit domains: digital gate-level netlists, analog/RF netlists and sizing descriptions, and PCB schematics or netlists. Digital RTL and other behavioral HDL remain in Section~5.1. Physical geometry is grouped under Layout in Section~5.4. In the current corpus, the systems in this family cover analog/RF circuit synthesis and PCB schematic generation, with circuit simulation or ERC providing downstream evidence. \Cref{tab:extra-objects-netlists} summarizes these systems and their object roles.

\minisection{Topology Generation and Sizing}
Analog topology generation and sizing selects circuit structure and device parameters that satisfy a natural-language performance specification.
\textit{Typical systems.} Artisan~\cite{chen2024artisan} decomposes op-amp topology and parameter design through a cascading dialogue. Atelier~\cite{shen2025atelier} extends this workflow to topology, sizing, evaluation, and backtracking around Spectre, while MenTeR~\cite{chen2025menter} adds diagram-aware retrieval and carries RF/analog requirements and PySpice candidates through circuit, testbench, and execution roles. AnalogCoder~\cite{lai2025analogcoder} instead uses a training-free prompting and tool-library workflow with a generic Level-1 MOS model, and LADAC~\cite{ladac2024} retrieves examples before revising candidates with SPICE feedback. Somayaji and Li~\cite{ns2025ai} transfer sizing knowledge across designs through two roles and a knowledge graph, whereas Wang et al.~\cite{wang2025multiagentgenerative} separate inverse-design data generation from optimization and check candidates with NGSPICE and Xyce. HRL-Topo~\cite{vijayaraghavan2026hrltopo} differs by learning a high-level structural-goal policy and a low-level incident-encoded-netlist policy for power converters, refined with a goal-conditioned PPO variant and NGSpice-derived evidence. Batch Bayesian optimization, AutoCkt, and transferable GCN-RL~\cite{lyu2018batchbo,settaluri2020autockt,wang2020gcnrl} provide the corresponding non-agentic references.
\textit{Validation and metrics.} A circuit simulator first rejects invalid topologies and parameterizations that miss the requested specifications. Among feasible netlists, PPA, figure of merit, and robustness determine the output delivered to the designer. Simulator calls, convergence, and wall-clock time measure the search cost under a matched budget.
\textit{Reported results.} On 24 AnalogCoder tasks, with at most five attempts per task, MenTeR reports average pass@1 and pass@5 rates of 84.2\% and 89.2\%. Removing diagram-aware retrieval or chain-of-stage reasoning lowers average pass@1 to 75.0\% or 70.0\%, respectively~\cite{chen2025menter}. HRL-Topo generates 500 designs per prompt and reports a 10--15\% improvement in overall success rate over flat generation and single-level refinement baselines~\cite{vijayaraghavan2026hrltopo}.

\minisection{PCB Schematic and Netlist Generation}
PCB schematic and netlist generation converts a natural-language circuit specification into connectivity consumable by board-design and simulation tools.
\textit{Typical systems.} PCBgen~\cite{zhang2026pcbgen} builds schematics and netlists from specifications and uses ERC plus a tightly budgeted SPICE loop. Its primary output is a connectivity and electrical netlist, placing it in this family. Manufacturability remains a downstream layout obligation.
\textit{Validation and metrics.} The board-design flow accepts generated connectivity only after ERC, pin-map, and circuit-behavior checks pass. DRC or routability evidence covers its use in later physical design, while correction rounds and human review time measure the effort required to deliver the schematic or netlist.
\textit{Reported results.} PCBgen demonstrates end-to-end generation of ERC-checked schematics and netlists, together with SPICE-guided parameter correction, on its evaluated board-design cases~\cite{zhang2026pcbgen}.

\begin{table*}[t!]
\begin{objectroletable}
\centering
\caption{Roles of handoff objects for NL$\rightarrow$Layout in Extra-stage systems. The Primary I/O column reports $I_p\rightarrow O_p$. Grouped columns use the role-first categories defined in \Cref{sec:framework-handoff}. A checkmark denotes at least one explicitly reported object in that category. A blank denotes none.}\label{tab:extra-objects-layout}
\begin{tabular}{@{}L{2.15cm}L{3.45cm}*{9}{C{0.695cm}}@{}}
\toprule
\rowcolor{tableheader}
\textbf{System} & \textbf{Primary I/O pair ($I_p \rightarrow O_p$)} & \multicolumn{3}{c}{$\boldsymbol{I_s}$: \textbf{Supporting Inputs}} & \multicolumn{3}{c}{$\boldsymbol{O_s}$: \textbf{Supporting Outputs}} & \multicolumn{3}{c}{$\boldsymbol{A_m}$: \textbf{Intermediate Artifacts}} \\
\cmidrule(lr){3-5}\cmidrule(lr){6-8}\cmidrule(lr){9-11}
\rowcolor{tableheader}
& & {\tiny\textbf{Constr.}} & \textbf{Ref.} & {\tiny\textbf{Knowl.}} & \textbf{Check} & \textbf{QoR} & {\tiny\textbf{Record}} & \textbf{Retr.} & \textbf{Cand.} & \textbf{Coord.} \\
\midrule
\revise{PANDA}~\cite{zhang2026panda} & NL spec.$\rightarrow$Analog layout & \cmark &  &  & \cmark & \cmark &  &  & \cmark & \cmark \\
Yang and Ren~\cite{yang2026code} & NL spec.$\rightarrow$GDSII & \cmark &  &  & \cmark &  &  &  & \cmark &  \\
AnAutomate~\cite{wang2026anautomate} & NL spec.$\rightarrow$Analog layout & \cmark &  & \cmark & \cmark & \cmark &  & \cmark & \cmark & \cmark \\
\bottomrule
\end{tabular}
\end{objectroletable}
\end{table*}

\subsection{NL-to-Layout}
Layout is used here as the umbrella physical-design category for generated geometry, including analog placement/routing and GDSII. \Cref{tab:extra-objects-layout} summarizes these systems and their object roles.

\minisection{IC Layout Generation}
IC layout generation converts textual constraints into physical geometry accepted by layout and signoff tools.
\textit{Typical systems.} Yang and Ren~\cite{yang2026code} generate Python that constructs rule-constrained GDSII and validate the result with DRC. The generated program is an auditable intermediate artifact, while the GDSII geometry and signoff report determine whether the consumer accepts the result.
\textit{Validation and metrics.} The layout and signoff tools must construct the generated geometry and accept it under the stated rule deck. DRC cleanliness and constraint satisfaction establish usability, area characterizes the result, and iterations and runtime measure the cost of reaching the accepted GDSII.
\textit{Reported results.} Yang and Ren report a diversity score of 12.009 while maintaining DRC-legal output, exceeding the evaluated vision-based layout generators~\cite{yang2026code}.

\minisection{Analog Layout and Post-Layout Closure}
Analog layout and post-layout closure converts circuit intent into geometry and iterates after extraction until physical and electrical constraints are met.
\textit{Typical systems.} PANDA~\cite{zhang2026panda} carries design intent through topology synthesis, sizing, placement, routing, extraction, and post-layout feedback. AnAutomate~\cite{wang2026anautomate} extends this pattern to a hierarchical full spec-to-layout flow: TopoSynth proposes and screens circuit topologies, AnaTest generates simulation infrastructure, and AnaFlow coordinates block-level sizing, learned constraints, ALIGN layout generation, and Siemens Calibre post-layout checks. Its primary output is the implemented analog layout/GDSII. Generated netlists, block constraints, candidate designs, and workflow state are supporting or intermediate objects. These paths require more extensive validation than schematic generation because a pre-layout SPICE target does not cover parasitics and geometry. The relevant non-agentic baseline is a conventional scripted analog-layout flow such as MAGICAL~\cite{chen2021magical}, using the same generators, technology, and simulation budget. Consumer acceptance draws on pre- and post-layout metrics, the extracted view, and DRC evidence.
\textit{Validation and metrics.} The physical-design flow accepts the layout only after DRC and extraction, and post-layout simulation must show that parasitics have not invalidated the requested specifications. Pre/post-layout QoR and area characterize the delivered result. Tool calls and closure time measure the effort beyond schematic-level success.
\textit{Reported results.} PANDA completes its reported end-to-end analog-layout cases in several hours, compared with several days for manual iterations, while retaining pre- and post-layout performance checks~\cite{zhang2026panda}.

\begin{figure*}[t!]
\centering
\includegraphics[width=\textwidth]{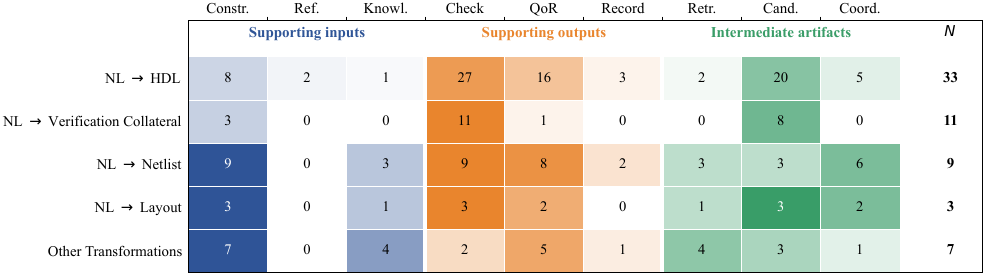}
\caption{Handoff-object role incidence across Extra-stage chapter groups. Each cell gives the number of systems reporting that role, and darker cells indicate a larger share of the systems in the corresponding row.}
\Description{An annotated heatmap compares nine supporting-input, supporting-output, and intermediate-artifact categories across five Extra-stage chapter groups containing 63 systems.}
\label{fig:extra-object-role-trends}
\end{figure*}

\begin{table*}[t!]
\begin{objectroletable}
\centering
\caption{Roles of handoff objects for Other Transformations in Extra-stage systems. The Primary I/O column reports $I_p\rightarrow O_p$. Grouped columns use the role-first categories defined in \Cref{sec:framework-handoff}. A checkmark denotes at least one explicitly reported object in that category. A blank denotes none.}\label{tab:extra-objects-other}
\begin{tabular}{@{}L{2.15cm}L{3.45cm}*{9}{C{0.695cm}}@{}}
\toprule
\rowcolor{tableheader}
\textbf{System} & \textbf{Primary I/O pair ($I_p \rightarrow O_p$)} & \multicolumn{3}{c}{$\boldsymbol{I_s}$: \textbf{Supporting Inputs}} & \multicolumn{3}{c}{$\boldsymbol{O_s}$: \textbf{Supporting Outputs}} & \multicolumn{3}{c}{$\boldsymbol{A_m}$: \textbf{Intermediate Artifacts}} \\
\cmidrule(lr){3-5}\cmidrule(lr){6-8}\cmidrule(lr){9-11}
\rowcolor{tableheader}
& & {\tiny\textbf{Constr.}} & \textbf{Ref.} & {\tiny\textbf{Knowl.}} & \textbf{Check} & \textbf{QoR} & {\tiny\textbf{Record}} & \textbf{Retr.} & \textbf{Cand.} & \textbf{Coord.} \\
\midrule
ChatArch~\cite{wu2025chatarch} & NL spec.$\rightarrow$Architecture & \cmark &  & \cmark &  & \cmark &  & \cmark & \cmark &  \\
\revise{Chat-A2}~\cite{zhu2026chata2} & NL spec.$\rightarrow$CPU configuration & \cmark &  &  &  & \cmark &  &  & \cmark &  \\
\revise{AgenticDSE}~\cite{zhu2026agenticdse} & NL spec.$\rightarrow$Chiplet configuration & \cmark &  &  &  & \cmark &  &  & \cmark &  \\
ChatLS~\cite{zheng2025chatls} & NL spec.$\rightarrow$Script & \cmark &  & \cmark & \cmark &  & \cmark & \cmark &  & \cmark \\
\revise{AgenticTCAD}~\cite{fan2026agentictcad} & NL spec.$\rightarrow$TCAD script & \cmark &  &  & \cmark &  &  &  &  &  \\
\revise{PCBgen3D}~\cite{di2026pcbgen3d} & NL spec.$\rightarrow$PCB 3D model & \cmark &  & \cmark &  & \cmark &  & \cmark &  &  \\
\revise{Smart-PCLib}~\cite{di2026smartpclib} & NL spec.$\rightarrow$PCB library & \cmark &  & \cmark &  & \cmark &  & \cmark &  &  \\
\bottomrule
\end{tabular}
\end{objectroletable}
\end{table*}


\begin{table*}[t!]
\begin{surveytable}
\centering
\fontsize{6.2pt}{6.6pt}\selectfont
\renewcommand{\arraystretch}{0.92}
\caption{Model-optimization and agent-harness mechanisms in Extra-stage systems. A checkmark denotes explicit implementation evidence under the definitions in \Cref{sec:framework-mechanisms}. A linked GitHub icon denotes public code.}\label{tab:extra-mechanisms}
\begin{tabular}{@{}L{2.50cm}C{0.65cm}C{0.76cm}C{0.76cm}C{1.20cm}C{0.82cm}C{1.15cm}C{1.25cm}L{1.55cm}@{}}
\toprule
\rowcolor{tableheader}
\textbf{System} & \textbf{Code} & \multicolumn{2}{c}{\textbf{Model Optimization}} & \multicolumn{4}{c}{\textbf{Agent Harness}} & \textbf{Source} \\
\cmidrule(lr){3-4}\cmidrule(lr){5-8}
\rowcolor{tableheader}
& & \textbf{SFT} & \textbf{RL} & \textbf{RAG/Mem.} & \textbf{Loop} & \textbf{Orchestr.} & \textbf{Search/Evo.} & \\
\midrule
\tablegroup{9}{NL$\rightarrow$HDL}
\revise{DeLorenzo et al.}~\cite{delorenzo2024makeeverymove} & -- & \cmark &  &  & \cmark &  & \cmark & \revise{arXiv'24} \\
\revise{VeriAssist}~\cite{huang2024veriassist} & -- &  &  &  & \cmark &  &  & \revise{arXiv'24} \\
\revise{Chang et al.}~\cite{chang2024data} & \href{https://github.com/aichipdesign/chipgptft}{\githubmark} & \cmark &  &  &  &  &  & \revise{DAC'24} \\
ChatCPU~\cite{wang2024chatcpu} & -- & \cmark &  &  &  &  &  & DAC'24 \\
LLM-AID$^{\dagger}$~\cite{firouzi2024llm} & -- &  &  &  &  & \cmark & \cmark & ICCAD'24 \\
\revise{OriGen}~\cite{cui2024origen} & -- &  &  &  & \cmark &  &  & \revise{ICCAD'24} \\
\revise{AutoVCoder}~\cite{mingzhe2024autovcoderasystemati} & \href{https://github.com/sjtu-zhao-lab/AutoVCoder}{\githubmark} & \cmark &  &  &  &  &  & \revise{ICCD'24} \\
\revise{ChatChisel}~\cite{liu2024chatchisel} & -- &  &  &  & \cmark &  &  & \revise{ISEDA'24} \\
\revise{Rome}~\cite{rome2024} & \href{https://github.com/ajn313/ROME-LLM}{\githubmark} & \cmark &  &  &  &  &  & \revise{MLCAD'24} \\
Blocklove et al.~\cite{blocklove2024evaluating} & -- &  &  &  & \cmark &  &  & LAD'24 \\
VerilogCoder~\cite{ho2025verilogcoder} & \href{https://github.com/SimpleHuracan/MixAgent}{\githubmark} &  &  &  & \cmark & \cmark &  & AAAI'25 \\
RTLSquad~\cite{wang2025rtlsquad} & -- &  &  & \cmark & \cmark & \cmark & \cmark & arXiv'25 \\
MAGE~\cite{zhao2025mage} & \href{https://github.com/stable-lab/MAGE}{\githubmark} &  &  &  &  & \cmark &  & DAC'25 \\
ReChisel~\cite{niu2025rechisel} & \href{https://github.com/niujuxin/rechisel}{\githubmark} &  &  &  & \cmark &  &  & DAC'25 \\
\revise{AIvril2}~\cite{ul2025eda} & -- &  &  &  & \cmark &  &  & DATE'25 \\
\revise{HaVen}~\cite{yiyao2025havenhallucinationmi} & \href{https://github.com/Intelligent-Computing-Research-Group/HaVen}{\githubmark} & \cmark &  &  &  &  &  & \revise{DATE'25} \\
\revise{LLM4GV}~\cite{dingyang2025llm4gvanllmbasedflex} & -- &  &  &  &  &  &  & \revise{DATE'25} \\
\revise{Firouzi et al.}~\cite{firouzi2025promptfabflex} & -- &  &  &  &  &  & \cmark & \revise{ICCAD'25} \\
MAHL$^{\dagger}$~\cite{tang2025mahl} & -- &  &  &  &  & \cmark &  & ICCAD'25 \\
VeriOpt~\cite{tasnia2025veriopt} & -- & \cmark &  &  & \cmark &  &  & ICCAD'25 \\
ITERTL~\cite{wu2025itertl} & -- & \cmark &  &  &  &  &  & ISCAS'25 \\
RTLCoder~\cite{liu2024rtlcoder} & \href{https://github.com/hkust-zhiyao/RTLCoder}{\githubmark} & \cmark &  &  &  &  &  & TCAD'25 \\
VRank~\cite{zhao2025vrank} & -- &  &  &  &  &  & \cmark & ISQED'25 \\
\revise{HiVeGen}~\cite{hivegen2024} & -- &  &  &  &  &  &  & \revise{ICLAD'25} \\
\revise{AutoChip}~\cite{blocklove2025autochip} & \href{https://github.com/shailja-thakur/AutoChip}{\githubmark} &  &  &  & \cmark &  &  & \revise{TODAES'25} \\
AutoSilicon~\cite{li2025autosilicon} & -- &  &  &  &  &  &  & TODAES'25 \\
\revise{Chang et al.}~\cite{chang2025data} & -- & \cmark &  &  & \cmark &  &  & TODAES'25 \\
REvolution~\cite{min2026revolution} & -- &  &  &  & \cmark &  & \cmark & ASP-DAC'26 \\
VFlow~\cite{wei2026vflow} & -- &  &  &  & \cmark &  & \cmark & ASP-DAC'26 \\
\revise{Lu et al.}~\cite{lu2026twotierverilog} & -- &  &  &  &  &  &  & \revise{DAC'26} \\
\revise{LLM4PQC}~\cite{perera2026llm4pqc} & -- &  &  &  &  &  &  & \revise{DATE'26} \\
\revise{CodeV}~\cite{yang2026codevempoweringllmsw} & \href{https://github.com/IPRC-DIP/CodeV}{\githubmark} & \cmark &  &  &  &  &  & \revise{TCAD'26} \\
VeriRAG~\cite{thangellamudi2026verirag} & \href{https://github.com/HArt-Git/VeriRAG}{\githubmark} &  &  & \cmark &  &  &  & ASP-DAC'26 \\
\tablegroup{9}{NL$\rightarrow$Verification Collateral}
\revise{AssertLLM}~\cite{yan2025assertllm} & -- &  &  &  & \cmark &  &  & \revise{ASP-DAC'25} \\
CorrectBench/\allowbreak ConfiBench~\cite{qiu2025correctbench,ruidi2026confibenchautomatict} & \href{https://github.com/AutoBench/ConfiBench}{\githubmark} &  &  &  & \cmark &  & \cmark & DATE'25/\allowbreak TODAES'26 \\
LAAG-RV~\cite{laagrv2024} & -- &  &  &  & \cmark &  &  & ITC India'24 \\
Hassan et al.~\cite{muhammad2024llmguidedformalverif} & -- &  &  &  & \cmark &  &  & DATE'24 \\
\revise{SuperSAGA}~\cite{paul2026supersaga} & -- &  &  &  & \cmark &  &  & \revise{ASP-DAC'26} \\
ChatSVA~\cite{fu2026chatsva} & -- &  &  &  &  &  &  & DAC'26 \\
\revise{LLM4SDC}~\cite{han2026llm4sdc} & -- &  &  &  &  &  &  & \revise{DAC'26} \\
\revise{Qing et al.}~\cite{qing2026llmstl} & -- &  &  &  & \cmark &  &  & \revise{DAC'26} \\
\revise{ReflectBench}~\cite{liu2026reflectbench} & -- &  &  &  & \cmark &  &  & \revise{DAC'26} \\
\revise{CoverAssert}~\cite{wang2026coverassert} & -- &  &  &  & \cmark &  &  & \revise{DATE'26} \\
DRC-Coder~\cite{chang2025drc} & -- &  &  &  &  &  &  & ISPD'25 \\
\tablegroup{9}{NL$\rightarrow$Netlist}
Artisan~\cite{chen2024artisan} & -- & \cmark &  &  &  & \cmark &  & DAC'24 \\
\revise{LADAC}~\cite{ladac2024} & -- &  &  & \cmark & \cmark &  &  & \revise{TechRxiv\'24} \\
AnalogCoder~\cite{lai2025analogcoder} & \href{https://github.com/laiyao1/AnalogCoder}{\githubmark} &  &  &  &  &  &  & AAAI'25 \\
\revise{Somayaji and Li}~\cite{ns2025ai} & -- &  &  & \cmark &  &  &  & ICCAD'25 \\
\revise{Wang et al.}~\cite{wang2025multiagentgenerative} & -- &  &  &  &  & \cmark &  & \revise{ICCAD'25} \\
Atelier~\cite{shen2025atelier} & -- & \cmark &  &  & \cmark & \cmark & \cmark & TCAD'25 \\
MenTeR~\cite{chen2025menter} & -- &  &  & \cmark & \cmark & \cmark &  & ICLAD'25 \\
HRL-Topo~\cite{vijayaraghavan2026hrltopo} & -- & \cmark & \cmark &  &  &  &  & DAC'26 \\
\revise{PCBgen}~\cite{zhang2026pcbgen} & -- &  &  &  &  & \cmark &  & \revise{DAC'26} \\
\tablegroup{9}{NL$\rightarrow$Layout}
\revise{PANDA}~\cite{zhang2026panda} & \href{https://github.com/PKU-IDEA/PANDA}{\githubmark} &  &  &  &  & \cmark & \cmark & \revise{DAC'26} \\
Yang and Ren~\cite{yang2026code} & -- &  &  &  &  &  & \cmark & ASP-DAC'26 \\
AnAutomate~\cite{wang2026anautomate} & -- &  &  & \cmark & \cmark & \cmark & \cmark & DAC'26 \\
\tablegroup{9}{Other Transformations}
ChatArch~\cite{wu2025chatarch} & -- &  &  & \cmark &  &  & \cmark & TODAES'25 \\
\revise{Chat-A2}~\cite{zhu2026chata2} & -- &  &  &  & \cmark &  & \cmark & \revise{ASP-DAC'26} \\
\revise{AgenticDSE}~\cite{zhu2026agenticdse} & \href{https://github.com/pku-pacific-lab-team/AgenticDSE}{\githubmark} &  &  &  & \cmark &  & \cmark & \revise{DAC'26} \\
ChatLS~\cite{zheng2025chatls} & -- &  &  & \cmark &  &  &  & DAC'25 \\
\revise{AgenticTCAD}~\cite{fan2026agentictcad} & -- &  &  &  &  &  &  & \revise{DATE'26} \\
\revise{PCBgen3D}~\cite{di2026pcbgen3d} & -- &  &  & \cmark &  &  &  & \revise{DAC'26} \\
\revise{Smart-PCLib}~\cite{di2026smartpclib} & -- &  &  & \cmark &  &  &  & \revise{DATE'26} \\
\bottomrule
\end{tabular}
\end{surveytable}
\end{table*}

\subsection{Other Transformations}
This residual group consolidates seven systems whose human-originated primary input--output pairs do not form a sufficiently large, uniform family. Their NL specifications produce configurations, executable EDA scripts, or component and library artifacts. The combined table preserves every exact input--output pair. The following task-level discussions keep their distinct consumers and validation criteria explicit. \Cref{tab:extra-objects-other} summarizes these systems and their object roles.

\minisection{Architecture and Chiplet Design-Space Exploration}
Architecture and chiplet design-space exploration maps workload objectives to candidate microarchitectural configurations and ranks them by measured or predicted QoR.
\textit{Typical systems.} Chat-A2~\cite{zhu2026chata2} assigns CPU bottleneck analysis and configuration changes to architecture-specialist agents. ChatArch~\cite{wu2025chatarch} combines retrieved microarchitecture knowledge with Bayesian optimization. AgenticDSE~\cite{zhu2026agenticdse} alternates design-space refinement, surrogate updating, and multi-phase Bayesian optimization for chiplet accelerators. Their papers compare against default configurations, black-box optimization, or learned surrogates instead of manual design alone. In these systems, agents revise the design space and interpret bottlenecks, while the simulator or surrogate ranks candidates.
\textit{Validation and metrics.} The architecture simulator or surrogate must evaluate each proposed configuration under the same workload and constraints. Pareto distance, hypervolume, diversity, and final performance, power, and area determine which configurations are returned. Evaluated points, model or simulator calls, and wall-clock time define the selection budget.
\textit{Reported results.} AgenticDSE reports up to a 36.9\% reduction in average distance to the real Pareto front and an 11.5\% increase in explored-front diversity over its XGBoost baseline~\cite{chen2016xgboost,zhu2026agenticdse}.

\minisection{EDA Script Generation}
EDA script generation converts natural-language requirements or documentation into executable inputs for a specific backend.
\textit{Typical systems.} ChatLS~\cite{zheng2025chatls} retrieves logic-synthesis documentation before customizing and executing scripts for Design Compiler, OpenROAD, or Genus. Its evaluation identifies Nangate45, so applicability depends on both tool documentation and cell-library context. AgenticTCAD~\cite{fan2026agentictcad} turns device requirements into Sentaurus TCAD scripts and simulated structures. Both systems produce executable scripts, while their backend-specific documentation, libraries, process data, and device models determine whether those scripts are applicable.
\textit{Validation and metrics.} The target EDA backend must parse and execute the generated script under the documented tool and library context. Simulation or formal results establish task completion, while source applicability and factuality qualify documentation-derived content. Latency, repair calls, and tool calls measure delivery cost, and retrieval relevance applies when retrieved sources support the script.
\textit{Reported results.} ChatLS reports retrieval precision, recall, and F1 of 0.85, 0.80, and 0.82 and obtains the best timing results among its evaluated script-generation models~\cite{zheng2025chatls}. AgenticTCAD reaches IRDS-2024 targets in 4.2 hours and 25 optimization iterations, compared with 7.1 days for human experts using commercial tools~\cite{fan2026agentictcad}.

\minisection{PCB Component and Library Generation}
PCB component and library generation converts specifications and datasheets into 3D models, schematic symbols, or footprints consumable by board-design tools.
\textit{Typical systems.} PCBgen3D~\cite{di2026pcbgen3d} converts datasheet text and drawings into parametric 3D components and checks semantic, visual, and numeric agreement. Smart-PCLib~\cite{di2026smartpclib} generates schematic symbols and footprints using specialized extraction, code, and verification agents. Their baselines include frozen LLM-only pipelines, general-purpose multimodal models, KiCad/PADS workflows, and human or vendor artifacts. Independent dimensional, pin-map, clearance, and library-rule checks determine whether the outputs are usable.
\textit{Validation and metrics.} A board-design tool must import the generated symbol, footprint, or component model before it can be reused. Pin-map and dimensional correctness, ERC/DRC, routability, and semantic or visual agreement establish technical acceptance. Task success, correction rounds, and human review time measure production effort.
\textit{Reported results.} On easy components, Smart-PCLib reports 99.8\% pin/pad-count correctness, 99.8\% pin-name match, and 97.1\% pad overlap. Its first two metrics remain above 98\% on the medium and hard groups~\cite{di2026smartpclib}.
\FloatBarrier

\subsection{Discussion}
\label{sec:extra-discussion}

\noindent\revise{\textbf{Handoff Object Observation.} \Cref{fig:extra-object-role-trends} shows checker outputs in 52 of 63 systems, candidate artifacts in 37, constraints in 30, and explicit reference inputs in only two. All 11 NL-to-Verification Collateral systems report checker outputs, while the nine NL-to-Netlist systems combine constraints and checking in every system, with QoR outputs in eight. Extra-stage systems thus commonly generate and check candidates from specifications without receiving an independent reference object.}

\noindent\revise{\textbf{Mechanism Observation.} Table~\ref{tab:extra-mechanisms} shows at least one agent harness mechanism in 44 of 63 Extra-stage systems (69.8\%) and an explicit loop in 28 (44.4\%), while 14 systems optimize model parameters. Executable feedback from a compiler, simulator, formal engine, or DRC checker supports repeated rejection and revision, whereas document- and specification-centered tasks less often provide an immediate acceptance check. Across the complete corpus, PRO-V-R1~\cite{zhao2026provr1}, drcAgent~\cite{ma2026drcagent}, and HRL-Topo~\cite{vijayaraghavan2026hrltopo} update an LLM or VLM policy through RL. PCBAgent~\cite{chen2025pcbagent} instead trains a separate placement policy while leaving its LLM unchanged, and black-box search ranks candidates without updating model parameters.}

\noindent\revise{\textbf{Validation Observation.} Extra-stage correctness is bounded by the information supplied in the initial specification. Systems such as AssertLLM and ChatSVA translate specifications into properties, so their formal results establish the generated properties but cannot recover requirements absent from the input~\cite{yan2025assertllm,fu2026chatsva}. CorrectBench's self-validation loop and ReflectBench's multi-agent consensus illustrate two ways of checking generated collateral~\cite{qiu2025correctbench,liu2026reflectbench}. Neither by itself supplies independently authored expected behavior. Hidden tests, human-written properties, or golden models provide a separate reference, while ambiguous requirements require explicit clarification or a record of the unresolved assumption.}

\section{Towards Trustworthy Agentic EDA: Some Research Directions}
\label{sec:challenges}

\begin{figure*}[t!]
\centering
\includegraphics[width=\textwidth]{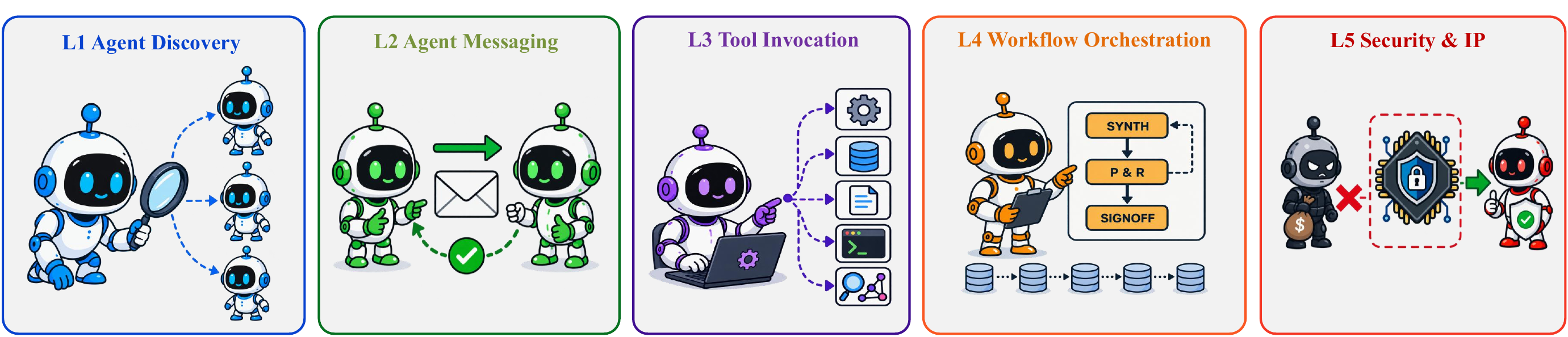}
\caption{Overview of the five-layer Agentic EDA Protocol Roadmap.}
\Description{Five adjacent cards identify the roadmap layers: L1 agent discovery, L2 agent messaging, L3 tool invocation, L4 workflow orchestration, and L5 security and IP.}
\label{fig:eacp-layers}
\end{figure*}

Sections~3--5 reveal two requirements for trustworthy agentic EDA. First, shared infrastructure must identify capabilities, exchange typed handoff objects, invoke tools, preserve state, and enforce security and IP boundaries. Second, research must establish sufficient independent evidence, semantic preservation, fair comparison, source applicability, and human authority. Section~6.1 develops the first as the five-layer Agentic EDA Protocol Roadmap in \Cref{fig:eacp-layers}; Section~6.2 presents the second as open questions. Together, they connect protocol design with dependable engineering decisions.

\subsection{Agentic EDA Protocol Roadmap}
\label{sec:eacp-directions}

\noindent\revise{\textbf{Limitations of existing standards.} General-purpose agent protocols already separate discovery, messaging, tool access, workflow description, and identity. The ACPs family~\cite{acps_community_2026} (reference implementation of GB/Z~185), Agent2Agent (A2A)~\cite{google_a2a_2025}, and Agent Network Protocol (ANP)~\cite{chang2025anp} provide precedents for layered agent communication~\cite{yang2025agentprotocols,ehtesham2025interop}. Model Context Protocol (MCP)~\cite{anthropic_mcp_spec_2024} supports tool access and messaging transport, OpenAPI~\cite{openapi_spec_2025} describes operation schemas, CWL/\allowbreak WDL~\cite{amstutz2016cwl,openwdl2024} describes portable workflows, and OAuth~2.0/\allowbreak OpenID Connect and SPIFFE/\allowbreak SPIRE~\cite{oauth2_rfc6749,spiffe_spec} provide identity mechanisms. These standards do not record the EDA-specific object roles, acceptance evidence, configuration fingerprints, dependencies, and PDK/IP policies required by the surveyed workflows. The proposed roadmap organizes these missing semantics into five layers, each pairing a recurring coordination gap with a required capability and a validation milestone.}

\noindent\revise{\textbf{Agent discovery (L1).} General-purpose discovery formats do not encode the process node, PDK, library, constraint, and tool assumptions that determine whether an agent can accept a requested handoff. In the corpus, prompts, wrappers, or configurations carry this scope. No shared manifest does~\cite{lai2025analogcoder,yang2025circuit,hu2025uvllm,fu2026chatsva}. The milestone is a versioned capability manifest and registry that verifies compatibility before dispatch and rejects incompatible agents with explainable verdicts.}

\noindent\revise{\textbf{Agent messaging (L2).} General-purpose message envelopes do not preserve EDA-specific context, object roles, constraint versions, or acceptance evidence, so the receiver must reconstruct what the sender should have carried~\cite{wu2024chateda,ghose2025orfs}. The milestone is a shared, versioned handoff message that distinguishes primary and supporting objects from candidates and workflow state, records requests, results, errors, and cancellations, and identifies the evidence used to accept the primary output.}

\noindent\revise{\textbf{Tool invocation (L3).} EDA tools expose checker results, QoR measurements, and generated artifacts through proprietary databases, native APIs, command-line output, and textual reports whose schemas vary across vendors and releases~\cite{wu2024chateda,ghose2025orfs,wang2024chatcpu}. The milestone is a typed operation descriptor with backend-specific adapters that return these supporting outputs with comparable units, artifact references, and error categories across tool versions.}

\noindent\revise{\textbf{Workflow orchestration (L4).} Multi-stage flows often pass plans, scripts, checkpoints, and candidates without a shared representation of their dependencies or configuration state. A later failure can invalidate earlier artifacts without triggering the required recovery~\cite{wu2024chateda,ghose2025orfs}. The milestone is a workflow representation in which each edge identifies its producer, consumer, primary and supporting objects, accepted configuration, and dependencies. Recorded evidence gates each transition, while recovery rules define when to retry, roll back, or revalidate.}

\noindent\revise{\textbf{Security and IP (L5).} EDA collaboration spans foundries, design houses, IP vendors, and tool providers with distinct trust domains. Technical compatibility does not authorize a handoff or make it accountable~\cite{synopsys_agentic_ai_2026,cadence_chipstack_pb_2026,siemens_fuse_agent_2026,talukdar2025llm}. The milestone is a set of authorization, disclosure, retention, and audit semantics that determines whether a handoff is authorized, appropriately disclosed, protected through its lifecycle, and admissible for organizational reuse.}

\subsection{Open Questions}
\label{sec:limitations}

\revise{A protocol can structure communication and workflow state, but it cannot determine whether evidence is sufficient, a checker is independent, or a comparison is fair. The corpus exposes recurring questions about acceptance evidence, semantic preservation, evaluation budgets, source applicability, human authority, and security. These questions define the research problems that must be resolved alongside the protocol roadmap.}

\revise{The benchmark releases and native units cataloged in \Cref{tab:benchmark-landscape} are a starting point for studying these questions, but a benchmark name and final score do not reconstruct an experiment. Reports should also preserve (i) the benchmark split and instance count; (ii) the primary input--output pair and the provenance and visibility of supporting objects; (iii) the model, sampling, and iteration policy; (iv) the EDA backend, technology/PDK, constraints, and seeds; (v) acceptance and downstream QoR results with variation across runs; and (vi) LLM tokens and calls, EDA-tool calls, wall-clock time, and human interventions. With this experimental context established, the following questions identify the evidence, controls, and engineering safeguards that trustworthy agentic EDA still requires.}

\begin{itemize}[noitemsep,topsep=2pt]
\item \textbf{What evidence is sufficient for closure?} A single passing verdict can leave semantic, corner-case, or downstream failures undetected. Compilation-guided RTL repair can establish syntax without functionality~\cite{tsai2024rtlfixer}, generated testbenches can pass simulation while sharing the generator's mistaken interpretation~\cite{qiu2025correctbench}, and DRC repair establishes rule compliance rather than functional intent~\cite{ma2026drcagent}. Research is needed on how compilation, simulation, coverage, formal properties, PPA, DRC, routed results, and human review should be combined, and on whether a harness should stop after a verdict, an iteration budget, or a measured residual-risk threshold.

\item \textbf{How can checking remain independent of generation?} When the same model interprets a specification and generates both an implementation and its collateral, correlated omissions can make the artifacts mutually consistent but jointly wrong. Assertion and reference-model generators such as AssertLLM and ChipModeler make this dependence explicit because generated collateral participates in the final verdict~\cite{yan2025assertllm,ye2026chipmodeler}. Open problems include recording collateral provenance, separating hidden evaluation material from agent-visible feedback, validating generated properties and reference models, and determining when cross-model review or human signoff provides sufficiently independent evidence.

\item \textbf{How should semantic preservation be verified across representations?} Synthesis success does not prove C/RTL equivalence, legal placement does not prove routed PPA, and script execution does not prove end-to-end constraint continuity. These limitations are visible in HLS-generation, macro-placement, and RTL-to-GDSII agents, whose reported checks cover different parts of the intended input--output relationship~\cite{chenwei2024hlspilotllmbasedhigh,yiming2026chipmapacollaborativ,wu2024chateda}. Scalable co-simulation, equivalence, invariant transfer, and downstream revalidation are needed for software-to-hardware, netlist-to-layout, and multi-stage workflow handoffs.

\item \textbf{What constitutes a fair agentic baseline?} An agent may receive a stronger LLM, more tool calls, privileged diagnostics, or a larger search budget than the compared method. Results reported on VerilogEval and RTLLM already show that pass rate depends strongly on task set, model, sampling, and feedback policy~\cite{liu2023verilogeval,lu2024rtllm,tsai2024rtlfixer}. Evaluation should pair a same-backbone direct-prompt baseline with the strongest task-specific non-agentic method, hold backend and budgets constant, and separately ablate planning, retrieval, memory, role decomposition, and feedback. Shared benchmark protocols are still missing for success rate, QoR, variance, cost, latency, human effort, and residual risk.

\item \textbf{How should validation effort scale with design and search cost?} LLM inference and repeated EDA-tool invocations become expensive on large designs, while context limits prevent an agent from observing the entire artifact or history. Current responses include decomposing the artifact through hierarchical RTL generation and searching over compact workflow descriptions, but neither provides a calibrated stopping rule~\cite{li2025autosilicon,wei2026vflow}. Open directions include hierarchical state summaries with verifiable links to source artifacts, risk-aware allocation of tool calls, incremental checking, and stopping policies parameterized by measured residual uncertainty and verification cost.

\item \textbf{How can external specifications and knowledge remain applicable?} Extra-stage systems must expose ambiguity and request clarification before selecting an unsupported interpretation. Retrieval-based systems must also establish that each source matches the active tool, version, PDK, library, and design context. ChatLS and VeriRAG condition generation on manuals or knowledge graphs, but their downstream checks do not establish the applicability of every retrieved statement~\cite{zheng2025chatls,thangellamudi2026verirag}. Needed mechanisms include clarification triggers, source-scope metadata, freshness policies, conflict detection, and uncertainty statements that identify what the receiver must still verify.

\item \textbf{When and how should humans intervene?} Human review is often the implicit final gate, but clarification, preference elicitation, exception approval, and signoff are distinct actions. LayoutCopilot separates designer intent from tool-mediated syntactic and functional validation in an interactive workflow~\cite{liu2025layoutcopilot}. Studies should measure human effort and specify who may revise a plan, override a tool verdict, accept residual risk, or authorize an irreversible action. The central design problem is to assign human judgment and accountability to decisions that cannot yet be represented mechanically.

\item \textbf{How can standardized handoffs preserve security and IP boundaries?} Each surveyed system defines bespoke wrappers, messages, and orchestration conventions, while cross-organization deployment must also control disclosure, retention, authorization, and audit. Hardware-security analyses of LLM-assisted design already identify prompt leakage, untrusted generated code, and verification exposure as concrete risks~\cite{talukdar2025llm}. The protocol roadmap addresses the common substrate, but reference implementations must demonstrate that standardized handoffs can enforce least privilege and provenance without exposing PDK data, proprietary IP, or confidential tool reports.
\end{itemize}

\revise{Together, these questions define a research agenda for trustworthy agentic EDA. It requires evidence that composes across stages, controlled and reproducible comparisons, uncertainty-aware orchestration, sources whose applicability is recorded, explicit human authority, and secure coordination across systems. The five roadmap layers address the shared representation and coordination substrate. Answering the evaluation and trust questions additionally requires community benchmarks and cross-system evaluations.}

\section{Conclusion}
\label{sec:conclu}

\revise{This survey applies a handoff-based lens to 115 agentic EDA systems and derives the Intra-stage, Inter-stage, and Extra-stage taxonomy from each system's primary input--output pair. Across the corpus, runtime agent harnesses, especially feedback loops around external EDA tools, are substantially more common than model-parameter optimization. Current work therefore concentrates on coupling models with executable EDA feedback. Two limitations constrain the available evidence. Generated verification collateral can be mutually consistent yet jointly wrong when the implementation and its checks inherit the same interpretation, and reported gains cannot be compared reliably across unmatched models, tool budgets, backends, and evaluation collateral. The five-layer Agentic EDA Protocol Roadmap organizes the missing shared infrastructure. Together, the taxonomy, object-role and mechanism inventories, benchmark catalog, and roadmap provide a common basis for this agenda: controlled baselines and replayable environments should isolate agent contributions, residual risk should accompany accepted handoffs, and cross-stage exchange should keep artifacts and supporting evidence traceable, auditable, and reusable throughout the design flow.}

\bibliographystyle{IEEEtran}
\bibliography{ref/ref}

\clearpage
\onecolumn

\end{document}